\PassOptionsToPackage{usenames,dvipsnames}{color}
\documentclass[manuscript=article,journal=jpcafh]{achemso}
\usepackage{graphicx} % Required for inserting images
\usepackage{chemformula} % Formula subscripts using \ch{}
\usepackage[T1]{fontenc} % Use modern font encodings
\usepackage[usenames,dvipsnames]{color}
\usepackage[colorlinks,linkcolor=blue,urlcolor=red,citecolor=red]{hyperref}
\usepackage{textcomp}
\usepackage{soul}
\usepackage{comment}
\usepackage{tabularx}
\usepackage{mathtools}
\usepackage{float}
\usepackage{amsmath}
\usepackage{amssymb}
\usepackage {ctable}
\usepackage{setspace}
\usepackage{mhchem}
\definecolor{LightCyan}{rgb}{0.88,1,1}
\DeclareUnicodeCharacter{2009}{\,}

\usepackage{makecell}
\usepackage{multirow}
\usepackage{newfloat}
\DeclareFloatingEnvironment[
    fileext=loa,
    listname=List of Algorithms,
    name=ALGORITHM,
    placement=tbhp,
]{algorithm}
\hypersetup{
colorlinks=true,
linkcolor=blue,
filecolor=magenta,      
urlcolor=red,
citecolor=red,
pdftitle={Frag-PBC},
}
\DeclareFloatingEnvironment[name=Box,within=none,placement=tbp]{CGBox}

\DeclarePairedDelimiter\ket{\lvert}{\rangle}

\DeclarePairedDelimiterX\braket[2]{\langle}{\rangle}{#1 \delimsize\vert #2}
\DeclarePairedDelimiterX\braopket[3]{\langle}{\rangle}{#1 \delimsize\vert #2 \delimsize\vert #3}

\mathchardef\mhyphen="2D

\newcommand*{\addFileDependency}[1]{
  \typeout{(#1)}
  \@addtofilelist{#1}
  \IfFileExists{#1}{}{\typeout{No file #1.}}
}
\makeatother

\usepackage{caption, lscape}
\title{Scalable Ab Initio Electronic Structure Methods with Near Chemical Accuracy for Main Group Chemistry}

\author{Wei, Y.}
\affiliation{Department of Chemistry, Columbia University, New York, New York 10027, United States}
\alsoaffiliation[Coauthor]
{Equal contribution}
\author{Debnath, S.}
\alsoaffiliation[Coauthor]
{Equal contribution}
\author{Weber, J. L.}
\author{Mahajan, A.}
\author{Reichman, D. R}
\author{Friesner, R. A.}
\email{raf8@columbia.edu}
\affiliation{Department of Chemistry, Columbia University, New York, New York 10027, United States}
\begin{document}

\begin{abstract}

This study evaluates the precision of widely recognized quantum chemical methodologies, CCSD(T), DLPNO-CCSD(T) and localized ph-AFQMC, for determining the thermochemistry of main group elements. DLPNO-CCSD(T) and localized ph-AFQMC, which offer greater scalability compared to canonical CCSD(T), have emerged over the last decade as pivotal in producing precise benchmark chemical data. Our investigation includes closed-shell, neutral molecules, focusing on their heat of formation and atomization energy sourced from four specific small molecule datasets. Firstly, we selected molecules from the G2 and G3 datasets, noted for their reliable experimental heat of formation data. Additionally, we incorporate molecules from the W4-11 and W4-17 sets, which provide high-level theoretical reference values for atomization energy at 0~K. Our findings reveal that both DLPNO-CCSD(T) and ph-AFQMC methods are capable of achieving a root-mean-square deviation (RMSD) of less than 1 kcal/mol across the combined dataset, aligning with the threshold for chemical accuracy. Moreover, we make efforts to confine the maximum deviations within 2 kcal/mol, a degree of precision that significantly broadens the applicability of these methods in fields such as biology and materials science.

\end{abstract}

\newpage
%%%%%%%%%%%%%%%%%%%%%%%%%%%%%%%%%%%%%%%%%%%%%%%%%%%%%%%%%%%%%%%%%%%%%
%% Start the main part of the manuscript here.
%%%%%%%%%%%%%%%%%%%%%%%%%%%%%%%%%%%%%%%%%%%%%%%%%%%%%%%%%%%%%%%%%%%%%
%\setcounter{section}{1}
\section{1. Introduction}

Quantum chemical methods have seen significant improvements in accuracy and computational efficiency when applied to the chemistry of main group elements over the last three decades. Density functional methods, which scale formally with system size $N$ as $N^4$ or $N^3$ and in practice as $N^2$ or even $N$ (for very large systems), can routinely be applied to systems with hundreds to thousands of atoms, with the best functionals providing an average unsigned error on the order of 2--3 kcal/mol for atomization energies of small molecules.\cite{mardirossian2017thirty} Alternatively, coupled cluster with perturbative triples (CCSD(T)) has been integrated into numerous quantum chemistry software packages in a robust manner, offering average error rates close to 1 kcal/mol for atomization energies of the same category of small molecules; however, the computational cost scales as $N^7$.\cite{raghavachari1989fifth,raghavachari1985augmented} Benchmark methodologies that incorporate higher-order coupled cluster terms and more elaborate treatment of core-valence interactions are capable of producing results reliably within 1 kcal/mol deviation from experimental values. \cite{karton2011w4} However, the high computational demand of these approaches limits their application to very small molecular systems.

While the accuracy that can be achieved with modern DFT approaches is extremely impressive, DFT calculations on large data sets reveal a significant number of outliers with errors significantly larger than the 2--3 kcal/mol cited above as the average unsigned error. Importantly, outliers can be obscured when large data sets are used and only the average errors are reported.\cite{mardirossian2017thirty} In a subsequent paper, we will examine in detail the outlier distribution obtained for a range of modern functionals when compared to benchmark methods and curated experimental results. For the present purposes it is sufficient to note that more work remains to be done to improve the robustness of DFT approaches across a wide range of chemistry, even for main group molecules. Moreover, systems containing transition metals can be prone to a higher incidence of outliers.\cite{quintal2006benchmark}

A consequence of the above observation is that high-level wavefunction-based approaches remain highly relevant in practical applications despite the significantly greater computational cost as compared to DFT calculations. When addressing the grand challenge of understanding the chemistry of complex systems via quantum chemistry, the initial step typically involves conducting a comprehensive set of DFT calculations to explore various possible structures and reaction mechanisms. However, in refining the results to select the correct reaction mechanism (for example) and in general to achieve chemical accuracy, the ability to do benchmark-level wavefunction-based calculations would be extremely valuable. Furthermore, accurate wavefunction calculations are the best path forward, via the production of benchmark training data sets,  to developing improved DFT methods in which the magnitude and frequency of outliers are substantially diminished.

However, in order for wavefunction-based methods to effectively address complex systems, an approach is needed which scales better with system size than the \(N^7\) of conventional CCSD(T). The past decade has seen the development of two notable methods that address this need, both leveraging the concept of orbital localization—a technique tracing back to Pulay’s work in the 1980s.\cite{saebo1985local} The first is localized coupled cluster (e.g. L-CCSD(T)), the most widely used implementation of which is the DLPNO algorithm of Neese and coworkers.\cite{neese_revisiting_2011,Neese2013dlpnoccsd,Neese2013dlpnoccsdt} The formal scaling of DLPNO-CCSD(T) is $N^3$~\cite{Neese2015dlpno} and extremely impressive timing and accuracy numbers over a wide range of systems (and particularly those restricted to main group chemistry) have been published in the past 5 years.\cite{kermani2023barrier,ballesteros2021coupled,santra2022s66x8,villot2022coupled,calbo2017dlpno,santra2022performance} DLPNO and related methods (for example other local CCSD(T) methods such as LNO-CCSD(T)~\cite{Rolik2013} and PNO-CCSD(T)~\cite{MaExplicity2018}) represent a revolution in quantum chemical technology as it is now possible to obtain something close to CCSD(T) quality results for systems containing on the order of 100 atoms.

The second approach is auxiliary field quantum Monte Carlo (AFQMC). The AFQMC algorithm was originally developed in the physics community, but it is only in the past 5 years that significant progress has been made in creating a scalable version of the methodology for the \textit{ab initio} study of molecules. There are several different implementations currently in use.~\cite{Shee2023, joonho_ipie,QMCPACK,sanshardice}. In the present paper, we will focus on two of these implementations. The first is a GPU implementation developed in our groups~\cite{SheeGPU} that exploits localized orbitals~\cite{Weber_Vuong_Devlaminck_Shee_Lee_Reichman_Friesner_2022} in a fashion similar to that employed in L-CCSD(T), reducing the formal scaling of AFQMC from $N^4$ to $N^3$. We refer to this approach as L-AFQMC. The second~\cite{AnkitTaming,AnkitSelected} is a CPU-based code optimized to enable systematic convergence of the bias in AFQMC energies to near-exact accuracy by using a large number of determinants in the trial function. We designate this implementation as W-AFQMC since it is based on use of the generalized Wick’s theorem. As yet, this method has not been formulated in a localized representation, although work in that direction is ongoing.\cite{kurian2023toward} To avoid high computational costs, here we employ W-AFQMC to resolve discrepancies identified in DLPNO-CCSD(T) and L-AFQMC data.

While the computational cost scaling of L-AFQMC and L-CCSD(T) with system size is similar, L-CCSD(T) is considerably faster due to a smaller prefactor. The advantage of AFQMC is that it is formally exact in the limit of the exact trial wavefunction, and in practice, multireference electronic states can often be readily converged due to the ease of utilizing a 
multiconfigurational trial function.\cite{neugebauer2023toward,shee2019achieving,rudshteyn2022calculation,rudshteyn2020predicting,AnkitSelected} This is more crucial for transition metal containing systems than for main group molecules, but there are still main group cases where AFQMC can achieve demonstrably greater accuracy with scalable trial wavefunctions.\cite{weber_silico_2021,lee2022twenty} While such trials allow systematic convergence of the bias in phaseless AFQMC (ph-AFQMC), this accuracy comes at the cost of greater computational expense. Designing protocols for generating trials that strike a desired balance between accuracy and cost is an active area of research. Based on our testing with the benchmark sets, we present two approaches with different cost-accuracy trade-offs. 

We believe that having two scalable benchmark methodologies with distinct theoretical frameworks offers substantial benefits. These benefits extend not only to the generation of data for evaluating and parameterizing DFT functionals but also to their direct application to challenging systems, like the manganese cluster in Photosystem II. As an example, in one of our recent papers we investigated reactions of organolithium systems relevant to lithium ion batteries.\cite{debnath2023accurate}  We performed DLPNO-CCSD(T) and L-AFQMC calculations to look at both reaction energies and barrier heights. The agreement between the two approaches was remarkably good (a few kcal/mol) across the various reactions that we investigated. We were therefore able to settle on benchmark numbers and use those to evaluate many different DFT functionals, discovering that only a few were able to reproduce the benchmark results reliably. We were then able to use the preferred functionals in computing energies for a large set of organolithium cluster geometries, which we then utilized in parametrizing a machine learning force field (MLFF) for carrying out simulations of lithium ion battery electrolytes.\cite{stevenson2023machine}  While there was no reason, \textit{a priori}, to doubt the performance of DLPNO-CCSD(T) for these systems, there is very little experimental data available for comparisons, and the validation by a second independent benchmark approach provided a much higher degree of confidence in the results than would otherwise have been possible. 

In the present paper, our goal is to provide an assessment of the accuracy of both DLPNO-CCSD(T) and L-AFQMC for main group chemistry atomization energies. We have chosen to focus on atomization energies because (a) a relatively large and reliable data set of benchmark experimental and theoretical values is available for a range of small molecules and (b) atomization energies are one of the most difficult properties for electronic structure methods to compute to high precision, due to the large changes in correlation energy upon atomization, and the relatively minimal cancellation of error. We combine four data sets: the G2\cite{Curtiss_Raghavachari_Redfern_Pople_1997,curtiss1991gaussian} and G3\cite{Curtiss_G3,curtiss2002gaussian} sets
of Pople and coworkers, which claim to have experimental atomization energies that are accurate to 1 kcal/mol or better, and the W4-11~\cite{Karton_W411} and W4-17~\cite{Karton_W417} (TAE - total atomization energy) data sets of Karton, Martin and coworkers, which employ very high level (and hence expensive) theoretical methods to achieve the same level of reliability. The W4-17 set, the latest iteration of the W4 sets, is an extension of the W4-11 set, as is G3 an extension of G2 with the addition of larger molecules. W4-17, however, is restricted to molecules with no more than 8 heavy atoms. On the other hand, the G3 data set contains larger molecules than the W4-17 heavy atom limit and therefore provides a test of how the quantum chemical methods perform for larger systems. There are (as far as we have been able to ascertain) no cases where the errors in the W4-11 and W4-17 results exceed the proposed error bars. In contrast, we have had to update a number of the G2 and G3 experimental values with more recent benchmark values. 

Our objective with regard to accuracy is to limit the outliers to a maximum of 2 kcal/mol deviation from the experimental or W4 theoretical values. While 2 kcal/mol is not what is generally meant by “chemical accuracy” (that terminology is conventionally reserved for a 1 kcal/mol accuracy level), it is likely to be sufficient accuracy for choosing among alternative reaction mechanisms in complex systems or parametrizing new functionals. An examination of the details of the W4-17 approach suggests that it is going to be difficult if not impossible (at least with current computing capabilities) to construct a \textit{scalable} approach that reliably achieves 1 kcal/mol precision. Our belief is that the maximum 2 kcal/mol level of error that we are aiming for will be good enough not only to analyze chemical reactions in complex systems but also for designing novel chemistries to address a variety of problems in biology and materials science. This is superior to any current DFT functional, where the average errors of the best functional for the current data set are in the range of 2--3 kcal/mol, but the maximum errors are on the order of 8--10 kcal/mol (as we will enumerate in a subsequent publication).

The paper is organized as follows. In Section 2, we discuss the data sets including our updating of a number of the reference values in the G2 and G3 sets. We then discuss computational methods including our scalable formulation of AFQMC, DLPNO-CCSD(T), and CCSD(T) methodology (enabling comparisons for the larger systems), basis sets, core-valence corrections, scalar relativistic corrections, extrapolation to the complete basis set limit, and treatment of atomic energies. Section 3 presents the results for all three methods as compared to the relevant experimental or theoretical benchmarks and discusses their implications. In Section 4, we conclude by summarizing our results and discussing future directions. In general, the performance of both DLPNO-CCSD(T) and L-AFQMC are quite robust, with only a few apparent outliers above our targeted 2 kcal/mol threshold. We investigate these outliers using more accurate trials with the W-AFQMC method. This enables us to identify outliers arising from the approximations in the methodology, as opposed to cases that are most likely errors or uncertainties in a few of the experimental reference data. Using this targeted convergence of ph-AFQMC, we are able to produce high-quality atomization energies while minimizing cost over a large dataset.

\section{2. Methods}

\subsection{2.1. Datasets}

In this study, we have limited our selection to closed-shell, neutral molecules, excluding carbenes. Future work will investigate open-shell systems and ions.  In total, 116 molecules are selected from G2 set and 73 molecules from the G3 set respectively. Moreover, the W4 sets also contain significant overlap with molecules from the G2 and G3 sets. After removing duplicates, we are left with 38 molecules from the W4-11 set and 32 molecules from the W4-17 set. The G3 and W4-17 extensions consist of generally larger molecules. Therefore, for the purpose of this study, such separation into “G2”, “G3”, “W4-11”, and “W4-17” is constructive, with the former two sets using experimental heat of formation as reference and the latter two using W4 theory as reference. In total, we have compiled a list of 259 unique molecules across all datasets, with the full list of molecules and the dataset into which each molecule is sorted given in the SI Section 1.

\subsection{2.2. Reference Values}

\begin{table}
    \centering
    \small
    \begin{tabular}{|c|l|c|c|c|} \hline 
         Molecule &Dataset&  Previous $\Delta_fH{(298 \,\text{K})}$&  Updated $\Delta_f H{(298 \,\text{K})}$& Source of update\\ \hline 
 AlCl$_3$
& G2& -139.7$\pm0.7$& -142.0&W4\\ \hline 
 AlF$_3$& G2& -289$\pm0.6$& -290.7&W4\\ \hline 
 CCl$_2$CCl$_2$& G2& -3.0$\pm0.7$& -5.1$\pm0.2$&ATcT\\\hline 
         CF$_2$CF$_2$
 &G2&  -157.4$\pm0.7$&  -161.1$\pm0.1$
& ATcT\\ \hline 
         CH$_2$CH-CN
 &G2&  43.2$\pm0.4$&  44.7$\pm0.2$
& ATcT\\ \hline 
         Cyclobutene
 &G2&  37.4$\pm0.4$&  38.5$\pm0.1$
& ATcT\\ \hline 
 Cyclopropene& G2& 66.2$\pm0.6$& 67.8$\pm0.1$&ATcT\\ \hline 
         LiF
 &G2&  -80.1&  -81.45$\pm2$
& CCCBDB\\ \hline 
         Vinyl chloride
 &G2&  8.9$\pm0.3$&  5.2$\pm0.06$
& ATcT\\ \hline 
         Azulene &G3&  69.1$\pm0.8$&  73.6
& CCCBDB \\ \hline 
         Benzoquinone &G3&  -29.4$\pm0.8$&  -28.7& CCCBDB \\ \hline 
          Tetramethylsilane&G3&  -55.7$\pm0.7$&  -51.7$\pm0.5$& ATcT\\ \hline
    \end{tabular}
    \caption{New vs old reference values for heat of formation. The reference values in the column `Previous  $\Delta_fH{(298 \,\text{K})}$' are the same as used in the original G2/G3 test set papers.\cite{curtiss1991gaussian,curtiss2002gaussian} All values are reported in kcal/mol. Where the source reports an experiment uncertainty, we have included the uncertainty in the table along with the value.}
    \label{tab:experiment_changes}
\end{table}

Reference values for the W4 sets are high-level theoretical atomization energies at 0 K, excluding zero-point energy (ZPE) (labeled as the property TAEe by Karton et al.~\cite{Karton_W417}). The reference values for the G2 and G3 sets are experimental heats of formation at 298 K, and the same as those used by Curtiss et al.~\cite{Curtiss_Raghavachari_Redfern_Pople_1997,Curtiss_G3}. However, there are some exceptions where we have found conflicting values, as summarized in Table~\ref{tab:experiment_changes}, where each of the sources of the updated reference is listed. The reported reference values from ATcT\cite{ATcT,atct1130} postdate the G2 and G3 papers. Furthermore, ATcT collates the most recent experiments and theory from various sources using a self-consistent approach and numerous different reactions\cite{Ruscic2004} and is readily updated~\cite{Welch2019}. Therefore, where ATcT data is available and conflicting with the reference used by Curtiss et al., we instead use the ATcT value. In addition, for the cases of AlCl$_3$ and AlF$_3$, we found disagreement between the experimental heats of formation and the W4 reported atomization energy. Moreover, the heat of formation values of these molecules are not present in the ATcT database. Thus, for the updated reference values of these molecules, we have converted the W4 atomization values into heats of formation at 298 K using temperature corrections from DFT (refer to Section 2.9 for details). For the three cases of LiF, azulene, and benzoquinone, where no ATcT or W4 value is available, we use the references reported by the NIST database CCCBDB\cite{cccbdb} instead of those reported by Curtiss. As shown in Table~\ref{tab:experiment_changes}, the two sources also give conflicting heats of formation for these three molecules. All of our benchmark wavefunction methods yield results that are within 2 kcal/mol of the latter source, rather than the former reference values. Although this choice is not based on information about the reference alone, CCSD(T) and AFQMC offer independent evaluations of the experimental data, especially in the case of conflict. In particular, we converge AFQMC with respect to the number of determinants for these cases using W-AFQMC. We refer the reader to Section 2.9 for the method of obtaining deviations against reference values.

\subsection{2.3. Phaseless AFQMC Formulation}
Provided that an initial state $\ket{\phi_i}$ has a nonzero overlap with the exact ground state of a system $\ket{\phi_0}$, then the the ground state can be projected from any such trial state as
\begin{equation}
    \ket{\phi_0} \propto \lim_{\tau \rightarrow \infty} e^{-\tau \hat{H}} \ket{\phi_i},
\end{equation}
where \(\tau\) is the imaginary time, and $\hat{H}$ is the electronic Hamiltonian under the Born-Oppenheimer approximation which can be written as a sum of one-electron and two-electron terms,
\begin{equation}
    \hat{H} = \hat{H}_1 + \hat{H}_2 = \sum_{pq}h_{pq}c_p^{\dagger}c_q + \frac{1}{2}\sum_{pqrs}V_{pqrs}c_p^{\dagger}c_q^{\dagger}c_sc_r.
\end{equation}
$h_{pq}$ are one-electron integrals and $V_{pqrs} = \left(pr |qs \right) = \braket[]{pq}{rs}{}{}$ (chemists' and physicists' notation respectively) are two-electron integrals. Numerically, we propagate
\begin{equation}
    \ket{\phi(\tau + \Delta \tau)} = e^{-\Delta \tau \hat{H}} \ket{\phi(\tau)},
\end{equation}
where \(\lvert\phi(0)\rangle=\lvert\phi_i\rangle\). The one-body and two-body terms in $\hat{H}$ can be split using the Trotter-Suzuki decomposition,
\begin{equation}
    e^{-\Delta \tau (\hat{H}_1 + \hat{H_2})} \approx e^{-\Delta \tau \hat{H}_1/2}e^{-\Delta \tau \hat{H}_2/2}e^{-\Delta \tau \hat{H}_1/2} + O(\Delta \tau^3),
\end{equation}
which introduces an error that scales with the timestep. The Hubbard-Stratonovic transformation and the phaseless approximation (see below) also induce timestep errors. We show in SI Section 2 that for this work, the timestep error converges at around $\Delta \tau=$0.005 Ha$^{-1}$ for frozen-core calculations, and at around 0.001 Ha$^{-1}$ for all-electron calculations. We note that in the calculation of AFQMC energy differences, there is some approximate cancellation of time-step errors.

The two-body operators can be decomposed as the sum of the square of one-body operators through Cholesky decomposition or density fitting. The Hubbard-Stratonovich transformation then converts an exponential with two-body operators into a multidimensional integral over fluctuating auxiliary fields, $x_\alpha$,
\begin{equation}
    e^{-\Delta \tau (\sum_\alpha \hat{L}_\alpha^2)/2} = \prod_\alpha \int_{-\infty}^{\infty}\frac{1}{\sqrt{2\pi}}e^{-x_\alpha^2/2}e^{\sqrt{\Delta \tau }x_\alpha \hat{L}_\alpha} \text{d}x_{\alpha} + O(\Delta \tau^2),
\end{equation}
and we arrive at
\begin{equation}
    \ket{\phi(\tau + \Delta \tau)} = e^{-\Delta \tau \hat{H}} \ket{\phi(\tau)} = \int \text{d}\mathbf{x} p(\mathbf{x}) \hat{B}(\mathbf{x}) \ket{\phi(\tau)},
\end{equation}
where $p(\mathbf{x})$ is a Gaussian probability density function and $\hat{B}(\mathbf{x})$ is a one-body propagator depending on the auxiliary fields $\mathbf{x}$. This multidimensional integral is evaluated using Monte Carlo importance sampling to obtain a stochastic representation of the wave function. For a more in-depth description of AFQMC, we refer the reader to these review articles~\cite{Shi2021,Motta2018}.

Due to the fermionic sign problem, the signal-to-noise ratio generally decays exponentially during the imaginary time propagation. It is possible to eliminate the sign problem using a constraint at the expense of a bias in the resulting energies. In this work, we use a constraint referred to as the phaseless approximation (ph-AFQMC), where the phase of the walkers is restricted according to a trial wavefunction. The bias induced by the trial wavefunction can be systematically reduced by improving its quality, for example, by increasing the active space or number of determinants included in the trial. The bias is formally zero in the limit of the exact trial (see the next subsection, Section 3.3, and SI Section 3 for trial details). We refer the reader to our previous work\cite{Weber_Vuong_Devlaminck_Shee_Lee_Reichman_Friesner_2022} for our approach to localized ph-AFQMC (L-AFQMC) which involves compressing the electron repulsion integrals in the localized orbital basis. Effectively, the scaling for the energy evaluation, the steepest scaling step with system size, is $N^2M + N_\text{det}N^2$ (with a prefactor depending on the compressed electron repulsion integral tensor), where $M$ is the number of basis functions, $N$ is the number of electrons, and $N_{\text{det}}$ is the number of determinants. See SI Section 4 for an estimate of localization error. For the practical deployment of L-AFQMC, we have developed two protocols, AFQMC 0 and AFQMC I, which are discussed in detail below as well as in Section 3.3. AFQMC I is a scalable AFQMC protocol (scaling $\sim N^3$) that achieves an accuracy comparable to that of DLPNO-CCSD(T), albeit with a significantly larger prefactor for the computation time. AFQMC 0 uses a black-box but less elaborate trial function, but is less accurate, particularly for molecules with significant multireference character.  

We also present results for a selected subset of molecules computed with another implementation of AFQMC~\cite{AnkitTaming,AnkitSelected}. This method, which uses a generalized Wick’s theorem approach to efficiently evaluate energies with mutlideterminantal trials, will be referred to in what follows as W-AFQMC.  The advantage of this approach is that it scales as $MN_{\text{det}} + N^2M^2$, which allows the use of a larger number of determinants in the trial wavefunction at an accessible computational cost. It enables one to converge the phaseless bias to the near-exact limit in a given basis set by increasing the number of active space orbitals and determinants included in trial wavefunctions. We use W-AFQMC to calculate energies of the outliers obtained from AFQMC I using on the order of 10,000 determinants and refer to the results in which the L-AFQMC outliers are replaced by the W-AFQMC results as AFQMC II (along with a select few other cases, see SI Section 3). This approach helps us to more confidently address the question, discussed in detail below, as to whether the AFQMC I outliers are due to the phaseless bias or more likely the result of errors in the reference data. We also run apparent DLPNO-CCSD(T) and CCSD(T) outliers with AFQMC II to assess their status.

\subsection{2.4. AFQMC Trial Generation}
In this work, we use a procedure to generate multi-determinant HCI (heat bath configuration interaction) and HCISCF trials for the entire dataset. The CAS family of trials (in this work, we use the HCI solver\cite{DICE1,DICE2}) provides a robust way of including static correlation in the reference of AFQMC and has been shown\cite{Shee2023,neugebauer2023toward} to generally perform more accurately than single determinant trials. Nonetheless, the selection of active space for these methods is non-trivial. Akin to multireference perturbation theory methods, a common practice is to use an active space, usually minimal, based on chemical intuition and to pick the leading determinants from the expansion in this active space. However, this does not systematically provide chemically accurate energies. Here, we generally follow a two-step process to select the active space in a relatively automated way that can be applied to large datasets. First, HCI is performed on a “valence” active space, selected based on the atomic composition of the molecule. See SI Section 3 and main text Section 3.3 for the considerations for selecting this active space. Using the spatial 1-RDM of the resulting state, we calculate the natural orbitals and their occupation numbers (NOONs) in the natural orbital basis. We choose a subset of active orbitals from this set based on a NOON threshold (\(\delta\)) as \(\delta \leq \text{NOON} \leq 2 - \delta\). This procedure is often used to flag the more correlated group of orbitals in quantum calculations. A second HCI calculation (or HCISCF for W-AFQMC trials, refer to the SI Section 3) is then performed with the second smaller active space, and this forms the final trial wavefunction. Generally, we choose the number of determinants necessary to retain 99.5\% of the CI weight from this final trial wavefunction unless indicated otherwise. Briefly, AFQMC 0 is fully automated and uses L-AFQMC, and thresholds are chosen to be loose, and for the outliers with AFQMC 0, AFQMC I combines trials with larger active spaces, and AFQMC II in turn combines trials with even larger active spaces run with W-AFQMC. This progression gives some indication of the number of determinants required to converge each molecule, but not much more than necessary. For more details, refer to the discussion in Section 3 and SI Section 3.3.

All L-AFQMC trials were generated with the PySCF package\cite{PYSCF} where we obtain the Hamiltonian and electron repulsion integrals. The HCI trials are obtained using Dice\cite{DICE1,DICE2,DICE_PYSCF} in conjunction with PYSCF. L-AFQMC energy is measured in blocks of size 0.1 Ha$^{-1}$ of 20 timesteps each (each timestep being 0.005 Ha$^{-1}$). In total, we propagate for between 2000 and 3000 such blocks for each molecule, with 1920 total walkers. W-AFQMC calculations used the same time-step of 0.005 Ha\(^{-1}\) for molecules containing only first row atoms, and used a time step of 0.0025 Ha\(^{-1}\) for those with heavier atoms. For W-AFQMC, we use 250 walkers and propagate for 1000 blocks of 50 timesteps each.

\subsection{2.5. CCSD(T) and DLPNO-CCSD(T)}
CCSD(T) and DLPNO-CCSD(T) calculations are carried out using the ORCA package~\cite{neese2012orca} using restricted Hartree-Fock (RHF) as the reference state. DLPNO-CCSD(T) correlation energies are extrapolated to the complete pair natural space (PNO) using the procedure in Altun et al.\cite{NeesePNO_extrap} and employing TightPNO thresholds\cite{Liakos_Sparta_Kesharwani_Martin_Neese_2015}, between T$_{\text{CutPNO}}$ thresholds of $10^{-6}$ and $10^{-7}$ for each basis set used. The matching auxiliary basis set is used if available, otherwise, the AutoAux~\cite{Stoychev_Auer_Neese_2017} functionality is used. Where linear dependence is encountered with AutoAux, we increase the even-tempered expansion factor for the s-shell from 1.8 to 2.0. 

\subsection{2.6. Basis Sets}
We use the following basis sets: aug-cc-pVXZ-DK (X = T, Q) for atoms with atomic number less than or equal to that of oxygen, and aug-cc-pCVXZ-DK (X = T, Q) for fluorine and heavier, obtained using the Basis Set Exchange database.\cite{dunning1989a,jong2001a,kendall1992a,prascher2011a,woon1993a,schuchardt2007a,pritchard2019a} This choice is motivated by the documented improvement of basis set convergence when using core-valence or tight-d functions in the basis set for second-row elements (Na-Cl)\cite{Martin1998_cpl,Martin1998_jpc, Bauschlicher1998, Bauschlicher1995} as well as fluorine\cite{Feller2003}. While the split-valence aug-cc-pVXZ basis sets are not designed for core-valence correlation~\cite{PetersonAccurate}, we find that all-electron calculations using these same basis sets can reach a respectable (especially for CCSD(T)) albeit overall inferior accuracy (especailly for AFQMC) to the frozen core calculations supplemented with MP2 core corrections, as discussed briefly in Section 3, and in more in detail in SI Section 5.

We extrapolate all single point energies to the complete basis set (CBS) limit according to the method of Neese and Valeev\cite{Neese_Extrapolation} for T/Q extrapolation for both Hartree-Fock and correlation energy, with $\alpha$ and $\beta$ matching the basis set used. We use the same coefficients $\alpha$ and $\beta$ for the core-valence sets aug-cc-pCVXZ-DK as the corresponding aug-cc-pVXZ-DK basis sets, where we use the aug-cc-pVXZ coefficients reported by Neese and Valeev. This CBS procedure is used by all the methods investigated, with the exception of the more expensive CCSD(T) and W-AFQMC where we use alternative schemes (see SI Section 3 and Section 6).

\subsection{2.7. Frozen Core Corrections}
Frozen core calculations are carried out according to freezing no orbitals for H-Be, 1s orbitals for B-Mg, and 1s and 2p orbitals for Al-Ar. We correct the core-valence energy using MP2,
\begin{equation}
   \Delta_{\text{CV}} = E_{\text{CC-MP2}}^{T} - E_{\text{FC-MP2}}^{T} 
\end{equation} 
where CC and FC denote core-correlated and frozen-core calculations, respectively. We used the aug-cc-pCVTZ-DK basis set for both calculations. Note that we freeze the 1s electrons in second-row atoms even in the CC calculations. In the following discussion, we focus on the frozen core calculations (along with the MP2 core correction above) for all four methodologies. All electron results for CCSD(T), DLPNO-CCSD(T), and L-AFQMC are presented and compared with the frozen core results in Tables S9 and S10 of the Supplementary Material. In general we do not see any deterioration in accuracy from the use of the frozen core, and recommend that this approach be used going forward for both AFQMC and CCSD(T) based calculations. 

%We have carried out both frozen core and all-electron calculations for CCSD(T), DLPNO-CCSD(T), AFQMC 0, AFQMC I, but not the more expensive W-AFQMC implementation where we use an ultralarge number of determinants and only frozen core is used. 

\subsection{2.8. Relativistic Effects}
Scalar relativistic effects are included through the DKH2\cite{Nakajima2000,Mayer2001,Wolf2002} Hamiltonian for DLPNO-CCSD(T) and CCSD(T), and X2C\cite{Liu2009} for ph-AFQMC, as X2C is not implemented in ORCA and DKH2 is not implemented in PySCF. The MP2 core-valence corrections follow the same relativistic corrections for each method, respectively (MP2 is carried out in ORCA for correcting DLPNO-CCSD(T) and CCSD(T) corrections and in PySCF for ph-AFQMC corrections). The difference between DKH2 and X2C energies are shown to be small for HF and MP2 and mostly cancelled out by atomic energies (see SI Section 7). Spin-orbit corrections to atomic energies are applied using the values from Curtiss et al.~\cite{Curtiss_Raghavachari_Redfern_Pople_1997}.

\subsection{2.9. Heat of Formation and Atomization}
Atomization energies ($\sum E_\text{atoms,\textit{g}}-E_\text{molecule,\textit{g}}$) and heats of formation are calculated according to the method in Ref.\cite{Curtiss_Raghavachari_Redfern_Pople_1997,Friesner_Knoll_Cao_2006,Goldfeld_Bochevarov_Friesner_2008}, including geometry optimization and thermochemical properties (ZPE, enthalpy, internal energy) using the DFT functional B3LYP~\cite{becke1993,lyp1988} and basis set 6-31G*~\cite{ditchfield1971a,gaussian09e01,hariharan1973a,francl1982a} using the Jaguar software package~\cite{Jaguar} with the maximum available grid point density. A few molecules required a higher-level geometry optimization (see SI Section 8). After optimization, DFT single-point energies were calculated with Jaguar. We note that the first step towards obtaining the heat of formation at 298 K is the atomization at 0 K, and the molecule temperature corrections (from DFT), atom temperature corrections (from experiment) and energy to change the atomic states from gas to standard state (from experiment, i.e. heat of formation of the single atom in the gaseous state) are added to achieve the heat of formation. Experimental values for the atomic heats of formation and temperature corrections are the same as that used by Curtiss et al.\cite{Curtiss_Raghavachari_Redfern_Pople_1997}, with the exception of the sulfur atom heat of formation where we use 66.18 kcal/mol from ATcT~\cite{ATcT} instead of 65.66 kcal/mol. 

We emphasize here that as opposed to atomization energy, the heat of formation is defined using standard states of atoms rather than the gas phase state. Nonetheless, these quantities are closely related and in the analysis we convert the W4 atomization energy at 0 K to heat of formation,
\begin{equation}
    \Delta_fH(0\,\text{K}) = \sum_\text{atoms}\Delta_fH_\text{atoms}(0\,\text{K})  - \Delta_a E(0\,\text{K}),
\end{equation}
where $\Delta_a E(0\,\text{K})$ is the atomization energy at 0 K. With the addition of aformentioned temperature correction terms from DFT ($H^{298\text{ K}}-H^{0\text{ K}}$ for the molecule) and experiment ($H^{298\text{ K}}-H^{0\text{ K}}$ for the atom), we obtain $\Delta_fH(298\,\text{K})$. Although these corrections are approximate (even though we expect the errors to be small), they cancel out when obtaining the deviation from the converted W4 heat of formation reference value as the same corrections are applied to the calculated atomization energy of the molecule. Effectively, the deviation $D$ becomes
\begin{equation}
    D = \Delta_f H_{\text{expt}}(298\,\text{K}) - \Delta_f H_{\text{calc}}(298\,\text{K})  
\end{equation}
or
\begin{equation}
    D = -(\Delta_a E_{\text{W4}}(0\,\text{K}) - \Delta_a E_{\text{calc}}(0\,\text{K}))  
\end{equation}
for G2/G3 and W4-11/W4-17 sets, respectively, where the only effect of the conversion of the W4 reference values to heat of formation is a change in sign of the deviation from the reference atomization energy. This simple change ensures we are comparing the same quantities.

\subsection{2.10. Atomic Energies}
The treatment of atoms is essential in achieving the targeted accuracy. An explicit near-exact treatment of core-valence correlation on par with valence correlation, as is done in W4-17, requires expensive core-valence corrections. In AFQMC, similar to other projection QMC approaches, the description of core-valence correlations requires onerous convergence of time-step errors. Furthermore, it has been shown that AFQMC atom energies can be difficult to calculate~\cite{lee2022twenty}. Hence, we converge the atomic energies with a large number of determinants in W-AFQMC. A less expensive and simpler alternative is to fit the values of the atoms to the experimental data, which benefits from cancellation of errors. We use this approach for the other benchmark methods. For small molecules, a relatively inexpensive version of the first approach can be shown to work quite well, however, for larger molecules, any difference in molecule versus atom accuracy for a given method is compounded.
%\textcolor{red}{Cite joonho paper and say atom energies very difficult and move part in intro to here leetwenty}

Atom energies for all methods are fit as free parameters according to the combined G2/G3 set experimental heats of formation (with the addition of AlCl, AlF, AlH, and AlH$_3$ in the W4-11 using the converted reference values to heat of formation); for each respective method we obtain a separate least squares multivariate linear regression fit with atomic energies as parameters, calculated heats of formation as dependent variables, and the loss function is the experimental vs calculated heats of formation,
\begin{equation}
    \Delta_fH_{\text{expt},\text{molecule}} - \Delta_fH_{\text{calc},\text{molecule}} = \sum_{\text{atom}}N_\text{atom} E_\text{atom} + c_\text{molecule},
    \label{eqn:atom_fit}
\end{equation}
where $\Delta_fH_{\text{expt},\text{molecule}}$ and $\Delta_fH_{\text{calc},\text{molecule}}$ are the experimental and calculated heats of formation for that molecule, and $N_\text{atom}$ is the number of the atom constituting the molecule. The energy of the atom, $E_\text{atom}$, is the fitting variable, and $c_\text{molecule}$ are the constant terms (for each molecule, i.e. independent of $E_\text{atom}$) such as the temperature terms, energy of the molecule, and heat of formation of the atoms, used when we minimize the left-hand side of the equation above. We have 189 such correlated equations for the 189 molecules in the G2 and G3 sets.

The initial guess for atom energies in the atomic energy fit is the \textit{ab initio} atom energies obtained through each method respectively. It is worth noting that a small error (i.e. slight imperfection in cancellation of error between atom and molecule energy) in the \textit{ab initio} atom energies is multiplied by the number of atoms in the molecule. While we do add spin-orbit corrections to atomic energies using the values from Curtiss et al.~\cite{Curtiss_Raghavachari_Redfern_Pople_1997}, this only applies to the initial guess and such atom-related corrections will be encompassed in the final fitted atom energy.
%and have corrected the sulfur atom heat of formation value from $65.66$ kcal/mol (as used in the original G2 paper~\cite{Curtiss_Raghavachari_Redfern_Pople_1997}) to 66.18 kcal/mol from ATcT~\cite{ATcT}, we also expect the atom fit to largely cancel errors from such atom-related corrections.

These fitted atom energies are used to calculate the atomization energies for the W4 set. Alternatively, using W-AFQMC, we show that by using around \(10^4\) determinants in a natural orbital active space the deviation from experiment is near-chemical accuracy without fitting atomic energies. However, we do not expect cancellation of error between molecules and atoms in this case, as we expect both molecular and atomic energies to be close to exact within this method. For the present, for methods that are not asymptotically exact like W-AFQMC, we thus recommend the atom fitting approach, even though i) the efficacy depends on the accuracy of a method across the entire dataset where fitting occurs, ii) the data set must be sufficiently large, and iii) the resultant atom energies are specific to the other conditions of the fit (basis set and electronic structure method). The third point also applies to \textit{ab inito} atom energies but to a less tailored extent. The reader is referred to SI Section 9 for the resulting fitted atomic energies.

\section{3. Results and Discussion}

\subsection{3.1. Overall Performance of CCSD(T) and AFQMC-based Methods}

\begin{figure}
\centering
\includegraphics[width=0.8\textwidth]{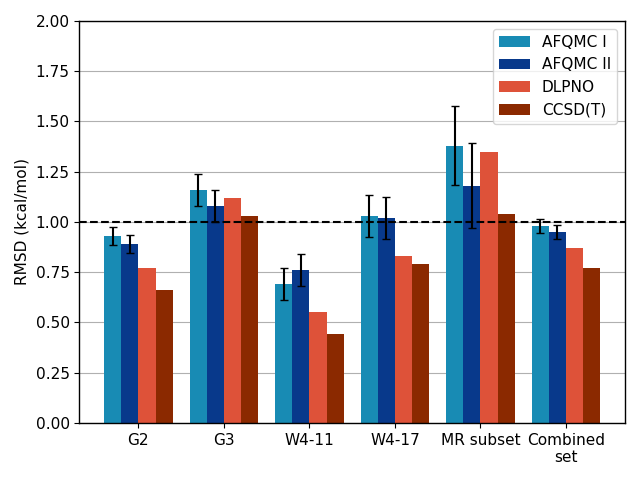}
\caption{ Root-mean-square deviations of the calculated heat of formation of G2, G3 and W4-11/17 datasets with respect to experiment (G2 and G3) or W4 reference values (W4-11, W4-17) for each benchmark method. `DLPNO' refers to DLPNO-CCSD(T). The number of molecules in our mutually exclusive definition of G2, G3, W4-11, and W4-17 are 116, 73, 38, and 32. The separated MR subset refers to the 10 molecules from Table S26, from a combination of G2, W4-11, and W4-17. The combined set consists of the total 259 molecules. The horizontal dashed line at 1 kcal/mol refers to the standard of chemical accuracy.}
\label{fig:RMSD_all}
\end{figure}

We begin by evaluating the overall performance of the four methods discussed in Section 2: CCSD(T), DLPNO-CCSD(T), AFQMC I, and AFQMC II, across our chosen datasets. Figure~\ref{fig:RMSD_all} displays the root mean square deviation (RMSD) in the enthalpies of formation calculated using these methods, as detailed in Section 2. For the combined dataset (represented by the far-right bar in Figure ~\ref{fig:RMSD_all}), the RMSD values for all methods fall within 1 kcal/mol, and as we will see below, all four methods have a very small number of outliers with deviations from benchmark experiments or computations greater than 2 kcal/mol.  We conclude that for main group chemistry, the localized (and hence scalable) versions of both coupled cluster and AFQMC achieve our target of reliably obtaining near-chemical accuracy for chemical transformations, sufficient for elucidating chemical reaction mechanisms in complex systems. Results along these lines have been presented previously for DLPNO-CCSD(T) (although not for as large and extensively curated a data set involving experimental and high-level theoretical references), but not for AFQMC.  The present exercise establishes AFQMC as a robust alternative benchmark quantum chemical methodology, albeit at a higher computational cost than DLPNO-CCSD(T). For the present systems, we find the scaling exponent with system size to be similar between DLPNO-CCSD(T) and L-AFQMC, and the high prefactor is responsible for the cost of L-AFQMC. See SI Section 10 for examples of the computational costs of each method. We refer the reader to more detailed demonstrations of AFQMC and DLPNO-CCSD(T) scaling in these articles~\cite{Weber_Vuong_Devlaminck_Shee_Lee_Reichman_Friesner_2022,kurian2023toward,Neese2015dlpno}.

With regard to the detailed results in Figure~\ref{fig:RMSD_all}, a few comments can be made. Firstly, full CCSD(T) displays the smallest RMS error across all four methods. This is most pronounced for the W4-11 data set, which is not surprising as the benchmark theory used to establish reference values is based on a coupled cluster approach. For the G3 data set, the difference is barely noticeable, reflecting likely performance when comparing with experimental data in practice.  

Secondly, the most difficult data set for all methods is, unsurprisingly, but not guaranteed, the subset of 10 cases that we have identified as “multireference” (MR). We classify molecules as multireference based on a set of diagnostics developed by Karton et. al.\cite{Karton_W411,Karton_W417}, as discussed in more detail in Section 11 of the Supplementary Material.  For the coupled-cluster based methods, only one molecule, ozone, stands out as displaying an error in excess of 2 kcal/mol.  Despite formally being a single reference methodology, the treatment of electron correlation via CCSD(T) appears to be powerful enough to handle many wavefunctions with nontrivial multireference character. DLPNO-CCSD(T) here displays a noticeable (although not large) degradation from full CCSD(T). For AFQMC, an improvement is obtained in the treatment of MR molecules by upgrading the trial function in the AFQMC II approach. Details of results for each quantum chemical method for all of the MR cases can be found in the SI Section 11.

\begin{figure}
    \centering
    \includegraphics[width=0.8\textwidth]{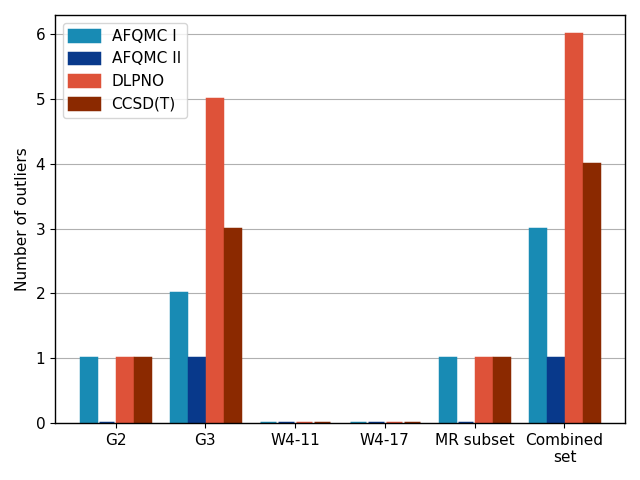}
    \caption{Number of outlier molecules for each method and dataset, where outlier is defined as having larger than 2 kcal/mol deviation in heat of formation from the reference value. `DLPNO' refers to DLPNO-CCSD(T).}
    \label{fig:outliers}
\end{figure}
We next analyze the outliers observed across the different data sets and methods, the number of which are summarized in Figure~\ref{fig:outliers} below. A number of interesting points can be made regarding the outliers, which are enumerated along with results for the various computational methods in Table~\ref{tab:dlpno_ccsdt_outliers}. Firstly, neither AFQMC I, AFQMC II nor the two coupled cluster based methods have any outliers for the W4-11 and W4-17 data sets, where the reference values are taken from ultrahigh level computation. Secondly, while CCSD(T) and DLPNO-CCSD(T) have a larger number of outliers than AFQMC II based on our (somewhat arbitrary) cutoff for the deviation from the reference of 2 kcal/mol, a perusal of the data in Table~\ref{tab:dlpno_ccsdt_outliers} shows that, with the exception of ozone (where we believe that the multireference character is great enough to cause significant errors in CCSD(T) and DLPNO-CCSD(T), which can then be reduced via a large trial function in AFQMC II), the computational results for the five remaining molecules are closer to each other than they are to the experimental reference data. A likely interpretation of the results is that the experiments have a residual error of a few kcal/mol (possibly as large as 3-4 kcal/mol for 3-butyn-2-one), and that in fact, the reliability of our scalable benchmark methods is higher than what is suggested in Figure~\ref{fig:outliers}. In contrast to a number of other cases that initially appeared to be outliers but were resolved by newer experiments, as discussed in Section 2.2, we were unable to find any relatively recent experimental data.  Note that the value of having two distinct computational methods which can be compared, suggested in the Introduction, is already manifested in this analysis.  Without the complementary AFQMC I and AFQMC II results, one might conclude that CCSD(T) has occasional outliers even for single-reference main group molecules and that higher-order treatments are required to achieve even the 2 kcal/mol accuracy threshold that we have set. 

\begin{table}
    \centering
    \begin{tabular}{|c|l|c|c|l|l|} \hline 
         Molecule&Dataset&  CCSD(T)& DLPNO-CCSD(T)& AFQMC I&AFQMC II
\\ \hline 
         Ozone&G2, MR&   -2.07& -3.24& -2.57(25)&-0.96(35)
\\ \hline 
         Pyrazine&G3&   -1.96& -2.52& -3.11(77)&-1.74(53)
\\ \hline 
         3-Butyn-2-one&G3&   -3.35& -3.57& -4.68(55)&-4.60(69)
\\ \hline 
         Cl$_2$O$_2$S&G3&   -3.53& -3.43& -1.98(69)&-1.99(78)
\\ \hline 
         Cyclooctatetraene&G3&   -1.99& -2.42& -1.43(65)&-1.49(104)
\\ \hline 
         Pyrimidine&G3&   2.40& 2.49& 1.76(55)&1.76(55)
\\\hline
    \end{tabular}
    \caption{All combined outliers for CCSD(T), DLPNO-CCSD(T), AFQMC I and AFQMC II and their respective deviations against reference heat of formation are reported in kcal/mol.}
    \label{tab:dlpno_ccsdt_outliers}
\end{table}

Having summarized the overall performance of our various methodologies, we next examine more carefully the differences between the scalable (DLPNO-CCSD(T) and AFQMC I) and benchmark (CCSD(T) and AFQMC II) versions of our two high level wavefunction based approaches. For the vast majority of molecules in the present data sets, equivalent results are obtained. However, it is useful to examine the cases where there are noticeable differences to see whether a systematic explanation is possible. 

\subsection{3.2. Comparison of CCSD(T) and DLPNO-CCSD(T) Results }

Table~\ref{tab:10_dlpno_ccsdt} below presents the 10 molecules with the largest deviations between the CCSD(T) and DLPNO-CCSD(T) results, in order of the size of the deviations (see SI Section 12 for a correlation plot). The interesting point here is that most of these molecules are classified as MR by our diagnostic criteria. This observation suggests that the DLPNO localization scheme may have more difficulties as the MR character of the wavefunction increases. Having said that, the deviations are in general quite small (and in some cases the DLPNO-CCSD(T) results are not clearly inferior to CCSD(T) when comparing with the reference value). We would view the question as to whether the performance of DLPNO-CCSD(T) for main group MR molecules is a significant source of concern (beyond the general question of the accuracy of the underlying CCSD(T) approximation) as a subject for future investigation. 

\begin{table}
    \centering
    \begin{tabular}{|c|l|c|c|c|c|} \hline 
         Molecule &Dataset&  DLPNO-CCSD(T)&  CCSD(T)&  Difference
& MR?\\ \hline 
         S$_4$ &W4-11&  -1.26&  0.66&  1.92
& Yes\\ \hline 
         N$_2$O$_4$ &W4-17&  0.03&  1.67&  1.64
& Yes\\ \hline 
         Ozone &G2&  -3.24&  -2.07&  1.17
& Yes\\ \hline 
         BN &W4-11&  -0.53&  0.85&  1.38
& Yes\\ \hline 
         ClF$_5$ &W4-17&  -0.93&  0.12&  1.05
& Yes\\ \hline 
         C$_2$ &W4-11&  -1.40&  -0.54&  0.86
& Yes\\ \hline 
         Ph-Cl &G3&  -0.60&  0.21&  0.81
& No\\ \hline 
         S$_3$ &W4-11&  0.05&  0.82&  0.77
& Yes\\ \hline 
         Ph-CH$_3$ &G3&  -1.15&  -0.42&  0.74
& No\\ \hline
 Benzoquinone &G3& -1.40& -0.69& 0.72&No\\\hline
    \end{tabular}
    \caption{DLPNO-CCSD(T) and CCSD(T) deviations in kcal/mol, against the reference heats of formation. The difference between DLPNO-CCSD(T) and CCSD(T) deviations are also reported in kcal/mol. The multireference (MR) criteria is according to Table S26.}
    \label{tab:10_dlpno_ccsdt}
\end{table}

\subsection{3.3. Detailed Discussion of ph-AFQMC Methodology}

In contrast to our coupled cluster based calculations, for which we utilize well established methods implemented by the Neese group in ORCA, the scalable AFQMC I protocol discussed above  required significant novel methodology development. We therefore discuss AFQMC I development and implementation in detail in what follows. Specification of the AFQMC II protocol given AFQMC I as a starting point is straightforward, using the general procedure described in Section 2.4.

We first perform an initial run of the 259 molecules in our datasets with relatively small initial active spaces (AS), including only the valence electrons and 4 orbitals per atom (excluding hydrogen). Additionally, we set a loose NOON (natural orbital occupation number) cutoff at 0.01, allowing for the selection of active orbitals in the second SHCI step with NOONs ranging from 0.01 to 1.99. The initial and final active spaces chosen for every molecule are listed in the SI and more details about the procedure can be found in SI Section 3. Although this relatively cheap procedure results in around 80\% of molecules being run with 1 determinant trials, and on average 2 determinants (maximum 70 determinants), it performs sufficiently well such that 88\% of the molecules achieve an unsigned deviation of less than 2 kcal/mol, and achieves an RMSD of 1.67 and MAD (mean absolute deviation) of 1.02 and across the entire combined dataset. We denote this procedure as AFQMC 0.

For the G2 dataset, the RMSD is 1.19 kcal/mol and the MAD is 0.88 kcal/mol. Similarly, for the G3 dataset, the RMSD and MAD are 1.27 kcal/mol and 0.99 kcal/mol, respectively, both of which are quite respectable. However, the performance of AFQMC 0 significantly declines for W4 datasets, with an RMSD of 3.04 kcal/mol for W4-11 and 1.70 kcal/mol for W4-17, with respective MADs of 1.26 kcal/mol and 1.33 kcal/mol. This reduced level of accuracy primarily stems from the enhanced presence of multireference molecules (8/10 from Table S26) in the W4 data sets. The RMSD is furthermore skewed by the presence of a few very large outliers. The 30 outliers for AFQMC 0 are enumerated in detail in Section 13 of the Supplementary Material. A few of the largest errors are listed below in Table~\ref{tab:init_afqmc_outliers}. 

\begin{table}
    \scriptsize
    \centering
    \begin{tabular}{|c|l|c|c|c|c|c|c|} \hline 
         Molecule &Dataset&  Deviation&  First CI AS&  TZ final AS&  QZ final AS& \thead{TZ final \\ \#dets} & \thead{QZ final \\ \#dets}
\\\hline
 BN&W4-11, MR& -10.51(61)& 4e+4e,8o& 1e+1e,2o& 1e+1e,2o& 2&2
\\\hline
        C$_2$&W4-11, MR& -14.62(43)& 4e+4e,8o& 1e+1e,2o& 1e+1e,2o& 2&2
\\ \hline 
         Ozone&G2, MR&  -5.05(51)& 9e+9e,12o& 2e+2e,3o& 2e+2e,3o&  3& 3
\\ \hline 
         3-Butyn-2-one&G3&  -4.27(57)& 11e+11e,20o& 1e+1e,2o& 1e+1e,2o&  2& 2
\\ \hline 
         ClCOF&W4-17&  3.00(72)& 12e+12e,16o& 1e+1e,1o& 1e+1e,1o&  1& 1
\\ \hline 
         Dioxetan2one&W4-17&  3.04(80)& 13e+13e,20o& 1e+1e,1o& 1e+1e,1o&  1& 1
\\ \hline 
         ClF$_5$&W4-17, MR&  3.13(104)& 21e+21e,24o& 1e+1e,2o& 1e+1e,1o&  2& 1
\\ \hline 
         OCS&G2&  3.25(67)& 8e+8e,12o& 1e+1e,1o& 1e+1e,1o&  1& 1
\\ \hline 
         LiF&G2&  3.38(39)& 4e+4e,5o& 1e+1e,1o& 1e+1e,1o&  1& 1
\\ \hline
 Pyrimidine&G3& 3.51(53)& 13e+13e,24o& 2e+2e,4o& 1e+1e,3o& 6&1
 \\ \hline 
         HClO$_4$&W4-17&  4.16(82)& 15e+15e,20o& 1e+1e,1o& 1e+1e,1o&  1& 1
\\\hline
    \end{tabular}
    \caption{The top outliers from AFQMC 0 protocol. Deviations (($\Delta_f H_\text{expt}(298\, \text{K}) - \Delta_f H_\text{calc} (298\,\text{K})$) for G2/G3 and $-$($\Delta_a E_\text{W4} (0\,\text{K}) - \Delta_a E_\text{calc} (0\,\text{K})$) for W4, see Section 2.9) are listed in kcal/mol with statistical errors in parentheses. After the first CI is performed with an active space (AS) based on orbital maps to the atoms of the molecules (refer Table S4) that returns the `First CI AS’ listed, the second AS (shown here as `TZ final AS’ and `QZ final AS’, as the NOONs have a slight basis set dependency due to approximations such as the SHCI solver) is chosen from those orbitals from the first AS that have NOONs of between 0.01 and 1.99. The final number of determinants is set by the number of determinants required to get to 99.5\% saturation of the CI coefficients.}
    \label{tab:init_afqmc_outliers}
\end{table}

In the AFQMC methodology, the standard approach to address outliers (including those of increasing MR character) is to create a better trial function, using for example an expanded active space as well as more determinants. To address the 17 outliers identified in the G2 and G3 sets and the 13 outliers in the W4-11 and W4-17 sets, we recalculated energies by expanding the valence space by one shell and applying a stricter NOON threshold. This approach reduced the number of outliers to 6 in G2/G3 (bicyclobutane, ozone, Li$_2$, LiF, pyrazine, 3-butyn-2-one) and 4 in W4-11/W4-17 (BN, C$_2$, N$_2$O$_4$, silole).

For these remaining outliers, we apply further modifications, starting with a further increase of the initial active space and the adjusting of the NOON threshold (see SI Section 3 for details). This procedure successfully reduced the list of outliers to only ozone, pyrazine, and 3-butyn-2-one for G2/G3 and none for W4-17. This improves the MAD from 1.02 kcal/mol to 0.78 kcal/mol and RMSD from 1.67 kcal/mol to 0.98 kcal/mol. As noted above, pyrazine and 3-butyn-2-one experimental values are potentially suspect, which suggests that ozone is the only real outlier in AFQMC I, which has among the highest multireference character in the dataset.

In summary, AFQMC I starts from AFQMC 0 and successively increases the active space for the outliers (starting with increasing the initial active space and tightening the NOON threshold if one wants to keep the CI \% retained similar) and hence, the number of determinants. The combination of the best trials fall under “AFQMC I” (see SI for a full list).  The aim behind running the dataset in a progressive fashion and only applying larger orbital maps (i.e. orbitals per atom) and stricter thresholds to outliers is a compromise. The goal is achieving useful AFQMC results with close to minimal number of determinants necessary for each molecule, while also reducing manual processes in the selection of active spaces to enable generation of a large amount of benchmark data. AFQMC 0, the fully automated protocol with loose thresholds, performs decently, but by using larger orbital maps for a small percentage of the molecules AFQMC I results in a large improvement of error over the entire dataset.

Under circumstances where the reference is unknown, typically having a converged energy with respect to, for example, determinants\cite{neugebauer2023toward} (See SI Section 3 for an example) gives confidence in the AFQMC benchmark value unless the CI expansion is qualitatively wrong. This process can become expensive, and based on our heuristic we have observed some guidelines for which type of trial and whether convergence is necessary for certain types of molecules. Observing the AFQMC 0 outliers, a few categories of molecules stand out: i) multireference molecules ii) small (< 4 atoms) molecules containing Li, F, or S atoms, iii) conjugated systems, iv) strained systems, and v) halogen oxoacids. The only exceptions to these are diethyl ketone (deviation 2.22 kcal/mol) and HNCO (deviation 2.63 kcal/mol). After an expanded valence space to two instead of one shell, and a 0.001 instead of 0.01 NOON threshold, the remaining real outliers as discussed above mostly fall into the multireference category. Therefore, for main group thermochemistry, for the above categories of outliers (except multireference), we recommend AFQMC 0 with these alternative thresholds. On the other hand, we still recommend that multireference molecules be converged with respect to the active space size and number of determinants.

Where performing calculations with more than 3600 determinants with L-AFQMC does not converge the absolute deviation to < 2 kcal/mol, we perform W-AFQMC calculations, using on the order of \(10^4\) determinants. As discussed above, we designate the resulting data set, in which the outlier results from AFQMC I are replaced by W-AFQMC derived values, as AFQMC II (along with a few other molecules with e.g. experimental discrepancies, see SI Section 3 Tables S6 and S7 for a full list). The net result is that only 3-butyn-2-one is an outlier for the entire dataset of 259 molecules with AFQMC II, with a deviation of -4.6 kcal/mol. As this value is within 1.5 kcal/mol of all of the other methods (L-AFQMC, CCSD(T), and DLPNO-CCSD(T)), in addition to this molecule not satisfying any of the multireference criteria, and furthermore having no experimental value from ATcT or theoretical value from W4, it seems highly likely that the experimental value may require updating. Furthermore, as noted above, the fact that CCSD(T), DLPNO-CCSD(T), and AFQMC I results are quite close to W-AFQMC results for the remaining outliers in Table 2 increases confidence that the discrepancies with the experimental reference values for these molecules are also due to experimental error.  The ability to perform W-AFQMC calculations for this subset of cases is critical to our suggestion that experimental error is a likely explanation for the deviations of the remaining methods.

Finally, we have investigated the accuracy of the AFQMC I (but not AFQMC II) protocols using all-electron calculations. While this is generally expected to be less accurate due to deficiencies in the aug-cc-pVXZ split-valence Dunning basis sets for correlating core electrons (and to some extent aug-cc-pCVXZ for correlating 1s), we find that overall the all-electron calculations still display an MAD of $\sim 1$ kcal/mol for the entire dataset, although with more outliers. Interestingly, although the timestep error is much larger for all electron than frozen core and does not cancel between atoms and molecules, the atom-fit for the same timestep (we used 0.005 Ha $^{-1}$) demonstrates an impressive cancellation of error even though more molecules require larger trials to be run in order to reduce the relative timestep error. We refer the reader to the SI Section 5 for a more detailed discussion of frozen vs non-frozen results, as well as timestep errors in SI Section 2. 

\section{4. Conclusion}

In this study, we have investigated the performance of three benchmark-level wavefunction approaches — CCSD(T), DLPNO-CCSD(T), and ph-AFQMC—in the context of main group element thermochemistry. The study highlights the ability of the more scalable DLPNO-CCSD(T) and localized ph-AFQMC to achieve accuracies remarkably close to canonical CCSD(T), showcasing their significance in the generation of accurate benchmark chemical data in recent years. The results demonstrate that both DLPNO-CCSD(T) and ph-AFQMC methods consistently deliver RMSDs of below 1 kcal/mol across these selected datasets, adhering to the standard of chemical accuracy, as well as a maximum error of 2 kcal/mol across the entire data set, excepting one or two cases. These above observations highlight the potential of ph-AFQMC as a robust benchmark method that is able to produce accurate results for the small molecules tested here, and is also promising for larger and more challenging systems.

The G2 and G3 test sets and the W4 sets are chosen on account of the readily available and accurate reference values. Further critical investigation of scalable benchmark methods such as DLPNO-CCSD(T) and AFQMC for larger systems is valuable, with the difficulty of such investigation being the lack of accurate experimental references and the computational expense of generating benchmark-level calculations for large datasets of such systems. In particular, having the AFQMC method at disposal would be valuable in cases where one does not expect CCSD(T) to perform well, for example multireference systems or non-equilibrium geometries and bond breaking. Regardless, the current studies in the literature are moving towards that direction.~\cite{neugebauer2023toward,Neese_Extrapolation}. The multireference diagnostic used in this study is by no means exhaustive, and investigation of multireference character for a comparison of DLPNO-CCSD(T), AFQMC and other scalable methods (for example, other L-CCSD(T) methods~\cite{SemidalasMOBH35} and composite methods~\cite{ChanSimple2023,SemidalasCanG42024}) is illustrative for the purposes of ascertaining the potential for evaluating challenging systems.

While we have shown that we can achieve an accuracy of < 2 kcal/mol for virtually all the molecules tested here by increasing the sophistication of the ph-AFQMC trial, finding the most compact trial wavefunction is a challenging multifaceted direction that is ongoing in the ph-AFQMC community. Furthermore, while we have semi-automated the trial generation for ph-AFQMC, there are still non-black-box elements. An algorithm that can find the most compact trial for every molecule in a black-box manner is highly desired but elusive at this point in time. Additionally, alternative AFQMC constraints and algorithms are also being explored in the literature to increase accuracy and decrease the computational cost. AFQMC is developing at a rapid rate, and moving forward, the improvements in implementation and protocol will cement this method as a powerful tool for electronic structure.

Finally, the thorough benchmarking conducted in this study is crucial for establishing benchmark datasets that evaluate the performance of DFT functionals. This will also aid the development of correction schemes aimed at enhancing the accuracy DFT by significantly reducing both the magnitude and frequency of outliers. We discuss this in detail in our subsequent work.

\begin{suppinfo}
The following files are available free of charge.
\begin{itemize}
\item Supporting Information: Molecule list, timestep errors, additional computational details, multireference diagnostics, additional outliers, atom energies, discussion for no frozen core (.pdf)
\item Supporting Information: Deviations from experiment for all molecules, trials for AFQMC 0 and AFQMC I for all molecules (.xlsx)
\item Supporting Information: xyz coordinates of all molecules (.zip)
\end{itemize}
\end{suppinfo}

\begin{acknowledgement}
We thank Hung T. Vuong for valuable discussions and critical contributions to the AFQMC code implementation. We thank Zach K. Goldsmith for assistance with computation and helpful discussions. We thank Hong-Zhou Ye for valuable insights. The computational work was supported by OLCF INCITE 2022 and 2024. The authors acknowledge support by Gates Ventures. A.M. and D.R.R. were partially supported by NSF CHE-2245592. W-AFQMC calculations were performed on the Delta system at the National Center for Supercomputing Applications through allocation CHE230028 from the Advanced Cyberinfrastructure Coordination Ecosystem: Services and Support (ACCESS) program, which is supported by National Science Foundation grants \#2138259, \#2138286, \#2138307, \#2137603, and \#2138296.
\end{acknowledgement}

\begin{tocentry}

\includegraphics[width=\textwidth]{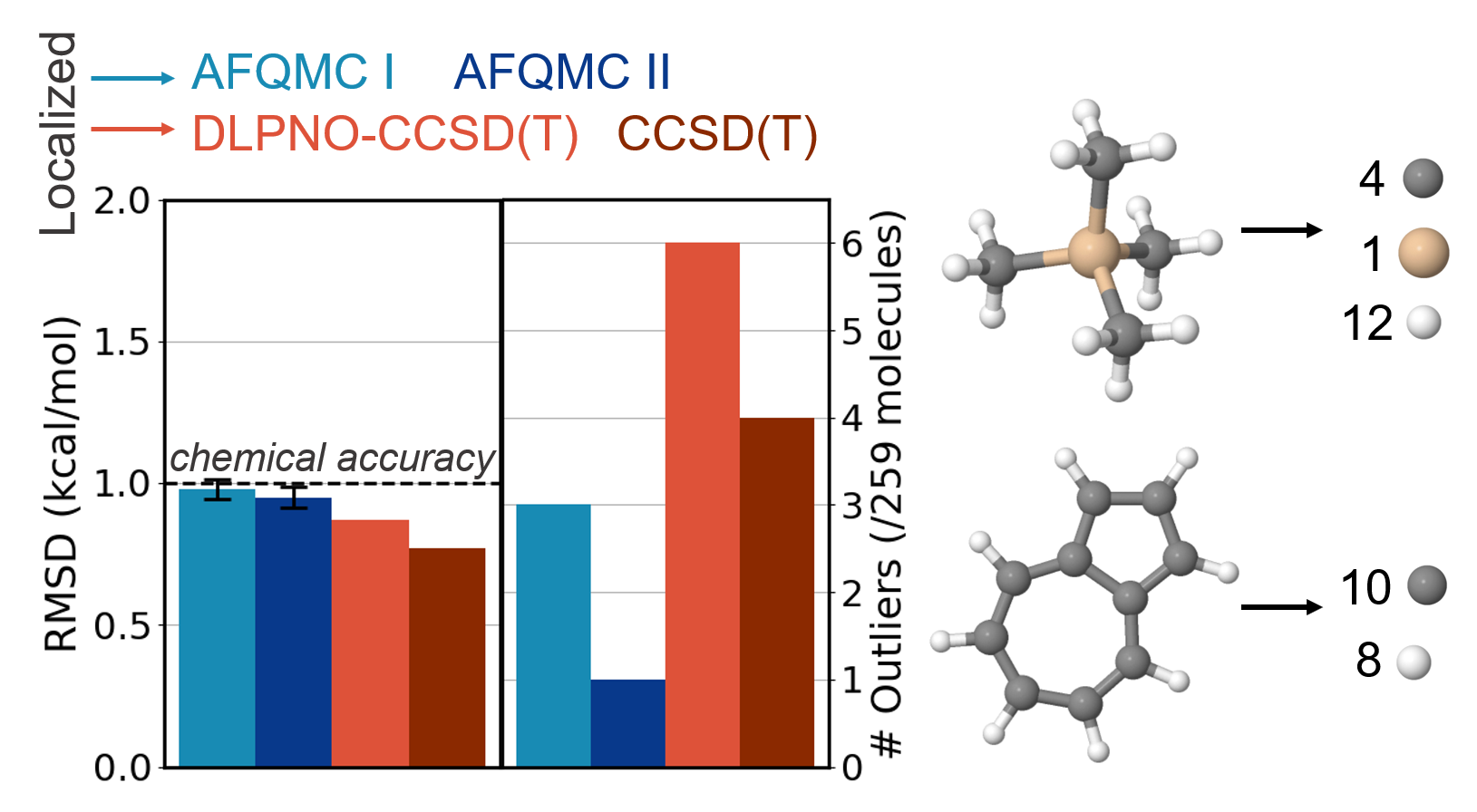}

\end{tocentry}

\bibliography{bibli}

\providecommand{\latin}[1]{#1}
\makeatletter
\providecommand{\doi}
  {\begingroup\let\do\@makeother\dospecials
  \catcode`\{=1 \catcode`\}=2 \doi@aux}
\providecommand{\doi@aux}[1]{\endgroup\texttt{#1}}
\makeatother
\providecommand*\mcitethebibliography{\thebibliography}
\csname @ifundefined\endcsname{endmcitethebibliography}  {\let\endmcitethebibliography\endthebibliography}{}
\begin{mcitethebibliography}{87}
\providecommand*\natexlab[1]{#1}
\providecommand*\mciteSetBstSublistMode[1]{}
\providecommand*\mciteSetBstMaxWidthForm[2]{}
\providecommand*\mciteBstWouldAddEndPuncttrue
  {\def\EndOfBibitem{\unskip.}}
\providecommand*\mciteBstWouldAddEndPunctfalse
  {\let\EndOfBibitem\relax}
\providecommand*\mciteSetBstMidEndSepPunct[3]{}
\providecommand*\mciteSetBstSublistLabelBeginEnd[3]{}
\providecommand*\EndOfBibitem{}
\mciteSetBstSublistMode{f}
\mciteSetBstMaxWidthForm{subitem}{(\alph{mcitesubitemcount})}
\mciteSetBstSublistLabelBeginEnd
  {\mcitemaxwidthsubitemform\space}
  {\relax}
  {\relax}

\bibitem[Mardirossian and Head-Gordon(2017)Mardirossian, and Head-Gordon]{mardirossian2017thirty}
Mardirossian,~N.; Head-Gordon,~M. Thirty years of density functional theory in computational chemistry: an overview and extensive assessment of 200 density functionals. \emph{Mol. Phys.} \textbf{2017}, \emph{115}, 2315--2372\relax
\mciteBstWouldAddEndPuncttrue
\mciteSetBstMidEndSepPunct{\mcitedefaultmidpunct}
{\mcitedefaultendpunct}{\mcitedefaultseppunct}\relax
\EndOfBibitem
\bibitem[Raghavachari \latin{et~al.}(1989)Raghavachari, Trucks, Pople, and Head-Gordon]{raghavachari1989fifth}
Raghavachari,~K.; Trucks,~G.~W.; Pople,~J.~A.; Head-Gordon,~M. A fifth-order perturbation comparison of electron correlation theories. \emph{Chem. Phys. Lett.} \textbf{1989}, \emph{157}, 479--483\relax
\mciteBstWouldAddEndPuncttrue
\mciteSetBstMidEndSepPunct{\mcitedefaultmidpunct}
{\mcitedefaultendpunct}{\mcitedefaultseppunct}\relax
\EndOfBibitem
\bibitem[Raghavachari(1985)]{raghavachari1985augmented}
Raghavachari,~K. An augmented coupled cluster method and its application to the first-row homonuclear diatomics. \emph{J. Chem. Phys.} \textbf{1985}, \emph{82}, 4607--4610\relax
\mciteBstWouldAddEndPuncttrue
\mciteSetBstMidEndSepPunct{\mcitedefaultmidpunct}
{\mcitedefaultendpunct}{\mcitedefaultseppunct}\relax
\EndOfBibitem
\bibitem[Karton \latin{et~al.}(2011)Karton, Daon, and Martin]{karton2011w4}
Karton,~A.; Daon,~S.; Martin,~J.~M. W4-11: A high-confidence benchmark dataset for computational thermochemistry derived from first-principles W4 data. \emph{Chem. Phys. Lett.} \textbf{2011}, \emph{510}, 165--178\relax
\mciteBstWouldAddEndPuncttrue
\mciteSetBstMidEndSepPunct{\mcitedefaultmidpunct}
{\mcitedefaultendpunct}{\mcitedefaultseppunct}\relax
\EndOfBibitem
\bibitem[Quintal \latin{et~al.}(2006)Quintal, Karton, Iron, Boese, and Martin]{quintal2006benchmark}
Quintal,~M.~M.; Karton,~A.; Iron,~M.~A.; Boese,~A.~D.; Martin,~J.~M. Benchmark study of DFT functionals for late-transition-metal reactions. \emph{J. Phys. Chem. A} \textbf{2006}, \emph{110}, 709--716\relax
\mciteBstWouldAddEndPuncttrue
\mciteSetBstMidEndSepPunct{\mcitedefaultmidpunct}
{\mcitedefaultendpunct}{\mcitedefaultseppunct}\relax
\EndOfBibitem
\bibitem[S{\ae}b{\o} and Pulay(1985)S{\ae}b{\o}, and Pulay]{saebo1985local}
S{\ae}b{\o},~S.; Pulay,~P. Local configuration interaction: An efficient approach for larger molecules. \emph{Chem. Phys. Lett.} \textbf{1985}, \emph{113}, 13--18\relax
\mciteBstWouldAddEndPuncttrue
\mciteSetBstMidEndSepPunct{\mcitedefaultmidpunct}
{\mcitedefaultendpunct}{\mcitedefaultseppunct}\relax
\EndOfBibitem
\bibitem[Neese and Valeev(2011)Neese, and Valeev]{neese_revisiting_2011}
Neese,~F.; Valeev,~E.~F. Revisiting the {Atomic} {Natural} {Orbital} {Approach} for {Basis} {Sets}: {Robust} {Systematic} {Basis} {Sets} for {Explicitly} {Correlated} and {Conventional} {Correlated} \textit{ab initio} {Methods}? \emph{J. Chem. Theory Comput.} \textbf{2011}, \emph{7}, 33--43\relax
\mciteBstWouldAddEndPuncttrue
\mciteSetBstMidEndSepPunct{\mcitedefaultmidpunct}
{\mcitedefaultendpunct}{\mcitedefaultseppunct}\relax
\EndOfBibitem
\bibitem[Riplinger and Neese(2013)Riplinger, and Neese]{Neese2013dlpnoccsd}
Riplinger,~C.; Neese,~F. {An efficient and near linear scaling pair natural orbital based local coupled cluster method}. \emph{J. Chem. Phys.} \textbf{2013}, \emph{138}, 034106\relax
\mciteBstWouldAddEndPuncttrue
\mciteSetBstMidEndSepPunct{\mcitedefaultmidpunct}
{\mcitedefaultendpunct}{\mcitedefaultseppunct}\relax
\EndOfBibitem
\bibitem[Riplinger \latin{et~al.}(2013)Riplinger, Sandhoefer, Hansen, and Neese]{Neese2013dlpnoccsdt}
Riplinger,~C.; Sandhoefer,~B.; Hansen,~A.; Neese,~F. {Natural triple excitations in local coupled cluster calculations with pair natural orbitals}. \emph{J. Chem. Phys.} \textbf{2013}, \emph{139}, 134101\relax
\mciteBstWouldAddEndPuncttrue
\mciteSetBstMidEndSepPunct{\mcitedefaultmidpunct}
{\mcitedefaultendpunct}{\mcitedefaultseppunct}\relax
\EndOfBibitem
\bibitem[Liakos and Neese(2015)Liakos, and Neese]{Neese2015dlpno}
Liakos,~D.~G.; Neese,~F. Is It Possible To Obtain Coupled Cluster Quality Energies at near Density Functional Theory Cost? Domain-Based Local Pair Natural Orbital Coupled Cluster vs Modern Density Functional Theory. \emph{J. Chem. Theory Comput.} \textbf{2015}, \emph{11}, 4054--4063\relax
\mciteBstWouldAddEndPuncttrue
\mciteSetBstMidEndSepPunct{\mcitedefaultmidpunct}
{\mcitedefaultendpunct}{\mcitedefaultseppunct}\relax
\EndOfBibitem
\bibitem[Kermani \latin{et~al.}(2023)Kermani, Li, Ottochian, Crescenzi, Janesko, Scalmani, Frisch, Ciofini, Adamo, and Truhlar]{kermani2023barrier}
Kermani,~M.~M.; Li,~H.; Ottochian,~A.; Crescenzi,~O.; Janesko,~B.~G.; Scalmani,~G.; Frisch,~M.~J.; Ciofini,~I.; Adamo,~C.; Truhlar,~D.~G. Barrier Heights for Diels--Alder Transition States Leading to Pentacyclic Adducts: A Benchmark Study of Crowded, Strained Transition States of Large Molecules. \emph{J. Phys. Chem. Letters} \textbf{2023}, \emph{14}, 6522--6531\relax
\mciteBstWouldAddEndPuncttrue
\mciteSetBstMidEndSepPunct{\mcitedefaultmidpunct}
{\mcitedefaultendpunct}{\mcitedefaultseppunct}\relax
\EndOfBibitem
\bibitem[Ballesteros \latin{et~al.}(2021)Ballesteros, Dunivan, and Lao]{ballesteros2021coupled}
Ballesteros,~F.; Dunivan,~S.; Lao,~K.~U. Coupled cluster benchmarks of large noncovalent complexes: The L7 dataset as well as DNA--ellipticine and buckycatcher--fullerene. \emph{J. Chem. Phys.} \textbf{2021}, \emph{154}, 154104\relax
\mciteBstWouldAddEndPuncttrue
\mciteSetBstMidEndSepPunct{\mcitedefaultmidpunct}
{\mcitedefaultendpunct}{\mcitedefaultseppunct}\relax
\EndOfBibitem
\bibitem[Santra \latin{et~al.}(2022)Santra, Semidalas, Mehta, Karton, and Martin]{santra2022s66x8}
Santra,~G.; Semidalas,~E.; Mehta,~N.; Karton,~A.; Martin,~J.~M. S66x8 noncovalent interactions revisited: new benchmark and performance of composite localized coupled-cluster methods. \emph{Phys. Chem. Chem. Phys.} \textbf{2022}, \emph{24}, 25555--25570\relax
\mciteBstWouldAddEndPuncttrue
\mciteSetBstMidEndSepPunct{\mcitedefaultmidpunct}
{\mcitedefaultendpunct}{\mcitedefaultseppunct}\relax
\EndOfBibitem
\bibitem[Villot \latin{et~al.}(2022)Villot, Ballesteros, Wang, and Lao]{villot2022coupled}
Villot,~C.; Ballesteros,~F.; Wang,~D.; Lao,~K.~U. Coupled cluster benchmarking of large noncovalent complexes in L7 and S12L as well as the C60 dimer, DNA--ellipticine, and HIV--indinavir. \emph{J. Phys. Chem. A} \textbf{2022}, \emph{126}, 4326--4341\relax
\mciteBstWouldAddEndPuncttrue
\mciteSetBstMidEndSepPunct{\mcitedefaultmidpunct}
{\mcitedefaultendpunct}{\mcitedefaultseppunct}\relax
\EndOfBibitem
\bibitem[Calbo \latin{et~al.}(2017)Calbo, Sancho-Garc{\'\i}a, Orti, and Arago]{calbo2017dlpno}
Calbo,~J.; Sancho-Garc{\'\i}a,~J.~C.; Orti,~E.; Arago,~J. DLPNO-CCSD (T) scaled methods for the accurate treatment of large supramolecular complexes. \emph{J. Comput. Chem.} \textbf{2017}, \emph{38}, 1869--1878\relax
\mciteBstWouldAddEndPuncttrue
\mciteSetBstMidEndSepPunct{\mcitedefaultmidpunct}
{\mcitedefaultendpunct}{\mcitedefaultseppunct}\relax
\EndOfBibitem
\bibitem[Santra and Martin(2022)Santra, and Martin]{santra2022performance}
Santra,~G.; Martin,~J.~M. Performance of localized-orbital coupled-cluster approaches for the conformational energies of longer n-alkane chains. \emph{J. Phys. Chem. A} \textbf{2022}, \emph{126}, 9375--9391\relax
\mciteBstWouldAddEndPuncttrue
\mciteSetBstMidEndSepPunct{\mcitedefaultmidpunct}
{\mcitedefaultendpunct}{\mcitedefaultseppunct}\relax
\EndOfBibitem
\bibitem[Rolik \latin{et~al.}(2013)Rolik, Szegedy, Ladjánszki, Ladóczki, and Kállay]{Rolik2013}
Rolik,~Z.; Szegedy,~L.; Ladjánszki,~I.; Ladóczki,~B.; Kállay,~M. {An efficient linear-scaling CCSD(T) method based on local natural orbitals}. \emph{The Journal of Chemical Physics} \textbf{2013}, \emph{139}, 094105\relax
\mciteBstWouldAddEndPuncttrue
\mciteSetBstMidEndSepPunct{\mcitedefaultmidpunct}
{\mcitedefaultendpunct}{\mcitedefaultseppunct}\relax
\EndOfBibitem
\bibitem[Ma and Werner(2018)Ma, and Werner]{MaExplicity2018}
Ma,~Q.; Werner,~H.-J. Explicitly correlated local coupled-cluster methods using pair natural orbitals. \emph{WIREs Computational Molecular Science} \textbf{2018}, \emph{8}, e1371\relax
\mciteBstWouldAddEndPuncttrue
\mciteSetBstMidEndSepPunct{\mcitedefaultmidpunct}
{\mcitedefaultendpunct}{\mcitedefaultseppunct}\relax
\EndOfBibitem
\bibitem[Shee \latin{et~al.}(2023)Shee, Weber, Reichman, Friesner, and Zhang]{Shee2023}
Shee,~J.; Weber,~J.~L.; Reichman,~D.~R.; Friesner,~R.~A.; Zhang,~S. {On the potentially transformative role of auxiliary-field quantum Monte Carlo in quantum chemistry: A highly accurate method for transition metals and beyond}. \emph{J. Chem. Phys.} \textbf{2023}, \emph{158}, 140901\relax
\mciteBstWouldAddEndPuncttrue
\mciteSetBstMidEndSepPunct{\mcitedefaultmidpunct}
{\mcitedefaultendpunct}{\mcitedefaultseppunct}\relax
\EndOfBibitem
\bibitem[Malone \latin{et~al.}(2023)Malone, Mahajan, Spencer, and Lee]{joonho_ipie}
Malone,~F.~D.; Mahajan,~A.; Spencer,~J.~S.; Lee,~J. ipie: A Python-Based Auxiliary-Field Quantum Monte Carlo Program with Flexibility and Efficiency on CPUs and GPUs. \emph{J. Chem. Theory Comput.} \textbf{2023}, \emph{19}, 109--121\relax
\mciteBstWouldAddEndPuncttrue
\mciteSetBstMidEndSepPunct{\mcitedefaultmidpunct}
{\mcitedefaultendpunct}{\mcitedefaultseppunct}\relax
\EndOfBibitem
\bibitem[Kim \latin{et~al.}(2018)Kim, Baczewski, Beaudet, Benali, Bennett, Berrill, Blunt, Borda, Casula, Ceperley, Chiesa, Clark, Clay, Delaney, Dewing, Esler, Hao, Heinonen, Kent, Krogel, Kylänpää, Li, Lopez, Luo, Malone, Martin, Mathuriya, McMinis, Melton, Mitas, Morales, Neuscamman, Parker, Flores, Romero, Rubenstein, Shea, Shin, Shulenburger, Tillack, Townsend, Tubman, Goetz, Vincent, Yang, Yang, Zhang, and Zhao]{QMCPACK}
Kim,~J.; Baczewski,~A.~D.; Beaudet,~T.~D.; Benali,~A.; Bennett,~M.~C.; Berrill,~M.~A.; Blunt,~N.~S.; Borda,~E. J.~L.; Casula,~M.; Ceperley,~D.~M. \latin{et~al.}  QMCPACK: an open source ab initio quantum Monte Carlo package for the electronic structure of atoms, molecules and solids. \emph{J. Phys. Condens. Matter} \textbf{2018}, \emph{30}, 195901\relax
\mciteBstWouldAddEndPuncttrue
\mciteSetBstMidEndSepPunct{\mcitedefaultmidpunct}
{\mcitedefaultendpunct}{\mcitedefaultseppunct}\relax
\EndOfBibitem
\bibitem[san()]{sanshardice}
Dice Repository. \url{https://github.com/sanshar/Dice} (accessed 2024-04-30)\relax
\mciteBstWouldAddEndPuncttrue
\mciteSetBstMidEndSepPunct{\mcitedefaultmidpunct}
{\mcitedefaultendpunct}{\mcitedefaultseppunct}\relax
\EndOfBibitem
\bibitem[Shee \latin{et~al.}(2018)Shee, Arthur, Zhang, Reichman, and Friesner]{SheeGPU}
Shee,~J.; Arthur,~E.~J.; Zhang,~S.; Reichman,~D.~R.; Friesner,~R.~A. Phaseless Auxiliary-Field Quantum Monte Carlo on Graphical Processing Units. \emph{J. Chem. Theory Comput.} \textbf{2018}, \emph{14}, 4109--4121\relax
\mciteBstWouldAddEndPuncttrue
\mciteSetBstMidEndSepPunct{\mcitedefaultmidpunct}
{\mcitedefaultendpunct}{\mcitedefaultseppunct}\relax
\EndOfBibitem
\bibitem[Weber \latin{et~al.}(2022)Weber, Vuong, Devlaminck, Shee, Lee, Reichman, and Friesner]{Weber_Vuong_Devlaminck_Shee_Lee_Reichman_Friesner_2022}
Weber,~J.~L.; Vuong,~H.; Devlaminck,~P.~A.; Shee,~J.; Lee,~J.; Reichman,~D.~R.; Friesner,~R.~A. A Localized-Orbital Energy Evaluation for Auxiliary-Field Quantum Monte Carlo. \emph{J. Chem. Theory Comput.} \textbf{2022}, \emph{18}, 3447–3459\relax
\mciteBstWouldAddEndPuncttrue
\mciteSetBstMidEndSepPunct{\mcitedefaultmidpunct}
{\mcitedefaultendpunct}{\mcitedefaultseppunct}\relax
\EndOfBibitem
\bibitem[Mahajan and Sharma(2021)Mahajan, and Sharma]{AnkitTaming}
Mahajan,~A.; Sharma,~S. Taming the Sign Problem in Auxiliary-Field Quantum Monte Carlo Using Accurate Wave Functions. \emph{J. Chem. Theory Comput.} \textbf{2021}, \emph{17}, 4786--4798\relax
\mciteBstWouldAddEndPuncttrue
\mciteSetBstMidEndSepPunct{\mcitedefaultmidpunct}
{\mcitedefaultendpunct}{\mcitedefaultseppunct}\relax
\EndOfBibitem
\bibitem[Mahajan \latin{et~al.}(2022)Mahajan, Lee, and Sharma]{AnkitSelected}
Mahajan,~A.; Lee,~J.; Sharma,~S. {Selected configuration interaction wave functions in phaseless auxiliary field quantum Monte Carlo}. \emph{J. Chem. Phys.} \textbf{2022}, \emph{156}, 174111\relax
\mciteBstWouldAddEndPuncttrue
\mciteSetBstMidEndSepPunct{\mcitedefaultmidpunct}
{\mcitedefaultendpunct}{\mcitedefaultseppunct}\relax
\EndOfBibitem
\bibitem[Kurian \latin{et~al.}(2023)Kurian, Ye, Mahajan, Berkelbach, and Sharma]{kurian2023toward}
Kurian,~J.~S.; Ye,~H.-Z.; Mahajan,~A.; Berkelbach,~T.~C.; Sharma,~S. Toward Linear Scaling Auxiliary-Field Quantum Monte Carlo with Local Natural Orbitals. \emph{J. Chem. Theory Comput.} \textbf{2023}, \emph{20}, 134--142\relax
\mciteBstWouldAddEndPuncttrue
\mciteSetBstMidEndSepPunct{\mcitedefaultmidpunct}
{\mcitedefaultendpunct}{\mcitedefaultseppunct}\relax
\EndOfBibitem
\bibitem[Neugebauer \latin{et~al.}(2023)Neugebauer, Vuong, Weber, Friesner, Shee, and Hansen]{neugebauer2023toward}
Neugebauer,~H.; Vuong,~H.~T.; Weber,~J.~L.; Friesner,~R.~A.; Shee,~J.; Hansen,~A. Toward Benchmark-Quality Ab Initio Predictions for 3d Transition Metal Electrocatalysts: A Comparison of CCSD (T) and ph-AFQMC. \emph{J. Chem. Theory Comput.} \textbf{2023}, \emph{19}, 6208--6225\relax
\mciteBstWouldAddEndPuncttrue
\mciteSetBstMidEndSepPunct{\mcitedefaultmidpunct}
{\mcitedefaultendpunct}{\mcitedefaultseppunct}\relax
\EndOfBibitem
\bibitem[Shee \latin{et~al.}(2019)Shee, Rudshteyn, Arthur, Zhang, Reichman, and Friesner]{shee2019achieving}
Shee,~J.; Rudshteyn,~B.; Arthur,~E.~J.; Zhang,~S.; Reichman,~D.~R.; Friesner,~R.~A. On achieving high accuracy in quantum chemical calculations of 3 d transition metal-containing systems: a comparison of auxiliary-field quantum monte carlo with coupled cluster, density functional theory, and experiment for diatomic molecules. \emph{J. Chem. Theory Comput.} \textbf{2019}, \emph{15}, 2346--2358\relax
\mciteBstWouldAddEndPuncttrue
\mciteSetBstMidEndSepPunct{\mcitedefaultmidpunct}
{\mcitedefaultendpunct}{\mcitedefaultseppunct}\relax
\EndOfBibitem
\bibitem[Rudshteyn \latin{et~al.}(2022)Rudshteyn, Weber, Coskun, Devlaminck, Zhang, Reichman, Shee, and Friesner]{rudshteyn2022calculation}
Rudshteyn,~B.; Weber,~J.~L.; Coskun,~D.; Devlaminck,~P.~A.; Zhang,~S.; Reichman,~D.~R.; Shee,~J.; Friesner,~R.~A. Calculation of metallocene ionization potentials via auxiliary field quantum Monte Carlo: Toward benchmark quantum chemistry for transition metals. \emph{J. Chem. Theory Comput.} \textbf{2022}, \emph{18}, 2845--2862\relax
\mciteBstWouldAddEndPuncttrue
\mciteSetBstMidEndSepPunct{\mcitedefaultmidpunct}
{\mcitedefaultendpunct}{\mcitedefaultseppunct}\relax
\EndOfBibitem
\bibitem[Rudshteyn \latin{et~al.}(2020)Rudshteyn, Coskun, Weber, Arthur, Zhang, Reichman, Friesner, and Shee]{rudshteyn2020predicting}
Rudshteyn,~B.; Coskun,~D.; Weber,~J.~L.; Arthur,~E.~J.; Zhang,~S.; Reichman,~D.~R.; Friesner,~R.~A.; Shee,~J. Predicting ligand-dissociation energies of 3d coordination complexes with auxiliary-field quantum Monte Carlo. \emph{J. Chem. Theory Comput.} \textbf{2020}, \emph{16}, 3041--3054\relax
\mciteBstWouldAddEndPuncttrue
\mciteSetBstMidEndSepPunct{\mcitedefaultmidpunct}
{\mcitedefaultendpunct}{\mcitedefaultseppunct}\relax
\EndOfBibitem
\bibitem[Weber \latin{et~al.}(2021)Weber, Churchill, Jockusch, Arthur, Pun, Zhang, Friesner, Campos, Reichman, and Shee]{weber_silico_2021}
Weber,~J.~L.; Churchill,~E.~M.; Jockusch,~S.; Arthur,~E.~J.; Pun,~A.~B.; Zhang,~S.; Friesner,~R.~A.; Campos,~L.~M.; Reichman,~D.~R.; Shee,~J. \textit{{In} silico} prediction of annihilators for triplet–triplet annihilation upconversion \textit{via} auxiliary-field quantum {Monte} {Carlo}. \emph{Chem. Sci.} \textbf{2021}, \emph{12}, 1068--1079\relax
\mciteBstWouldAddEndPuncttrue
\mciteSetBstMidEndSepPunct{\mcitedefaultmidpunct}
{\mcitedefaultendpunct}{\mcitedefaultseppunct}\relax
\EndOfBibitem
\bibitem[Lee \latin{et~al.}(2022)Lee, Pham, and Reichman]{lee2022twenty}
Lee,~J.; Pham,~H.~Q.; Reichman,~D.~R. Twenty years of auxiliary-field quantum Monte Carlo in quantum chemistry: An overview and assessment on main group chemistry and bond-breaking. \emph{J. Chem. Theory Comput.} \textbf{2022}, \emph{18}, 7024--7042\relax
\mciteBstWouldAddEndPuncttrue
\mciteSetBstMidEndSepPunct{\mcitedefaultmidpunct}
{\mcitedefaultendpunct}{\mcitedefaultseppunct}\relax
\EndOfBibitem
\bibitem[Debnath \latin{et~al.}(2023)Debnath, Neufeld, Jacobson, Rudshteyn, Weber, Berkelbach, and Friesner]{debnath2023accurate}
Debnath,~S.; Neufeld,~V.~A.; Jacobson,~L.~D.; Rudshteyn,~B.; Weber,~J.~L.; Berkelbach,~T.~C.; Friesner,~R.~A. Accurate quantum chemical reaction energies for lithium-mediated electrolyte decomposition and evaluation of density functional approximations. \emph{J. Phys. Chem. A} \textbf{2023}, \emph{127}, 9178--9184\relax
\mciteBstWouldAddEndPuncttrue
\mciteSetBstMidEndSepPunct{\mcitedefaultmidpunct}
{\mcitedefaultendpunct}{\mcitedefaultseppunct}\relax
\EndOfBibitem
\bibitem[Stevenson \latin{et~al.}(2023)Stevenson, Agarwal, and Jacobson]{stevenson2023machine}
Stevenson,~J.; Agarwal,~G.; Jacobson,~L. Machine learning force field ranking of candidate solid electrolyte interphase structures in Li-ion batteries. \emph{ChemRxiv} \textbf{2023}, DOI:10.26434/chemrxiv-2023-gwnl8 (2023-12-07 Version 1).\relax
\mciteBstWouldAddEndPunctfalse
\mciteSetBstMidEndSepPunct{\mcitedefaultmidpunct}
{}{\mcitedefaultseppunct}\relax
\EndOfBibitem
\bibitem[Curtiss \latin{et~al.}(1997)Curtiss, Raghavachari, Redfern, and Pople]{Curtiss_Raghavachari_Redfern_Pople_1997}
Curtiss,~L.~A.; Raghavachari,~K.; Redfern,~P.~C.; Pople,~J.~A. Assessment of Gaussian-2 and density functional theories for the computation of enthalpies of formation. \emph{J. Chem. Phys.} \textbf{1997}, \emph{106}, 1063–1079\relax
\mciteBstWouldAddEndPuncttrue
\mciteSetBstMidEndSepPunct{\mcitedefaultmidpunct}
{\mcitedefaultendpunct}{\mcitedefaultseppunct}\relax
\EndOfBibitem
\bibitem[Curtiss \latin{et~al.}(1991)Curtiss, Raghavachari, Trucks, and Pople]{curtiss1991gaussian}
Curtiss,~L.~A.; Raghavachari,~K.; Trucks,~G.~W.; Pople,~J.~A. Gaussian-2 theory for molecular energies of first-and second-row compounds. \emph{J. Chem. Phys.} \textbf{1991}, \emph{94}, 7221--7230\relax
\mciteBstWouldAddEndPuncttrue
\mciteSetBstMidEndSepPunct{\mcitedefaultmidpunct}
{\mcitedefaultendpunct}{\mcitedefaultseppunct}\relax
\EndOfBibitem
\bibitem[Curtiss \latin{et~al.}(2000)Curtiss, Raghavachari, Redfern, and Pople]{Curtiss_G3}
Curtiss,~L.~A.; Raghavachari,~K.; Redfern,~P.~C.; Pople,~J.~A. {Assessment of Gaussian-3 and density functional theories for a larger experimental test set}. \emph{J. Chem. Phys.} \textbf{2000}, \emph{112}, 7374--7383\relax
\mciteBstWouldAddEndPuncttrue
\mciteSetBstMidEndSepPunct{\mcitedefaultmidpunct}
{\mcitedefaultendpunct}{\mcitedefaultseppunct}\relax
\EndOfBibitem
\bibitem[Curtiss and Raghavachari(2002)Curtiss, and Raghavachari]{curtiss2002gaussian}
Curtiss,~L.~A.; Raghavachari,~K. Gaussian-3 and related methods for accurate thermochemistry. \emph{Theor. Chem. Acc.} \textbf{2002}, \emph{108}, 61--70\relax
\mciteBstWouldAddEndPuncttrue
\mciteSetBstMidEndSepPunct{\mcitedefaultmidpunct}
{\mcitedefaultendpunct}{\mcitedefaultseppunct}\relax
\EndOfBibitem
\bibitem[Karton \latin{et~al.}(2011)Karton, Daon, and Martin]{Karton_W411}
Karton,~A.; Daon,~S.; Martin,~J.~M. W4-11: A high-confidence benchmark dataset for computational thermochemistry derived from first-principles W4 data. \emph{Chem. Phys. Lett.} \textbf{2011}, \emph{510}, 165--178\relax
\mciteBstWouldAddEndPuncttrue
\mciteSetBstMidEndSepPunct{\mcitedefaultmidpunct}
{\mcitedefaultendpunct}{\mcitedefaultseppunct}\relax
\EndOfBibitem
\bibitem[Karton \latin{et~al.}(2017)Karton, Sylvetsky, and Martin]{Karton_W417}
Karton,~A.; Sylvetsky,~N.; Martin,~J. M.~L. W4-17: A diverse and high-confidence dataset of atomization energies for benchmarking high-level electronic structure methods. \emph{J. Comput. Chem.} \textbf{2017}, \emph{38}, 2063--2075\relax
\mciteBstWouldAddEndPuncttrue
\mciteSetBstMidEndSepPunct{\mcitedefaultmidpunct}
{\mcitedefaultendpunct}{\mcitedefaultseppunct}\relax
\EndOfBibitem
\bibitem[ATc()]{ATcT}
Active Thermochemical Tables. \url{https://atct.anl.gov/} (accessed 2024-04-30)\relax
\mciteBstWouldAddEndPuncttrue
\mciteSetBstMidEndSepPunct{\mcitedefaultmidpunct}
{\mcitedefaultendpunct}{\mcitedefaultseppunct}\relax
\EndOfBibitem
\bibitem[Ruscic and Bross()Ruscic, and Bross]{atct1130}
Ruscic,~B.; Bross,~D. Active Thermochemical Tables (ATcT) Thermochemical Values ver. 1.130. \url{https://atct.anl.gov/Thermochemical%20Data/version%201.130/index.php} (accessed 2024-04-30)\relax
\mciteBstWouldAddEndPuncttrue
\mciteSetBstMidEndSepPunct{\mcitedefaultmidpunct}
{\mcitedefaultendpunct}{\mcitedefaultseppunct}\relax
\EndOfBibitem
\bibitem[Ruscic \latin{et~al.}(2004)Ruscic, Pinzon, Morton, von Laszevski, Bittner, Nijsure, Amin, Minkoff, and Wagner]{Ruscic2004}
Ruscic,~B.; Pinzon,~R.~E.; Morton,~M.~L.; von Laszevski,~G.; Bittner,~S.~J.; Nijsure,~S.~G.; Amin,~K.~A.; Minkoff,~M.; Wagner,~A.~F. Introduction to Active Thermochemical Tables: Several “Key” Enthalpies of Formation Revisited. \emph{J. Phys. Chem. A} \textbf{2004}, \emph{108}, 9979--9997\relax
\mciteBstWouldAddEndPuncttrue
\mciteSetBstMidEndSepPunct{\mcitedefaultmidpunct}
{\mcitedefaultendpunct}{\mcitedefaultseppunct}\relax
\EndOfBibitem
\bibitem[Welch \latin{et~al.}(2019)Welch, Dawes, Bross, and Ruscic]{Welch2019}
Welch,~B.~K.; Dawes,~R.; Bross,~D.~H.; Ruscic,~B. An Automated Thermochemistry Protocol Based on Explicitly Correlated Coupled-Cluster Theory: The Methyl and Ethyl Peroxy Families. \emph{The Journal of Physical Chemistry A} \textbf{2019}, \emph{123}, 5673--5682, PMID: 31244124\relax
\mciteBstWouldAddEndPuncttrue
\mciteSetBstMidEndSepPunct{\mcitedefaultmidpunct}
{\mcitedefaultendpunct}{\mcitedefaultseppunct}\relax
\EndOfBibitem
\bibitem[ccc()]{cccbdb}
Computational Chemistry Comparison and Benchmark DataBase. \url{https://cccbdb.nist.gov/} (accessed 2024-04-30)\relax
\mciteBstWouldAddEndPuncttrue
\mciteSetBstMidEndSepPunct{\mcitedefaultmidpunct}
{\mcitedefaultendpunct}{\mcitedefaultseppunct}\relax
\EndOfBibitem
\bibitem[Shi and Zhang(2021)Shi, and Zhang]{Shi2021}
Shi,~H.; Zhang,~S. {Some recent developments in auxiliary-field quantum Monte Carlo for real materials}. \emph{J. Chem. Phys.} \textbf{2021}, \emph{154}, 024107\relax
\mciteBstWouldAddEndPuncttrue
\mciteSetBstMidEndSepPunct{\mcitedefaultmidpunct}
{\mcitedefaultendpunct}{\mcitedefaultseppunct}\relax
\EndOfBibitem
\bibitem[Motta and Zhang(2018)Motta, and Zhang]{Motta2018}
Motta,~M.; Zhang,~S. Ab initio computations of molecular systems by the auxiliary-field quantum Monte Carlo method. \emph{Wiley Interdiscip. Rev.: Comput. Mol. Sci.} \textbf{2018}, \emph{8}, e1364\relax
\mciteBstWouldAddEndPuncttrue
\mciteSetBstMidEndSepPunct{\mcitedefaultmidpunct}
{\mcitedefaultendpunct}{\mcitedefaultseppunct}\relax
\EndOfBibitem
\bibitem[Sharma \latin{et~al.}(2017)Sharma, Holmes, Jeanmairet, Alavi, and Umrigar]{DICE1}
Sharma,~S.; Holmes,~A.~A.; Jeanmairet,~G.; Alavi,~A.; Umrigar,~C.~J. Semistochastic Heat-Bath Configuration Interaction Method: Selected Configuration Interaction with Semistochastic Perturbation Theory. \emph{J. Chem. Theory Comput.} \textbf{2017}, \emph{13}, 1595--1604\relax
\mciteBstWouldAddEndPuncttrue
\mciteSetBstMidEndSepPunct{\mcitedefaultmidpunct}
{\mcitedefaultendpunct}{\mcitedefaultseppunct}\relax
\EndOfBibitem
\bibitem[Holmes \latin{et~al.}(2016)Holmes, Tubman, and Umrigar]{DICE2}
Holmes,~A.~A.; Tubman,~N.~M.; Umrigar,~C.~J. Heat-Bath Configuration Interaction: An Efficient Selected Configuration Interaction Algorithm Inspired by Heat-Bath Sampling. \emph{J. Chem. Theory Comput.} \textbf{2016}, \emph{12}, 3674--3680\relax
\mciteBstWouldAddEndPuncttrue
\mciteSetBstMidEndSepPunct{\mcitedefaultmidpunct}
{\mcitedefaultendpunct}{\mcitedefaultseppunct}\relax
\EndOfBibitem
\bibitem[Sun \latin{et~al.}(2020)Sun, Zhang, Banerjee, Bao, Barbry, Blunt, Bogdanov, Booth, Chen, Cui, Eriksen, Gao, Guo, Hermann, Hermes, Koh, Koval, Lehtola, Li, Liu, Mardirossian, McClain, Motta, Mussard, Pham, Pulkin, Purwanto, Robinson, Ronca, Sayfutyarova, Scheurer, Schurkus, Smith, Sun, Sun, Upadhyay, Wagner, Wang, White, Whitfield, Williamson, Wouters, Yang, Yu, Zhu, Berkelbach, Sharma, Sokolov, and Chan]{PYSCF}
Sun,~Q.; Zhang,~X.; Banerjee,~S.; Bao,~P.; Barbry,~M.; Blunt,~N.~S.; Bogdanov,~N.~A.; Booth,~G.~H.; Chen,~J.; Cui,~Z.-H. \latin{et~al.}  {Recent developments in the PySCF program package}. \emph{J. Chem. Phys.} \textbf{2020}, \emph{153}, 024109\relax
\mciteBstWouldAddEndPuncttrue
\mciteSetBstMidEndSepPunct{\mcitedefaultmidpunct}
{\mcitedefaultendpunct}{\mcitedefaultseppunct}\relax
\EndOfBibitem
\bibitem[Smith \latin{et~al.}(2017)Smith, Mussard, Holmes, and Sharma]{DICE_PYSCF}
Smith,~J. E.~T.; Mussard,~B.; Holmes,~A.~A.; Sharma,~S. Cheap and Near Exact CASSCF with Large Active Spaces. \emph{J. Chem. Theory Comput.} \textbf{2017}, \emph{13}, 5468--5478\relax
\mciteBstWouldAddEndPuncttrue
\mciteSetBstMidEndSepPunct{\mcitedefaultmidpunct}
{\mcitedefaultendpunct}{\mcitedefaultseppunct}\relax
\EndOfBibitem
\bibitem[Neese(2012)]{neese2012orca}
Neese,~F. The ORCA program system. \emph{Wiley Interdiscip. Rev.: Comput. Mol. Sci.} \textbf{2012}, \emph{2}, 73--78\relax
\mciteBstWouldAddEndPuncttrue
\mciteSetBstMidEndSepPunct{\mcitedefaultmidpunct}
{\mcitedefaultendpunct}{\mcitedefaultseppunct}\relax
\EndOfBibitem
\bibitem[Altun \latin{et~al.}(2020)Altun, Neese, and Bistoni]{NeesePNO_extrap}
Altun,~A.; Neese,~F.; Bistoni,~G. Extrapolation to the Limit of a Complete Pair Natural Orbital Space in Local Coupled-Cluster Calculations. \emph{J. Chem. Theory Comput.} \textbf{2020}, \emph{16}, 6142–6149\relax
\mciteBstWouldAddEndPuncttrue
\mciteSetBstMidEndSepPunct{\mcitedefaultmidpunct}
{\mcitedefaultendpunct}{\mcitedefaultseppunct}\relax
\EndOfBibitem
\bibitem[Liakos \latin{et~al.}(2015)Liakos, Sparta, Kesharwani, Martin, and Neese]{Liakos_Sparta_Kesharwani_Martin_Neese_2015}
Liakos,~D.~G.; Sparta,~M.; Kesharwani,~M.~K.; Martin,~J. M.~L.; Neese,~F. Exploring the Accuracy Limits of Local Pair Natural Orbital Coupled-Cluster Theory. \emph{J. Chem. Theory Comput.} \textbf{2015}, \emph{11}, 1525–1539\relax
\mciteBstWouldAddEndPuncttrue
\mciteSetBstMidEndSepPunct{\mcitedefaultmidpunct}
{\mcitedefaultendpunct}{\mcitedefaultseppunct}\relax
\EndOfBibitem
\bibitem[Stoychev \latin{et~al.}(2017)Stoychev, Auer, and Neese]{Stoychev_Auer_Neese_2017}
Stoychev,~G.~L.; Auer,~A.~A.; Neese,~F. Automatic Generation of Auxiliary Basis Sets. \emph{J. Chem. Theory Comput.} \textbf{2017}, \emph{13}, 554–562\relax
\mciteBstWouldAddEndPuncttrue
\mciteSetBstMidEndSepPunct{\mcitedefaultmidpunct}
{\mcitedefaultendpunct}{\mcitedefaultseppunct}\relax
\EndOfBibitem
\bibitem[Dunning(1989)]{dunning1989a}
Dunning,~T.~H. Gaussian basis sets for use in correlated molecular calculations. I. The atoms boron through neon and hydrogen. \emph{J. Chem. Phys.} \textbf{1989}, \emph{90}, 1007--1023\relax
\mciteBstWouldAddEndPuncttrue
\mciteSetBstMidEndSepPunct{\mcitedefaultmidpunct}
{\mcitedefaultendpunct}{\mcitedefaultseppunct}\relax
\EndOfBibitem
\bibitem[de~Jong \latin{et~al.}(2001)de~Jong, Harrison, and Dixon]{jong2001a}
de~Jong,~W.~A.; Harrison,~R.~J.; Dixon,~D.~A. Parallel Douglas-Kroll energy and gradients in NWChem: Estimating scalar relativistic effects using Douglas-Kroll contracted basis sets. \emph{J. Chem. Phys.} \textbf{2001}, \emph{114}, 48\relax
\mciteBstWouldAddEndPuncttrue
\mciteSetBstMidEndSepPunct{\mcitedefaultmidpunct}
{\mcitedefaultendpunct}{\mcitedefaultseppunct}\relax
\EndOfBibitem
\bibitem[Kendall \latin{et~al.}(1992)Kendall, Dunning, and Harrison]{kendall1992a}
Kendall,~R.~A.; Dunning,~T.~H.; Harrison,~R.~J. Electron affinities of the first-row atoms revisited. Systematic basis sets and wave functions. \emph{J. Chem. Phys.} \textbf{1992}, \emph{96}, 6796--6806\relax
\mciteBstWouldAddEndPuncttrue
\mciteSetBstMidEndSepPunct{\mcitedefaultmidpunct}
{\mcitedefaultendpunct}{\mcitedefaultseppunct}\relax
\EndOfBibitem
\bibitem[Prascher \latin{et~al.}(2011)Prascher, Woon, Peterson, Dunning, and Wilson]{prascher2011a}
Prascher,~B.~P.; Woon,~D.~E.; Peterson,~K.~A.; Dunning,~T.~H.; Wilson,~A.~K. Gaussian basis sets for use in correlated molecular calculations. VII. Valence, core-valence, and scalar relativistic basis sets for Li, Be, Na, and Mg. \emph{Theor. Chem. Acc.} \textbf{2011}, \emph{128}, 69--82\relax
\mciteBstWouldAddEndPuncttrue
\mciteSetBstMidEndSepPunct{\mcitedefaultmidpunct}
{\mcitedefaultendpunct}{\mcitedefaultseppunct}\relax
\EndOfBibitem
\bibitem[Woon and Dunning(1993)Woon, and Dunning]{woon1993a}
Woon,~D.~E.; Dunning,~T.~H. Gaussian basis sets for use in correlated molecular calculations. III. The atoms aluminum through argon. \emph{J. Chem. Phys.} \textbf{1993}, \emph{98}, 1358--1371\relax
\mciteBstWouldAddEndPuncttrue
\mciteSetBstMidEndSepPunct{\mcitedefaultmidpunct}
{\mcitedefaultendpunct}{\mcitedefaultseppunct}\relax
\EndOfBibitem
\bibitem[Schuchardt \latin{et~al.}(2007)Schuchardt, Didier, Elsethagen, Sun, Gurumoorthi, Chase, Li, and Windus]{schuchardt2007a}
Schuchardt,~K.~L.; Didier,~B.~T.; Elsethagen,~T.; Sun,~L.; Gurumoorthi,~V.; Chase,~J.; Li,~J.; Windus,~T.~L. Basis Set Exchange: A Community Database for Computational Sciences. \emph{J. Chem. Inf. Model.} \textbf{2007}, \emph{47}, 1045--1052\relax
\mciteBstWouldAddEndPuncttrue
\mciteSetBstMidEndSepPunct{\mcitedefaultmidpunct}
{\mcitedefaultendpunct}{\mcitedefaultseppunct}\relax
\EndOfBibitem
\bibitem[Pritchard \latin{et~al.}(2019)Pritchard, Altarawy, Didier, Gibsom, and Windus]{pritchard2019a}
Pritchard,~B.~P.; Altarawy,~D.; Didier,~B.; Gibsom,~T.~D.; Windus,~T.~L. A New Basis Set Exchange: An Open, Up-to-date Resource for the Molecular Sciences Community. \emph{J. Chem. Inf. Model.} \textbf{2019}, \emph{59}, 4814--4820\relax
\mciteBstWouldAddEndPuncttrue
\mciteSetBstMidEndSepPunct{\mcitedefaultmidpunct}
{\mcitedefaultendpunct}{\mcitedefaultseppunct}\relax
\EndOfBibitem
\bibitem[Martin and Uzan(1998)Martin, and Uzan]{Martin1998_cpl}
Martin,~J.~M.; Uzan,~O. Basis set convergence in second-row compounds. The importance of core polarization functions. \emph{Chem. Phys. Lett.} \textbf{1998}, \emph{282}, 16--24\relax
\mciteBstWouldAddEndPuncttrue
\mciteSetBstMidEndSepPunct{\mcitedefaultmidpunct}
{\mcitedefaultendpunct}{\mcitedefaultseppunct}\relax
\EndOfBibitem
\bibitem[Martin(1998)]{Martin1998_jpc}
Martin,~J. M.~L. {Basis set convergence study of the atomization energy, geometry, and anharmonic force field of SO2: The importance of inner polarization functions}. \emph{J. Chem. Phys.} \textbf{1998}, \emph{108}, 2791--2800\relax
\mciteBstWouldAddEndPuncttrue
\mciteSetBstMidEndSepPunct{\mcitedefaultmidpunct}
{\mcitedefaultendpunct}{\mcitedefaultseppunct}\relax
\EndOfBibitem
\bibitem[Bauschlicher and Ricca(1998)Bauschlicher, and Ricca]{Bauschlicher1998}
Bauschlicher,~C.~W.; Ricca,~A. Atomization Energies of SO and SO2:  Basis Set Extrapolation Revisited. \emph{J. Phys. Chem. A} \textbf{1998}, \emph{102}, 8044--8050\relax
\mciteBstWouldAddEndPuncttrue
\mciteSetBstMidEndSepPunct{\mcitedefaultmidpunct}
{\mcitedefaultendpunct}{\mcitedefaultseppunct}\relax
\EndOfBibitem
\bibitem[Bauschlicher and Partridge(1995)Bauschlicher, and Partridge]{Bauschlicher1995}
Bauschlicher,~C.~W.; Partridge,~H. The sensitivity of B3LYP atomization energies to the basis set and a comparison of basis set requirements for CCSD(T) and B3LYP. \emph{Chem. Phys. Lett.} \textbf{1995}, \emph{240}, 533--540\relax
\mciteBstWouldAddEndPuncttrue
\mciteSetBstMidEndSepPunct{\mcitedefaultmidpunct}
{\mcitedefaultendpunct}{\mcitedefaultseppunct}\relax
\EndOfBibitem
\bibitem[Feller and Dixon(2003)Feller, and Dixon]{Feller2003}
Feller,~D.; Dixon,~D.~A. Coupled Cluster Theory and Multireference Configuration Interaction Study of FO, F2O, FO2, and FOOF. \emph{J. Phys. Chem. A} \textbf{2003}, \emph{107}, 9641--9651\relax
\mciteBstWouldAddEndPuncttrue
\mciteSetBstMidEndSepPunct{\mcitedefaultmidpunct}
{\mcitedefaultendpunct}{\mcitedefaultseppunct}\relax
\EndOfBibitem
\bibitem[Peterson and Dunning(2002)Peterson, and Dunning]{PetersonAccurate}
Peterson,~K.~A.; Dunning,~J.,~Thom~H. {Accurate correlation consistent basis sets for molecular core–valence correlation effects: The second row atoms Al–Ar, and the first row atoms B–Ne revisited}. \emph{J. Chem. Phys.} \textbf{2002}, \emph{117}, 10548--10560\relax
\mciteBstWouldAddEndPuncttrue
\mciteSetBstMidEndSepPunct{\mcitedefaultmidpunct}
{\mcitedefaultendpunct}{\mcitedefaultseppunct}\relax
\EndOfBibitem
\bibitem[Neese and Valeev(2011)Neese, and Valeev]{Neese_Extrapolation}
Neese,~F.; Valeev,~E.~F. Revisiting the Atomic Natural Orbital Approach for Basis Sets: Robust Systematic Basis Sets for Explicitly Correlated and Conventional Correlated ab initio Methods? \emph{J. Chem. Theory Comput.} \textbf{2011}, \emph{7}, 33--43\relax
\mciteBstWouldAddEndPuncttrue
\mciteSetBstMidEndSepPunct{\mcitedefaultmidpunct}
{\mcitedefaultendpunct}{\mcitedefaultseppunct}\relax
\EndOfBibitem
\bibitem[Nakajima and Hirao(2000)Nakajima, and Hirao]{Nakajima2000}
Nakajima,~T.; Hirao,~K. {The higher-order Douglas–Kroll transformation}. \emph{J. Chem. Phys.} \textbf{2000}, \emph{113}, 7786--7789\relax
\mciteBstWouldAddEndPuncttrue
\mciteSetBstMidEndSepPunct{\mcitedefaultmidpunct}
{\mcitedefaultendpunct}{\mcitedefaultseppunct}\relax
\EndOfBibitem
\bibitem[Mayer \latin{et~al.}(2001)Mayer, Krüger, and Rösch]{Mayer2001}
Mayer,~M.; Krüger,~S.; Rösch,~N. {A two-component variant of the Douglas–Kroll relativistic linear combination of Gaussian-type orbitals density-functional method: Spin–orbit effects in atoms and diatomics}. \emph{J. Chem. Phys.} \textbf{2001}, \emph{115}, 4411--4423\relax
\mciteBstWouldAddEndPuncttrue
\mciteSetBstMidEndSepPunct{\mcitedefaultmidpunct}
{\mcitedefaultendpunct}{\mcitedefaultseppunct}\relax
\EndOfBibitem
\bibitem[Wolf \latin{et~al.}(2002)Wolf, Reiher, and Hess]{Wolf2002}
Wolf,~A.; Reiher,~M.; Hess,~B.~A. {The generalized Douglas–Kroll transformation}. \emph{J. Chem. Phys.} \textbf{2002}, \emph{117}, 9215--9226\relax
\mciteBstWouldAddEndPuncttrue
\mciteSetBstMidEndSepPunct{\mcitedefaultmidpunct}
{\mcitedefaultendpunct}{\mcitedefaultseppunct}\relax
\EndOfBibitem
\bibitem[Liu and Peng(2009)Liu, and Peng]{Liu2009}
Liu,~W.; Peng,~D. {Exact two-component Hamiltonians revisited}. \emph{J. Chem. Phys.} \textbf{2009}, \emph{131}, 031104\relax
\mciteBstWouldAddEndPuncttrue
\mciteSetBstMidEndSepPunct{\mcitedefaultmidpunct}
{\mcitedefaultendpunct}{\mcitedefaultseppunct}\relax
\EndOfBibitem
\bibitem[Friesner \latin{et~al.}(2006)Friesner, Knoll, and Cao]{Friesner_Knoll_Cao_2006}
Friesner,~R.~A.; Knoll,~E.~H.; Cao,~Y. A localized orbital analysis of the thermochemical errors in hybrid density functional theory: Achieving chemical accuracy via a simple empirical correction scheme. \emph{J. Chem. Phys.} \textbf{2006}, \emph{125}, 124107\relax
\mciteBstWouldAddEndPuncttrue
\mciteSetBstMidEndSepPunct{\mcitedefaultmidpunct}
{\mcitedefaultendpunct}{\mcitedefaultseppunct}\relax
\EndOfBibitem
\bibitem[Goldfeld \latin{et~al.}(2008)Goldfeld, Bochevarov, and Friesner]{Goldfeld_Bochevarov_Friesner_2008}
Goldfeld,~D.~A.; Bochevarov,~A.~D.; Friesner,~R.~A. Localized orbital corrections applied to thermochemical errors in density functional theory: The role of basis set and application to molecular reactions. \emph{J. Chem. Phys.} \textbf{2008}, \emph{129}, 214105\relax
\mciteBstWouldAddEndPuncttrue
\mciteSetBstMidEndSepPunct{\mcitedefaultmidpunct}
{\mcitedefaultendpunct}{\mcitedefaultseppunct}\relax
\EndOfBibitem
\bibitem[Becke(1993)]{becke1993}
Becke,~A.~D. Density‐functional thermochemistry. III. The role of exact exchange. \emph{J. Chem. Phys.} \textbf{1993}, \emph{98}, 5648--5652\relax
\mciteBstWouldAddEndPuncttrue
\mciteSetBstMidEndSepPunct{\mcitedefaultmidpunct}
{\mcitedefaultendpunct}{\mcitedefaultseppunct}\relax
\EndOfBibitem
\bibitem[Lee \latin{et~al.}(1988)Lee, Yang, and Parr]{lyp1988}
Lee,~C.; Yang,~W.; Parr,~R.~G. Development of the {Colle-Salvetti} correlation-energy formula into a functional of the electron density. \emph{Phys. Rev. B} \textbf{1988}, \emph{37}, 785--789\relax
\mciteBstWouldAddEndPuncttrue
\mciteSetBstMidEndSepPunct{\mcitedefaultmidpunct}
{\mcitedefaultendpunct}{\mcitedefaultseppunct}\relax
\EndOfBibitem
\bibitem[Ditchfield \latin{et~al.}(1971)Ditchfield, Hehre, and Pople]{ditchfield1971a}
Ditchfield,~R.; Hehre,~W.~J.; Pople,~J.~A. Self-Consistent Molecular-Orbital Methods. IX. An Extended Gaussian-Type Basis for Molecular-Orbital Studies of Organic Molecules. \emph{J. Chem. Phys.} \textbf{1971}, \emph{54}, 724--728\relax
\mciteBstWouldAddEndPuncttrue
\mciteSetBstMidEndSepPunct{\mcitedefaultmidpunct}
{\mcitedefaultendpunct}{\mcitedefaultseppunct}\relax
\EndOfBibitem
\bibitem[Frisch \latin{et~al.}()Frisch, Trucks, Schlegel, Scuseria, Robb, Cheeseman, Scalmani, Barone, Mennucci, Petersson, Nakatsuji, Caricato, Li, Hratchian, Izmaylov, Bloino, Zheng, Sonnenberg, Hada, Ehara, Toyota, Fukuda, Hasegawa, Ishida, Nakajima, Honda, Kitao, Nakai, Vreven, Montgomery, Peralta, Ogliaro, Bearpark, Heyd, Brothers, Kudin, Staroverov, Kobayashi, Normand, Raghavachari, Rendell, Burant, Iyengar, Tomasi, Cossi, Rega, Millam, Klene, Knox, Cross, Bakken, Adamo, Jaramillo, Gomperts, Stratmann, Yazyev, Austin, Cammi, Pomelli, Ochterski, Martin, Morokuma, Zakrzewski, Voth, Salvador, Dannenberg, Dapprich, Daniels, Farkas, Foresman, Ortiz, Cioslowski, and Fox]{gaussian09e01}
Frisch,~M.~J.; Trucks,~G.~W.; Schlegel,~H.~B.; Scuseria,~G.~E.; Robb,~M.~A.; Cheeseman,~J.~R.; Scalmani,~G.; Barone,~V.; Mennucci,~B.; Petersson,~G.~A. \latin{et~al.}  Gaussian 09 Revision E.01. Gaussian, Inc., Wallingford, CT, 2009\relax
\mciteBstWouldAddEndPuncttrue
\mciteSetBstMidEndSepPunct{\mcitedefaultmidpunct}
{\mcitedefaultendpunct}{\mcitedefaultseppunct}\relax
\EndOfBibitem
\bibitem[Hariharan and Pople(1973)Hariharan, and Pople]{hariharan1973a}
Hariharan,~P.~C.; Pople,~J.~A. The influence of polarization functions on molecular orbital hydrogenation energies. \emph{Theor. Chim. Acta} \textbf{1973}, \emph{28}, 213--222\relax
\mciteBstWouldAddEndPuncttrue
\mciteSetBstMidEndSepPunct{\mcitedefaultmidpunct}
{\mcitedefaultendpunct}{\mcitedefaultseppunct}\relax
\EndOfBibitem
\bibitem[Francl \latin{et~al.}(1982)Francl, Pietro, Hehre, Binkley, Gordon, DeFrees, and Pople]{francl1982a}
Francl,~M.~M.; Pietro,~W.~J.; Hehre,~W.~J.; Binkley,~J.~S.; Gordon,~M.~S.; DeFrees,~D.~J.; Pople,~J.~A. Self-consistent molecular orbital methods. XXIII. A polarization-type basis set for second-row elements. \emph{J. Chem. Phys.} \textbf{1982}, \emph{77}, 3654--3665\relax
\mciteBstWouldAddEndPuncttrue
\mciteSetBstMidEndSepPunct{\mcitedefaultmidpunct}
{\mcitedefaultendpunct}{\mcitedefaultseppunct}\relax
\EndOfBibitem
\bibitem[Bochevarov \latin{et~al.}(2013)Bochevarov, Harder, Hughes, Greenwood, Braden, Philipp, Rinaldo, Halls, Zhang, and Friesner]{Jaguar}
Bochevarov,~A.~D.; Harder,~E.; Hughes,~T.~F.; Greenwood,~J.~R.; Braden,~D.~A.; Philipp,~D.~M.; Rinaldo,~D.; Halls,~M.~D.; Zhang,~J.; Friesner,~R.~A. Jaguar: A high-performance quantum chemistry software program with strengths in life and materials sciences. \emph{Int. J. Quantum Chem} \textbf{2013}, \emph{113}, 2110--2142\relax
\mciteBstWouldAddEndPuncttrue
\mciteSetBstMidEndSepPunct{\mcitedefaultmidpunct}
{\mcitedefaultendpunct}{\mcitedefaultseppunct}\relax
\EndOfBibitem
\bibitem[Semidalas and Martin(2022)Semidalas, and Martin]{SemidalasMOBH35}
Semidalas,~E.; Martin,~J.~M. The MOBH35 Metal–Organic Barrier Heights Reconsidered: Performance of Local-Orbital Coupled Cluster Approaches in Different Static Correlation Regimes. \emph{Journal of Chemical Theory and Computation} \textbf{2022}, \emph{18}, 883--898, PMID: 35045709\relax
\mciteBstWouldAddEndPuncttrue
\mciteSetBstMidEndSepPunct{\mcitedefaultmidpunct}
{\mcitedefaultendpunct}{\mcitedefaultseppunct}\relax
\EndOfBibitem
\bibitem[Chan and Ho(2023)Chan, and Ho]{ChanSimple2023}
Chan,~B.; Ho,~J. Simple Composite Approach to Efficiently Estimate Basis Set Limit CCSD(T) Harmonic Frequencies and Reaction Thermochemistry. \emph{The Journal of Physical Chemistry A} \textbf{2023}, \emph{127}, 10026--10031, PMID: 37970798\relax
\mciteBstWouldAddEndPuncttrue
\mciteSetBstMidEndSepPunct{\mcitedefaultmidpunct}
{\mcitedefaultendpunct}{\mcitedefaultseppunct}\relax
\EndOfBibitem
\bibitem[Semidalas and Martin(2024)Semidalas, and Martin]{SemidalasCanG42024}
Semidalas,~E.; Martin,~J. M.~L. Can G4-like composite Ab Initio methods accurately predict vibrational harmonic frequencies? \emph{Molecular Physics} \textbf{2024}, \emph{122}, e2263593\relax
\mciteBstWouldAddEndPuncttrue
\mciteSetBstMidEndSepPunct{\mcitedefaultmidpunct}
{\mcitedefaultendpunct}{\mcitedefaultseppunct}\relax
\EndOfBibitem
\end{mcitethebibliography}


\providecommand{\latin}[1]{#1}
\makeatletter
\providecommand{\doi}
  {\begingroup\let\do\@makeother\dospecials
  \catcode`\{=1 \catcode`\}=2 \doi@aux}
\providecommand{\doi@aux}[1]{\endgroup\texttt{#1}}
\makeatother
\providecommand*\mcitethebibliography{\thebibliography}
\csname @ifundefined\endcsname{endmcitethebibliography}  {\let\endmcitethebibliography\endthebibliography}{}
\begin{mcitethebibliography}{24}
\providecommand*\natexlab[1]{#1}
\providecommand*\mciteSetBstSublistMode[1]{}
\providecommand*\mciteSetBstMaxWidthForm[2]{}
\providecommand*\mciteBstWouldAddEndPuncttrue
  {\def\EndOfBibitem{\unskip.}}
\providecommand*\mciteBstWouldAddEndPunctfalse
  {\let\EndOfBibitem\relax}
\providecommand*\mciteSetBstMidEndSepPunct[3]{}
\providecommand*\mciteSetBstSublistLabelBeginEnd[3]{}
\providecommand*\EndOfBibitem{}
\mciteSetBstSublistMode{f}
\mciteSetBstMaxWidthForm{subitem}{(\alph{mcitesubitemcount})}
\mciteSetBstSublistLabelBeginEnd
  {\mcitemaxwidthsubitemform\space}
  {\relax}
  {\relax}

\bibitem[Neugebauer \latin{et~al.}(2023)Neugebauer, Vuong, Weber, Friesner, Shee, and Hansen]{neugebauer2023toward}
Neugebauer,~H.; Vuong,~H.~T.; Weber,~J.~L.; Friesner,~R.~A.; Shee,~J.; Hansen,~A. Toward Benchmark-Quality Ab Initio Predictions for 3d Transition Metal Electrocatalysts: A Comparison of CCSD (T) and ph-AFQMC. \emph{J. Chem. Theory Comput.} \textbf{2023}, \emph{19}, 6208--6225\relax
\mciteBstWouldAddEndPuncttrue
\mciteSetBstMidEndSepPunct{\mcitedefaultmidpunct}
{\mcitedefaultendpunct}{\mcitedefaultseppunct}\relax
\EndOfBibitem
\bibitem[Rudshteyn \latin{et~al.}(2022)Rudshteyn, Weber, Coskun, Devlaminck, Zhang, Reichman, Shee, and Friesner]{rudshteyn2022calculation}
Rudshteyn,~B.; Weber,~J.~L.; Coskun,~D.; Devlaminck,~P.~A.; Zhang,~S.; Reichman,~D.~R.; Shee,~J.; Friesner,~R.~A. Calculation of metallocene ionization potentials via auxiliary field quantum Monte Carlo: Toward benchmark quantum chemistry for transition metals. \emph{J. Chem. Theory Comput.} \textbf{2022}, \emph{18}, 2845--2862\relax
\mciteBstWouldAddEndPuncttrue
\mciteSetBstMidEndSepPunct{\mcitedefaultmidpunct}
{\mcitedefaultendpunct}{\mcitedefaultseppunct}\relax
\EndOfBibitem
\bibitem[Lee \latin{et~al.}(2022)Lee, Pham, and Reichman]{JoonhoTwenty}
Lee,~J.; Pham,~H.~Q.; Reichman,~D.~R. Twenty Years of Auxiliary-Field Quantum Monte Carlo in Quantum Chemistry: An Overview and Assessment on Main Group Chemistry and Bond-Breaking. \emph{J. Chem. Theory Comput.} \textbf{2022}, \emph{18}, 7024--7042\relax
\mciteBstWouldAddEndPuncttrue
\mciteSetBstMidEndSepPunct{\mcitedefaultmidpunct}
{\mcitedefaultendpunct}{\mcitedefaultseppunct}\relax
\EndOfBibitem
\bibitem[Martin(1996)]{MartinCBS}
Martin,~J.~M. Ab initio total atomization energies of small molecules — towards the basis set limit. \emph{Chem. Phys. Lett.} \textbf{1996}, \emph{259}, 669--678\relax
\mciteBstWouldAddEndPuncttrue
\mciteSetBstMidEndSepPunct{\mcitedefaultmidpunct}
{\mcitedefaultendpunct}{\mcitedefaultseppunct}\relax
\EndOfBibitem
\bibitem[Weber \latin{et~al.}(2022)Weber, Vuong, Devlaminck, Shee, Lee, Reichman, and Friesner]{Weber_Vuong_Devlaminck_Shee_Lee_Reichman_Friesner_2022}
Weber,~J.~L.; Vuong,~H.; Devlaminck,~P.~A.; Shee,~J.; Lee,~J.; Reichman,~D.~R.; Friesner,~R.~A. A Localized-Orbital Energy Evaluation for Auxiliary-Field Quantum Monte Carlo. \emph{J. Chem. Theory Comput.} \textbf{2022}, \emph{18}, 3447–3459\relax
\mciteBstWouldAddEndPuncttrue
\mciteSetBstMidEndSepPunct{\mcitedefaultmidpunct}
{\mcitedefaultendpunct}{\mcitedefaultseppunct}\relax
\EndOfBibitem
\bibitem[Wang \latin{et~al.}(2023)Wang, He, Taylor, Lorpaiboon, Mun, and Ho]{WangMolecular2023}
Wang,~M.; He,~X.; Taylor,~M.; Lorpaiboon,~W.; Mun,~H.; Ho,~J. Molecular Geometries and Vibrational Contributions to Reaction Thermochemistry Are Surprisingly Insensitive to the Choice of Basis Sets. \emph{Journal of Chemical Theory and Computation} \textbf{2023}, \emph{19}, 5036--5046, PMID: 37463146\relax
\mciteBstWouldAddEndPuncttrue
\mciteSetBstMidEndSepPunct{\mcitedefaultmidpunct}
{\mcitedefaultendpunct}{\mcitedefaultseppunct}\relax
\EndOfBibitem
\bibitem[Bakowies and von Lilienfeld(2021)Bakowies, and von Lilienfeld]{BakowiesDensity2021}
Bakowies,~D.; von Lilienfeld,~O.~A. Density Functional Geometries and Zero-Point Energies in Ab Initio Thermochemical Treatments of Compounds with First-Row Atoms (H, C, N, O, F). \emph{Journal of Chemical Theory and Computation} \textbf{2021}, \emph{17}, 4872--4890, PMID: 34260240\relax
\mciteBstWouldAddEndPuncttrue
\mciteSetBstMidEndSepPunct{\mcitedefaultmidpunct}
{\mcitedefaultendpunct}{\mcitedefaultseppunct}\relax
\EndOfBibitem
\bibitem[Martin and de~Oliveira(1999)Martin, and de~Oliveira]{MartinW11999}
Martin,~J. M.~L.; de~Oliveira,~G. {Towards standard methods for benchmark quality ab initio thermochemistry—W1 and W2 theory}. \emph{The Journal of Chemical Physics} \textbf{1999}, \emph{111}, 1843--1856\relax
\mciteBstWouldAddEndPuncttrue
\mciteSetBstMidEndSepPunct{\mcitedefaultmidpunct}
{\mcitedefaultendpunct}{\mcitedefaultseppunct}\relax
\EndOfBibitem
\bibitem[Semidalas and Martin(2024)Semidalas, and Martin]{SemidalasCanG42024}
Semidalas,~E.; Martin,~J. M.~L. Can G4-like composite Ab Initio methods accurately predict vibrational harmonic frequencies? \emph{Molecular Physics} \textbf{2024}, \emph{122}, e2263593\relax
\mciteBstWouldAddEndPuncttrue
\mciteSetBstMidEndSepPunct{\mcitedefaultmidpunct}
{\mcitedefaultendpunct}{\mcitedefaultseppunct}\relax
\EndOfBibitem
\bibitem[Chan and Ho(2023)Chan, and Ho]{ChanSimple2023}
Chan,~B.; Ho,~J. Simple Composite Approach to Efficiently Estimate Basis Set Limit CCSD(T) Harmonic Frequencies and Reaction Thermochemistry. \emph{The Journal of Physical Chemistry A} \textbf{2023}, \emph{127}, 10026--10031, PMID: 37970798\relax
\mciteBstWouldAddEndPuncttrue
\mciteSetBstMidEndSepPunct{\mcitedefaultmidpunct}
{\mcitedefaultendpunct}{\mcitedefaultseppunct}\relax
\EndOfBibitem
\bibitem[Nelson \latin{et~al.}(2023)Nelson, Glick, and Sherrill]{NelsonApproximating2023}
Nelson,~P.~M.; Glick,~Z.~L.; Sherrill,~C.~D. {Approximating large-basis coupled-cluster theory vibrational frequencies using focal-point approximations}. \emph{The Journal of Chemical Physics} \textbf{2023}, \emph{159}, 094104\relax
\mciteBstWouldAddEndPuncttrue
\mciteSetBstMidEndSepPunct{\mcitedefaultmidpunct}
{\mcitedefaultendpunct}{\mcitedefaultseppunct}\relax
\EndOfBibitem
\bibitem[Karton \latin{et~al.}(2011)Karton, Daon, and Martin]{Karton_W411}
Karton,~A.; Daon,~S.; Martin,~J.~M. W4-11: A high-confidence benchmark dataset for computational thermochemistry derived from first-principles W4 data. \emph{Chem. Phys. Lett.} \textbf{2011}, \emph{510}, 165--178\relax
\mciteBstWouldAddEndPuncttrue
\mciteSetBstMidEndSepPunct{\mcitedefaultmidpunct}
{\mcitedefaultendpunct}{\mcitedefaultseppunct}\relax
\EndOfBibitem
\bibitem[Karton \latin{et~al.}(2017)Karton, Sylvetsky, and Martin]{Karton_W417}
Karton,~A.; Sylvetsky,~N.; Martin,~J. M.~L. W4-17: A diverse and high-confidence dataset of atomization energies for benchmarking high-level electronic structure methods. \emph{J. Comput. Chem.} \textbf{2017}, \emph{38}, 2063--2075\relax
\mciteBstWouldAddEndPuncttrue
\mciteSetBstMidEndSepPunct{\mcitedefaultmidpunct}
{\mcitedefaultendpunct}{\mcitedefaultseppunct}\relax
\EndOfBibitem
\bibitem[Curtiss \latin{et~al.}(1997)Curtiss, Raghavachari, Redfern, and Pople]{Curtiss_Raghavachari_Redfern_Pople_1997}
Curtiss,~L.~A.; Raghavachari,~K.; Redfern,~P.~C.; Pople,~J.~A. Assessment of Gaussian-2 and density functional theories for the computation of enthalpies of formation. \emph{J. Chem. Phys.} \textbf{1997}, \emph{106}, 1063–1079\relax
\mciteBstWouldAddEndPuncttrue
\mciteSetBstMidEndSepPunct{\mcitedefaultmidpunct}
{\mcitedefaultendpunct}{\mcitedefaultseppunct}\relax
\EndOfBibitem
\bibitem[acc()]{accessCI}
ACCESS Allocations: Exchange Calculator. \url{https://allocations.access-ci.org/exchange_calculator} (accessed 2024-04-30)\relax
\mciteBstWouldAddEndPuncttrue
\mciteSetBstMidEndSepPunct{\mcitedefaultmidpunct}
{\mcitedefaultendpunct}{\mcitedefaultseppunct}\relax
\EndOfBibitem
\bibitem[Lee and Taylor(1989)Lee, and Taylor]{T1_diagnostic}
Lee,~T.~J.; Taylor,~P.~R. A diagnostic for determining the quality of single‐reference electron correlation methods. \emph{Int. J. Quantum Chem} \textbf{1989}, \emph{36}, 199 – 207, Cited by: 1978; All Open Access, Green Open Access\relax
\mciteBstWouldAddEndPuncttrue
\mciteSetBstMidEndSepPunct{\mcitedefaultmidpunct}
{\mcitedefaultendpunct}{\mcitedefaultseppunct}\relax
\EndOfBibitem
\bibitem[{Janssen} and {Nielsen}(1998){Janssen}, and {Nielsen}]{D1_diagnostic}
{Janssen},~C.~L.; {Nielsen},~I. M.~B. {New diagnostics for coupled-cluster and M{\o}ller Plesset perturbation theory}. \emph{Chem. Phys. Lett.} \textbf{1998}, \emph{290}, 423--430\relax
\mciteBstWouldAddEndPuncttrue
\mciteSetBstMidEndSepPunct{\mcitedefaultmidpunct}
{\mcitedefaultendpunct}{\mcitedefaultseppunct}\relax
\EndOfBibitem
\bibitem[Pulay and Hamilton(1988)Pulay, and Hamilton]{Pulay1988}
Pulay,~P.; Hamilton,~T.~P. {UHF natural orbitals for defining and starting MC‐SCF calculations}. \emph{J. Chem. Phys.} \textbf{1988}, \emph{88}, 4926--4933\relax
\mciteBstWouldAddEndPuncttrue
\mciteSetBstMidEndSepPunct{\mcitedefaultmidpunct}
{\mcitedefaultendpunct}{\mcitedefaultseppunct}\relax
\EndOfBibitem
\bibitem[Lee and Head-Gordon(2019)Lee, and Head-Gordon]{Lee2019}
Lee,~J.; Head-Gordon,~M. Distinguishing artificial and essential symmetry breaking in a single determinant: approach and application to the C60{,} C36{,} and C20 fullerenes. \emph{Phys. Chem. Chem. Phys.} \textbf{2019}, \emph{21}, 4763--4778\relax
\mciteBstWouldAddEndPuncttrue
\mciteSetBstMidEndSepPunct{\mcitedefaultmidpunct}
{\mcitedefaultendpunct}{\mcitedefaultseppunct}\relax
\EndOfBibitem
\bibitem[Langhoff and Davidson(1974)Langhoff, and Davidson]{Langhoff1974}
Langhoff,~S.~R.; Davidson,~E.~R. Configuration interaction calculations on the nitrogen molecule. \emph{Int. J. Quantum Chem.} \textbf{1974}, \emph{8}, 61--72\relax
\mciteBstWouldAddEndPuncttrue
\mciteSetBstMidEndSepPunct{\mcitedefaultmidpunct}
{\mcitedefaultendpunct}{\mcitedefaultseppunct}\relax
\EndOfBibitem
\bibitem[Wang \latin{et~al.}(2015)Wang, Manivasagam, and Wilson]{Wang2015}
Wang,~J.; Manivasagam,~S.; Wilson,~A.~K. Multireference Character for 4d Transition Metal-Containing Molecules. \emph{J. Chem. Theory Comput.} \textbf{2015}, \emph{11}, 5865--5872\relax
\mciteBstWouldAddEndPuncttrue
\mciteSetBstMidEndSepPunct{\mcitedefaultmidpunct}
{\mcitedefaultendpunct}{\mcitedefaultseppunct}\relax
\EndOfBibitem
\bibitem[Weigend and Ahlrichs(2005)Weigend, and Ahlrichs]{weigend2005a}
Weigend,~F.; Ahlrichs,~R. Balanced basis sets of split valence, triple zeta valence and quadruple zeta valence quality for H to Rn: Design and assessment of accuracy. \emph{Phys. Chem. Chem. Phys.} \textbf{2005}, \emph{7}, 3297\relax
\mciteBstWouldAddEndPuncttrue
\mciteSetBstMidEndSepPunct{\mcitedefaultmidpunct}
{\mcitedefaultendpunct}{\mcitedefaultseppunct}\relax
\EndOfBibitem
\bibitem[Adamo and Barone(1999)Adamo, and Barone]{Adamo1999}
Adamo,~C.; Barone,~V. {Toward reliable density functional methods without adjustable parameters: The PBE0 model}. \emph{J. Chem. Phys.} \textbf{1999}, \emph{110}, 6158--6170\relax
\mciteBstWouldAddEndPuncttrue
\mciteSetBstMidEndSepPunct{\mcitedefaultmidpunct}
{\mcitedefaultendpunct}{\mcitedefaultseppunct}\relax
\EndOfBibitem
\end{mcitethebibliography}

\end{document}

\tableofcontents

\newpage
\section{1. List of Molecules}
\addcontentsline{toc}{section}{1. List of Molecules}

{
\footnotesize
\begin{longtable}{|p{4cm}p{3cm}p{4cm}p{4cm}p{0cm}|}
\caption{Full list of molecules, the name in SI tables and xyz file name (.xyz files provided in a separate .zip along with the SI), the mutually exclusive datasets sorted in reporting for this work, the common chemical name, and the chemical formula.} % needs to go inside longtable environment
\label{tab:All_molecules_trials}
\endfirsthead
\endhead
\hline
Name in SI Tables	&	Dataset	&	Common Name	&	Formula	& \\ \hline
2-butyne	&	G2	&	2-Butyne	&	C4H6	& \\ \hline
Acetaldehyde	&	G2	&	Acetaldehyde	&	CH3CHO	& \\ \hline
Acethylene	&	G2	&	Ethyne	&	C2H2	& \\ \hline
Acetone	&	G2	&	Acetone	&	CH3COCH3	& \\ \hline
AlCl3	&	G2	&	Aluminum trichloride	&	AlCl3	& \\ \hline
AlF3	&	G2	&	Aluminum trifluoride	&	AlF3	& \\ \hline
Allene	&	G2	&	Allene	&	CH2=C=CH2	& \\ \hline
Aziridine	&	G2	&	Aziridine	&	C2H4NH	& \\ \hline
BCl3	&	G2	&	Trichloroborane	&	BCl3	& \\ \hline
Benzene	&	G2	&	Benzene	&	C6H6	& \\ \hline
BF3	&	G2	&	Trifluoroborane	&	BF3	& \\ \hline
Bicyclo-1-1-0-butane	&	G2	&	Bicyclo-1-1-0-butane	&	C4H6	& \\ \hline
CCl2CCl2	&	G2	&	Perchloroethene	&	CCl2CCl2	& \\ \hline
CCl4	&	G2	&	Perchloromethane	&	CCl4	& \\ \hline
CF2CF2	&	G2	&	Perfluoroethene	&	CF2CF2	& \\ \hline
CF3-CN	&	G2	&	2,2,2-Trifluoroacetonitrile	&	CF3-CN	& \\ \hline
CF4	&	G2	&	Perfluoromethane	&	CF4	& \\ \hline
CH2CH-CN	&	G2	&	Acrylonitrile	&	CH2CH-CN	& \\ \hline
CH2Cl2	&	G2	&	Dichloromethane	&	CH2Cl2	& \\ \hline
CH2F2	&	G2	&	Difluoromethane	&	CH2F2	& \\ \hline
CH3-CH2-CH2-Cl	&	G2	&	Propyl chloride	&	CH3-CH2-CH2-Cl	& \\ \hline
CH3-CH2-Cl	&	G2	&	Ethyl chloride	&	CH3-CH2-Cl	& \\ \hline
CH3-CH2-O-CH3	&	G2	&	Methoxyethane	&	CH3-CH2-O-CH3	& \\ \hline
CH3-CH2-SH	&	G2	&	Ethanethiol	&	CH3-CH2-SH	& \\ \hline
CH3-CN	&	G2	&	Acetonitrile	&	CH3-CN	& \\ \hline
CH3-O-CH3	&	G2	&	Methoxymethane	&	CH3-O-CH3	& \\ \hline
CH3-O-NO	&	G2	&	Methyl nitrite	&	CH3-O-NO	& \\ \hline
CH3-S-CH3	&	G2	&	Dimethyl sulfphide	&	CH3-S-CH3	& \\ \hline
CH3-SH	&	G2	&	Methanethiol	&	CH3-SH	& \\ \hline
CH3-SiH3	&	G2	&	Methyl silane	&	CH3-SiH3	& \\ \hline
CH3CFO	&	G2	&	Acetyl fluoride	&	CH3CFO	& \\ \hline
CH3Cl	&	G2	&	Chloromethane 	&	CH3Cl	& \\ \hline
CH3COCl	&	G2	&	Acetyl chloride	&	CH3COCl	& \\ \hline
CH3CONH2	&	G2	&	Acetamide	&	CH3CONH2	& \\ \hline
CH3COOH	&	G2	&	Acetic acid	&	CH3COOH	& \\ \hline
CH3NO2	&	G2	&	Nitromethane	&	CH3NO2	& \\ \hline
CH4	&	G2	&	Methane	&	CH4	& \\ \hline
CHCl3	&	G2	&	Chloromethane 	&	CHCl3	& \\ \hline
Cl2	&	G2	&	Dichlorine	&	Cl2	& \\ \hline
ClF	&	G2	&	Fluorine chloride	&	ClF	& \\ \hline
CLF3	&	G2	&	Chloride trifluoride	&	CLF3	& \\ \hline
ClNO	&	G2	&	Nitrosyl chloride	&	ClNO	& \\ \hline
CO	&	G2	&	Carbon monoxide	&	CO	& \\ \hline
CO2	&	G2	&	Carbon dioxide	&	CO2	& \\ \hline
CS	&	G2	&	Carbon monosulfide	&	CS	& \\ \hline
CS2	&	G2	&	Carbon disulfide	&	CS2	& \\ \hline
Cyanogen	&	G2	&	Cyanogen	&	NCCN	& \\ \hline
Cyclobutane	&	G2	&	Cyclobutane	&	C4H8	& \\ \hline
Cyclobutene	&	G2	&	Cyclobutene	&	C4H6	& \\ \hline
Cyclopropane	&	G2	&	Cyclopropane	&	C3H6	& \\ \hline
Cyclopropene	&	G2	&	Cyclopropene	&	C3H4	& \\ \hline
Dimethylamine	&	G2	&	Dimethylamine	&	(CH3)2NH	& \\ \hline
Dimethylsulfoxide	&	G2	&	Dimethylsulfoxide	&	(CH3)2SO	& \\ \hline
Ethane	&	G2	&	Ethane	&	C2H6	& \\ \hline
Ethanol	&	G2	&	Ethanol	&	CH3CH2OH	& \\ \hline
Ethenone	&	G2	&	Ethenone	&	H 2C=C=O	& \\ \hline
Ethylene	&	G2	&	Ethylene	&	C2H4	& \\ \hline
F2	&	G2	&	Difluroine	&	F2	& \\ \hline
F2O	&	G2	&	Hypofluorous anhydride	&	F2O	& \\ \hline
Furan	&	G2	&	Furan	&	C4H4O	& \\ \hline
Glyoxal	&	G2	&	Oxaldehyde	&	HCOCOH	& \\ \hline
H2	&	G2	&	Dihydrogen	&	H2	& \\ \hline
H2CO	&	G2	&	Formaldehyde	&	H2CO	& \\ \hline
H2NNH2	&	G2	&	Hydrazine	&	H2NNH2	& \\ \hline
H2O	&	G2	&	Water	&	H2O	& \\ \hline
HCF3	&	G2	&	Trifluromethane	&	HCF3	& \\ \hline
HCl	&	G2	&	Hydrogen chloride	&	HCl	& \\ \hline
HCN	&	G2	&	Hydrogen cyanide	&	HCN	& \\ \hline
HCOOCH3	&	G2	&	Methyl formate	&	HCOOCH3	& \\ \hline
HCOOH	&	G2	&	Formic acid	&	HCOOH	& \\ \hline
HF	&	G2	&	Hydrofluoric acid	&	HF	& \\ \hline
HOCl	&	G2	&	Hypochlorous acid	&	HOCl	& \\ \hline
HOOH	&	G2	&	Hydrogen peroxide	&	HOOH	& \\ \hline
Isobutane	&	G2	&	Isobutane	&	C4H10	& \\ \hline
Isobutene	&	G2	&	Isobutene	&	C4H8	& \\ \hline
Isopropyl-alcohol	&	G2	&	Isopropanol	&	(CH3)2CHOH	& \\ \hline
Ketene	&	G2	&	Ketene	&	C2H2O	& \\ \hline
Li2	&	G2	&	Dilithium	&	Li2	& \\ \hline
LiF	&	G2	&	Lithium fluroide	&	LiF	& \\ \hline
LiH	&	G2	&	Lithium hydride	&	LiH	& \\ \hline
Methanol	&	G2	&	Methanol	&	H3COH	& \\ \hline
Methylamine	&	G2	&	Methylamine	&	CH3NH2	& \\ \hline
Methylene-cyclopropane	&	G2	&	Methylene-cyclopropane	&	C4H6	& \\ \hline
N2	&	G2	&	Dinitrogen	&	N2	& \\ \hline
Na2	&	G2	&	Disodium	&	Na2	& \\ \hline
NaCl	&	G2	&	Sodium chloride	&	NaCl	& \\ \hline
NF3	&	G2	&	Nitrogen trifluoride	&	NF3	& \\ \hline
NH3	&	G2	&	Ammonia	&	NH3	& \\ \hline
NNO	&	G2	&	Nitrous oxide	&	NNO	& \\ \hline
OCS-m1	&	G2	&	Carbon oxide sulfide	&	OCS-m1	& \\ \hline
Oxirane	&	G2	&	Oxirane	&	C2H4O	& \\ \hline
Ozone	&	G2	&	Ozone	&	O3	& \\ \hline
P2	&	G2	&	Diphosphorus	&	P2	& \\ \hline
PF3	&	G2	&	Triflurophosphorus	&	PF3	& \\ \hline
PH3	&	G2	&	Phosphine	&	PH3	& \\ \hline
Propane	&	G2	&	Propane	&	C3H8	& \\ \hline
Propene-CS	&	G2	&	Propylene	&	CH3CH=CH2	& \\ \hline
Propyne	&	G2	&	Propyne	&	CH3CCH	& \\ \hline
Pyridine	&	G2	&	Pyridine	&	C5H5N	& \\ \hline
Pyrole	&	G2	&	Pyrrole	&	C4H5N	& \\ \hline
SH2	&	G2	&	Hydrogen sulfide	&	SH2	& \\ \hline
Si2H6	&	G2	&	Disilane	&	Si2H6	& \\ \hline
SiCl4	&	G2	&	Silicon tetrachloride	&	SiCl4	& \\ \hline
SiF4	&	G2	&	Silicon tetrafluoride	&	SiF4	& \\ \hline
SiH4	&	G2	&	Silane	&	SiH4	& \\ \hline
SiO	&	G2	&	Silicon monoxide	&	SiO	& \\ \hline
SO2	&	G2	&	Sulfur dioxide	&	SO2	& \\ \hline
Spiropentane	&	G2	&	Spiropentane	&	C5H8	& \\ \hline
Thiooxirane	&	G2	&	Thiirane	&	C2H4S	& \\ \hline
Thiophene	&	G2	&	Thiophene	&	C4H4S	& \\ \hline
Trans-1-3-butadiene	&	G2	&	1-3-butadiene	&	C4H6	& \\ \hline
Trans-butane	&	G2	&	Butane	&	C4H10	& \\ \hline
Trans-ethylamine	&	G2	&	Ethylamine	&	CH3CH2NH2	& \\ \hline
Trimethyl-amine	&	G2	&	Trimethyl amine	&	N(CH3)3	& \\ \hline
Vinyl-chloride	&	G2	&	Vinyl chloride	&	H2C$=$CHCl	& \\ \hline
Vynil-fluoride	&	G2	&	Vinyl fluoride	&	H2C$=$CHF	& \\ \hline
1,3-cyclohexadiene	&	G3	&	1,3-Cyclohexadiene	&	(C2H4)(CH)4	& \\ \hline
1,3-DiFluorobenzene	&	G3	&	1,3-Difluorobenzene	&	C6H4F2	& \\ \hline
1,4-DiFluorobenzene	&	G3	&	1,4-Difluorobenzene	&	C6H4F2	& \\ \hline
2-methyl	&	G3	&	2-Methylthiophene	&	CH3C4H3S	& \\ \hline
2,5-Dihydrothiophene	&	G3	&	2,5-dihydrothiophene	&	C6H6S	& \\ \hline
3-methyl	&	G3	&	3-methylpentane	&	C6H14	& \\ \hline
Acetic	&	G3	&	Acetic anhydride	&	C4H6O3	& \\ \hline
azulene	&	G3	&	Azulene	&	C10H8	& \\ \hline
benzoquinone	&	G3	&	Benzoquinone	&	C6H4O2	& \\ \hline
c2f6	&	G3	&	Hexafluoroethane	&	C2F6	& \\ \hline
C4H4N2	&	G3	&	Pyrazine	&	C4H4N2	& \\ \hline
C4H6	&	G3	&	Methyl allene	&	CH3-CH=C=CH2	& \\ \hline
C4H6O	&	G3	&	Divinyl ether	&	O(CH=CH2)2	& \\ \hline
C4H8O2	&	G3	&	1,4-Dioxane	&	C4H8O2	& \\ \hline
C5H8	&	G3	&	Isoprene	&	C5H8	& \\ \hline
C6H12	&	G3	&	Cyclohexane	&	C6H12	& \\ \hline
C6H5-CH3	&	G3	&	Toluene	&	PhCH3	& \\ \hline
C6H5-NH2	&	G3	&	Aniline	&	PhNH2	& \\ \hline
C6H5-OH	&	G3	&	Phenol	&	PhOH	& \\ \hline
cf3cl	&	G3	&	Chlorotrifluoromethane	&	CF3Cl	& \\ \hline
CH3\_2CH-CHO	&	G3	&	Isobutanal	&	(CH3)2CH-CHO	& \\ \hline
CH3\_2CH-CN	&	G3	&	Isobutane nitrile	&	(CH3)2CH-CN	& \\ \hline
CH3\_2CH-O-CH\_CH3\_2	&	G3	&	Diisopropyl ether	&	(CH3)2CH-O-CH(CH3)2	& \\ \hline
CH3\_3C-NH2	&	G3	&	t-Butyl amine	&	(CH3)3C-NH2	& \\ \hline
CH3\_3C-O-CH3	&	G3	&	t-Butyl methyl ether	&	(CH3)3C-O-CH3	& \\ \hline
CH3\_3C-SH	&	G3	&	t-Butanethiol	&	(CH3)3C-SH	& \\ \hline
CH3-C\_O\_-CCH	&	G3	&	3-butyn-2-one	&	C4H4O	& \\ \hline
CH3-C\_O\_-O-CH\_CH3\_2	&	G3	&	Isopropyl acetate	&	CH3-COOCH(CH3)2	& \\ \hline
CH3-C\_O\_-OCH3	&	G3	&	Methyl acetate	&	CH3-COOCH3	& \\ \hline
CH3-CH\_OCH3\_2	&	G3	&	1,1-dimethoxy ethane	&	CH3-CH(OCH3)2	& \\ \hline
CH3-CH2-CH\_CH3\_-NO2	&	G3	&	Nitro-s-butane	&	CH3-CH2-CH(CH3)-NO2	& \\ \hline
CH3-CH2-CO-CH2-CH3	&	G3	&	Diethyl ketone	&	CH3-CH2-CO-CH2-CH3	& \\ \hline
CH3-CH2-O-CH2-CH3	&	G3	&	Diethyl ether	&	CH3-CH2-O-CH2-CH3	& \\ \hline
CH3-CH2-S-S-CH2-CH3	&	G3	&	Diethyl disulfide	&	CH3-CH2-S-S-CH2-CH3	& \\ \hline
CH3-CHCH-CHO	&	G3	&	Crotonaldehyde	&	CH3-CHCH-CHO	& \\ \hline
CH3-CO-CH2-CH3	&	G3	&	Methyl ethyl ketone	&	CH3-CO-CH2-CH3	& \\ \hline
Chlorobenzene	&	G3	&	Chlorobenzene	&	PhCl	& \\ \hline
Cl2O2S	&	G3	&	Chlorosulfinyl hypochlorite	&	Cl2O2S	& \\ \hline
Cl2S2	&	G3	&	Disulfur dichloride	&	Cl2S2	& \\ \hline
cyclooctatetraene	&	G3	&	1,3,5,7-Cyclooctatetraene	&	C8H8	& \\ \hline
cyclopentane	&	G3	&	Cyclopentane	&	C5H10	& \\ \hline
cyclopentanone	&	G3	&	Cyclopentanone	&	C5H8O	& \\ \hline
dimethyl	&	G3	&	Dimethyl sulfone	&	(CH3)2SO2	& \\ \hline
Fluorobenzene	&	G3	&	Fluorobenzene	&	PhF	& \\ \hline
n-Butyl	&	G3	&	n-Butyl chloride	&	CH3(CH2)3Cl	& \\ \hline
n-heptane	&	G3	&	n-Heptane	&	C7H16	& \\ \hline
n-hexane	&	G3	&	n-Hexane	&	C6H14	& \\ \hline
N-methyl	&	G3	&	N-Methyl pyrrole	&	C5H7N	& \\ \hline
n-octane	&	G3	&	n-Octane	&	C8H18	& \\ \hline
n-pentane	&	G3	&	n-Pentane	&	C5H12	& \\ \hline
Naphthalene	&	G3	&	Naphthalene	&	C10H8	& \\ \hline
NC-CH2-CH2-CN	&	G3	&	Succinonitrile	&	NC-CH2-CH2-CN	& \\ \hline
Neopentane	&	G3	&	Neopentane	&	C(CH3)4	& \\ \hline
P4	&	G3	&	Tetraphosphorus	&	P4	& \\ \hline
para-cyclohexadiene	&	G3	&	1,4-Cyclohexadiene	&	C6H8	& \\ \hline
PCl3	&	G3	&	Phosphorus trichloride	&	PCl3	& \\ \hline
PCl5	&	G3	&	Phosphorus pentachloride	&	PCl5	& \\ \hline
Perhydropyridine	&	G3	&	Piperidine	&	C5H10NH	& \\ \hline
pf5	&	G3	&	Phosphorus pentafluoride	&	PF5	& \\ \hline
POCl3	&	G3	&	Phosphorus oxychloride	&	POCl3	& \\ \hline
pyrimidine	&	G3	&	Pyrimidine	&	C4H4N2	& \\ \hline
SCl2	&	G3	&	Sulfur dichloride	&	SCl2	& \\ \hline
sf6	&	G3	&	Sulfur hexafluoride	&	SF6	& \\ \hline
SiCl2	&	G3	&	Dichlorosilane	&	SiCl2	& \\ \hline
SO3	&	G3	&	Sulfur trioxide	&	SO3	& \\ \hline
t-butanol	&	G3	&	t-Butanol	&	(CH3)3COH	& \\ \hline
t-Butyl	&	G3	&	t-Butyl chloride	&	(CH3)3CCl	& \\ \hline
tetrahydrofuran	&	G3	&	Tetrahydrofuran	&	C4H8O	& \\ \hline
Tetrahydropyran	&	G3	&	Tetrahydropyran	&	C5H10O	& \\ \hline
Tetrahydropyrrole	&	G3	&	Tetrahydropyrrole	&	C4H8NH	& \\ \hline
Tetrahydrothiophene	&	G3	&	Tetrahydrothiophene	&	C4H8S	& \\ \hline
Tetrahydrothiopyran	&	G3	&	Tetrahydrothiopyran	&	C5H10S	& \\ \hline
Tetramethylsilane	&	G3	&	Tetramethylsilane	&	(CH3)4Si	& \\ \hline
alcl	&	W4-11	&	Aluminum monochloride	&	AlCl	& \\ \hline
alf	&	W4-11	&	Aluminum monofluoride	&	AlF	& \\ \hline
alh	&	W4-11	&	Aluminum monohydride	&	AlH	& \\ \hline
alh3	&	W4-11	&	Aluminum trihydride	&	AlH3	& \\ \hline
b2h6	&	W4-11	&	Diborane	&	B2H6	& \\ \hline
bf	&	W4-11	&	Beryllium monofluoride	&	BF	& \\ \hline
bh	&	W4-11	&	Beryllium monohydride	&	BH	& \\ \hline
bh3	&	W4-11	&	Borane	&	BH3	& \\ \hline
bhf2	&	W4-11	&	Difluoroborane	&	BHF2	& \\ \hline
c-hono	&	W4-11	&	cis-Nitrous acid	&	c-HONO	& \\ \hline
c-n2h2	&	W4-11	&	cis-Diazine	&	c-N2H2	& \\ \hline
ch2nh	&	W4-11	&	Methanimine	&	CH2NH	& \\ \hline
ch3f	&	W4-11	&	Fluoromethane	&	CH3F	& \\ \hline
clcn	&	W4-11	&	Nitryl chloride	&	ClCN	& \\ \hline
dioxirane	&	W4-11	&	Dioxirane	&	CH2O2	& \\ \hline
f2co	&	W4-11	&	Carbonyl fluoride	&	F2CO	& \\ \hline
fccf	&	W4-11	&	Difluoroacetylene	&	FCCF	& \\ \hline
hccf	&	W4-11	&	Fluoroacetylene	&	HCCF	& \\ \hline
hcno	&	W4-11	&	Formonitrile oxide	&	HCNO	& \\ \hline
hcof	&	W4-11	&	Formyl fluoride	&	HCOF	& \\ \hline
hnco	&	W4-11	&	Isocyanic acid	&	HNCO	& \\ \hline
hnnn	&	W4-11	&	Hydrogen azide	&	HNNN	& \\ \hline
hno	&	W4-11	&	Nitrosyl hydride	&	HNO	& \\ \hline
hocn	&	W4-11	&	Cyanic acid	&	HOCN	& \\ \hline
hof	&	W4-11	&	Hypofluorous acid	&	HOF	& \\ \hline
nh2cl	&	W4-11	&	Chloramine	&	NH2Cl	& \\ \hline
oxirene	&	W4-11	&	Oxirene	&	C2H2O	& \\ \hline
s2o	&	W4-11	&	Disulfur monoxide	&	S2O	& \\ \hline
sih3f	&	W4-11	&	Fluorosilane	&	SiH3F	& \\ \hline
t-hono	&	W4-11	&	trans-Nitrous acid	&	t-HONO	& \\ \hline
t-n2h2	&	W4-11	&	trans-Diazine	&	t-N2H2	& \\ \hline
c2h5f	&	W4-11	&	Fluoroethane	&	C2H5F	& \\ \hline
bn	&	W4-11	&	Boron nitride	&	BN	& \\ \hline
c2	&	W4-11	&	Dicarbon	&	C2	& \\ \hline
cl2o	&	W4-11	&	Dichlorine monoxide	&	Cl2O	& \\ \hline
foof	&	W4-11	&	Dioxygen difluoride	&	FOOF	& \\ \hline
s3	&	W4-11	&	Trisulfur	&	S3	& \\ \hline
s4-c2v	&	W4-11	&	Tetrasulfur (C2v)	&	S4 (C2v)	& \\ \hline
beta-lactim	&	W4-17	&	Beta-lactim	&	(C3H4N)OH	& \\ \hline
borole	&	W4-17	&	Borole	&	C4H5B	& \\ \hline
c2cl2	&	W4-17	&	Dichloroacetylene	&	C2Cl2	& \\ \hline
c2cl6	&	W4-17	&	Hexachloroethane	&	C2Cl6	& \\ \hline
c2clh	&	W4-17	&	Chloroethyne	&	C2ClH	& \\ \hline
ccl2o	&	W4-17	&	Phosgene	&	CCl2O	& \\ \hline
cf2cl2	&	W4-17	&	Dichlorodifluoromethane	&	CF2Cl2	& \\ \hline
ch2clf	&	W4-17	&	Chlorofluoromethane	&	CH2ClF	& \\ \hline
ch3ph2	&	W4-17	&	Methylphosphine	&	CH2PH2	& \\ \hline
cis-c2f2cl2	&	W4-17	&	cis-Dichlorodifluoroethene	&	c-CF2Cl2	& \\ \hline
clcof	&	W4-17	&	Carbonyl chloride fluoride	&	ClCOF	& \\ \hline
cyclobutadiene	&	W4-17	&	Cyclobutadiene	&	C4H4	& \\ \hline
cyclopentadiene	&	W4-17	&	Cyclopentadiene	&	H6C5	& \\ \hline
dioxetan2one	&	W4-17	&	1,3-Dioxetan-2-one	&	H2C2O3	& \\ \hline
dioxetane	&	W4-17	&	Dioxetane	&	H4C2O2	& \\ \hline
dithiotane	&	W4-17	&	1,3-Dithiotane	&	H4C2S2	& \\ \hline
fno	&	W4-17	&	Nitrosyl fluoride	&	FNO	& \\ \hline
formamide	&	W4-17	&	Formamide	&	CH3NO	& \\ \hline
formic-anhydride	&	W4-17	&	Formic-anhydride	&	H2C2O3	& \\ \hline
hclo4	&	W4-17	&	Perchloric acid	&	HClO4	& \\ \hline
hoclo2	&	W4-17	&	Chloric acid	&	HOClO2	& \\ \hline
hoclo	&	W4-17	&	Chlorous acid	&	HOClO	& \\ \hline
n2o4	&	W4-17	&	Dinitrogen tetraoxide	&	N2O4	& \\ \hline
nh2f	&	W4-17	&	Fluoroamine	&	NH2F	& \\ \hline
nh2oh	&	W4-17	&	Hydroxylamine	&	NH2OH	& \\ \hline
oxadiazole	&	W4-17	&	Oxadiazole	&	H2C2N2O	& \\ \hline
oxetane	&	W4-17	&	Oxetane	&	H6C3O	& \\ \hline
silole	&	W4-17	&	Silole	&	H6C4Si	& \\ \hline
tetrahedrane	&	W4-17	&	Tetrahedrane	&	C4H4	& \\ \hline
trans-c2f2cl2	&	W4-17	&	trans-Dichlorodifluoroethene	&	t-CF2Cl2	& \\ \hline
clf5	&	W4-17	&	Chlorine pentafluoride	&	ClF5	& \\ \hline
cloocl	&	W4-17	&	Chlorine peroxide	&	ClOOCl	& \\ \hline
\end{longtable}
}

\newpage
\section{2. Timestep Errors}
\addcontentsline{toc}{section}{2. Timestep Errors}

\begin{table}[h]
	\centering
	\caption{AFQMC atomization energies in kcal/mol for the molecule \ce{FOOF} with all electron calculation. Bottom of the table shows various reference atomization energy values in kcal/mol.}
	\begin{tabular}{ccc}
		\toprule
		%& \multicolumn{2}{c}{\textbf{CCSD}} & \multicolumn{2}{c}{\textbf{CCSD(T)}} \\
		%\cmidrule(lr){2-3} \cmidrule(lr){4-5}
		Time step (a.u.) & Atomization energy & \(\sim\) Core hrs cost\\
		\midrule
		0.0050  & 142.5(4) & 2500 \\
		0.0025 & 144.1(4) & 5000 \\
		0.0010  & 145.3(6) & 10500 \\
		\hline
		\hline
		ATcT reference & 146.4 
& \\
		W4 & 146.0 & \\
	\end{tabular}
	\label{tab:foof}
\end{table}
\begin{figure}
	\centering
	\includegraphics[width=0.8\textwidth]{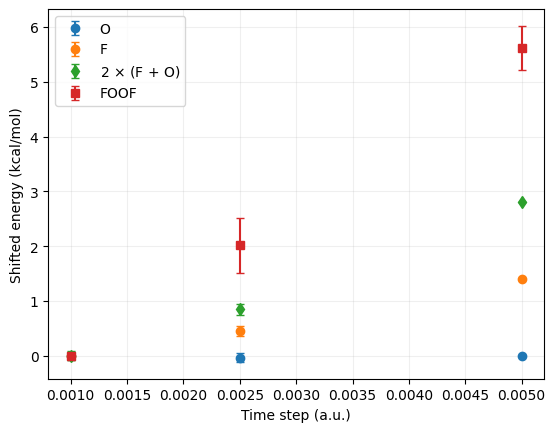}
	\caption{All electron AFQMC CBS energies (with respect to the corresponding time step 0.001 a.u. values) as a function of time step for FOOF.}
	\label{fig:time_step}
\end{figure}
As a demonstration, we show the timestep error for the molecule FOOF, where we calculate both the atom and molecule energies using W-AFQMC with 10000 determinants for the molecule and 10000 determinants for the atoms. Instead of the MP2 basis correction to get to the CBS limit from aug-cc-pVTZ-DK and aug-cc-pVQZ-DK (see main text Section 2.6), for this exercise this molecule is small enough to run aug-cc-pVTZ-DK and aug-cc-pVQZ-DK directly. The convergence with timestep is shown in Table~\ref{tab:foof}. The computational scaling with timestep is linear.

Looking at Figure \ref{fig:time_step}, all electron atomic energies display a less pronounced dependence on the AFQMC time-step compared to the molecular energy. This results in a notable time-step error in atomization energies. Note that there are multiple sources of time step errors in AFQMC: Trotter decomposition, Hubbard-Stratonovich transformation, hybrid and phaseless approximations, with the last two being trial dependent. Calculations with frozen core electrons or those using basis sets without tight core functions exhibit a much weaker time-step dependence. Indeed, the timestep error for the molecule can reach the order of 10 kcal/mol by using 0.005 Ha$^{-1}$ instead of 0.0025 Ha${^{-1}}$ when using a single determinant trial without frozen core. In contrast, for most cases a timestep of 0.005 Ha${^{-1}}$ is sufficient when freezing core electrons. For example, Table~\ref{tab:CCl4_timestep} shows the single point energies using L-AFQMC all-electron calculation, with a 1 determinant trial for CCl4. Real space methods, like diffusion Monte Carlo, also struggle with time step issues, and the field often resorts to pseudopotentials to avoid substantial costs. AFQMC, as an orbital space method, offers the convenient option of freezing core electrons (not possible in DMC) and adding composite perturbative corrections from cheaper methods.

\begin{table}
    \centering
    \small
    \begin{tabular}{ccll} \hline
         Timestep&  SPE all-electron 1 det&SPE frozen core 1 det&SPE all-electron 200 dets\\ \hline
         0.00125
&   -1884.01412(60) &-1882.64117(63) &
-1884.01443(66)\\
 0.0025
& -1884.01220(78) & -1882.64127(48) &
-1884.01136(59)\\
 0.0050& -1883.99762(61) & -1882.64129(58) &
-1883.99929(53)\\
 0.010& -1883.96231(74) & -1882.64146(45) &-1883.96526(56)\\ \hline
    \end{tabular}
    \caption{Single point energies (SPEs) for CCl4 L-AFQMC run with a 1 determinant trial for all-electron and frozen-core (without MP2 core-valence corrections), for the "TZ" basis set as used in the main text. For comparison, SPE with a 200 determinant trial is also shown for all-electron, displaying slightly lower timestep erorr. All values are reported in atomic units.}
    \label{tab:CCl4_timestep}
\end{table}

\newpage
\section{3. AFQMC Procedure in Detail}
\addcontentsline{toc}{section}{3. AFQMC Procedure in Detail}

\subsection{L-AFQMC --- AFQMC 0, AFQMC I}

In selecting the first active space, the  number of orbitals for each atom is chosen according to one of the schemes in Table~\ref{tab:orbital maps}, and for the molecule the number of active orbitals is the sum of that for the constituent atoms. The number of active electrons per atom is the number of valence electrons starting from the last noble gas (H:1, He:2, Li:1, Be:2, B:3 ...). This automated procedure mimics "chemical intuition", where the orbital maps represent "electron shells" — for example, 4 orbitals from the last noble gas constitute the s and p shells in the first and second rows (where the 1st row starts at Li, as commonly denoted in the literature). We find that this first choice of active space also often results in a relatively large energy separation between the active and inactive orbitals in the canonical restricted Hartree Fock basis, which is another way that has been used in the literature to choose active orbitals~\cite{neugebauer2023toward,rudshteyn2022calculation} and which we consider for manual selection of active spaces for some molecules in the all-electron calculations (see SI Section 5), but not for AFMQC 0 which is fully automated. For AFQMC 0, orbital map I from Table~\ref{tab:orbital maps} is used for every molecule.

\begin{table}
    \centering
    \begin{tabular}{|c|c|c|c|} \hline 
         \thead{First AS \\ Orbital Maps }&  H&  1st row& 2nd row
 \\ \hline 
         I&  0&  4& 4
 \\ \hline 
         II&  1&  8& 10
 \\ \hline 
         III&  1&  17& 19 \\ \hline
 IV& 1& 23& 23\\\hline
 V& 1& 27& 23\\\hline
    \end{tabular}
    \caption{Number of active orbitals that are chosen for each constituent atom in the molecule, for the first round of SHCI. 1st row refers to atoms from lithium to neon, and 2nd row refers to atoms from sodium to argon. AFQMC 0 is fully automated and only uses orbital map I. AFQMC I aims to achieve an economical balance between orbital maps I, II, III, and in a small number of cases, IV or V, as well as the NOON thresholds that determine the active space to run a second and final SHCI.}
    \label{tab:orbital maps}
\end{table}

AFQMC 0 selects the second active space from orbitals that are between 0.01 and 1.99 occupancy from NOONs. We used a SHCI threshold of \(\epsilon_1=10^{-4}\) for the first CASCI, and an \(\epsilon_1=10^{-5}\) for the second CASCI.

\begin{table}
    \footnotesize
    \centering
    \begin{tabular}{|ccccccccc|}
         \hline
         Valence orbital map&  NOON&  Initial AS&  Final AS&  Dets&  CI \%&  CBS Energy&  Sterr & Deviation\\
          & & & & & & (Ha) & (kcal/mol) & (kcal/mol) \\ \hline
         I&  0.01&  4e+4e, 8o&  1e+1e, 2o&  2&  99.50\%&  -75.82118&  0.43& -14.62\\ \hline
         II&  0.001&  4e+4e, 16o&  4e+4e, 8o&  62&  99.50\%&  -75.834882&  0.34& -6.02\\ \hline
         III&  0.005&  4e+4e, 34o&  4e+4e, 12o&  289&  99.50\%&  -75.84212&  0.20& -1.48\\ \hline
         III&  0.001&  4e+4e, 34o&  4e+4e, 17o&  771&  99.50\%&  -75.84274&  0.13& -1.09\\ \hline
    \end{tabular}
    \caption{Convergence of C2 with respect to active space and number of determinants. Refer to Table~\ref{tab:orbital maps} for orbital map meanings. Deviation is reported in kcal/mol from the reference heat of formation (see main text Section 2.9 for details).}
    \label{tab:c2_convergence}
\end{table}

In cases where we seek to obtain a large final active space, we use a larger number of orbitals for the first SHCI. We take the outliers from AFQMC 0 and use orbital map II, and if that does not the converge the result then we move on top orbital map III. The systematic progression is as follows: AFQMC 0 (orbital map I, NOON threshold 0.01) $\rightarrow$ orbital map II, NOON threshold 0.001 $\rightarrow$ orbital map III, NOON threshold 0.005. Only BN, LiF and Li2 required orbital maps IV and V. AFQMC 0 is deliberately chosen to be loose, while the following progressions prioritize using a larger first active space and the NOON threshold were chosen based on how many determinants can be realistically used with L-AFQMC (which for this work we cap at 2000 or 3000) where we still maintain close to 99\% of the CI weight. Table~\ref{tab:All_molecules_trials} show the trials required for each molecule, where "I, 0.01" is always the AFQMC 0 trial, and for AFQMC I and AFQMC II, the outliers are replaced with more sophisticated trials, where available. The trial shown is the smallest orbital map required (out of the list in Table~\ref{tab:orbital maps}) to achieve < 2kcal/mol deviation, and the combination of these trials comprise AFQMC I. The few molecules which have thresholds that deviate from the "default" NOON thresholds of 0.001 or 0.005 for orbital map II and III respectively are either limited by the number of determinants that our computational resources allow (if higher) or require more determinants (if lower). See Table~\ref{tab:c2_convergence} for an example. As mentioned in the main text, we run a few additional W-AFQMC trials other than the AFQMC I outliers, to check the updated experimental values or geometry discrepancies (see main text Section 2.2 and Table~\ref{tab:W-AFQMC} below).

{
\tiny
\begin{longtable}{|p{2cm} p{0.85cm}p{1.2cm}p{1.2cm}p{1.4cm}p{1.15cm}p{1.15cm}p{1.25cm} p{1.15cm}p{1.15cm}p{1.25cm}p{0cm}|}
\caption{The full list of molecules and trials (valence orbital map, NOON threshold, and CI\% retained if not 99.5\%) are reported, as well as if the molecule is an outlier (>2 kcal/mol deviation from the reference value), and the number of determinants for each trial. Unless indicated otherwise, the trials shown are for L-AFQMC, and "I, 0.01" is the AFQMC 0 protocol, while AFQMC I uses more sophisticated trials for the outliers of AFQMC 0. The number of determinants shown are for the TZ basis set.} % needs to go inside longtable environment
\label{tab:All_molecules_trials}
\endfirsthead
\endhead
\hline
Molecule	&	Datasets	&	AFQMC 0 &	AFQMC I	&	AFQMC II	&	AFQMC 0 &	AFQMC I	&	AFQMC II	& AFQMC 0 &	AFQMC I	&	AFQMC II &\\ 
	&		&	&	&	&	outlier? &	outlier?	&	outlier?	& dets &	dets &	dets &\\ \hline
2-butyne	&	G2	&	I, 0.01	&	I, 0.01	&	I, 0.01	&		&		&		&	1	&	1	&	1	& \\ \hline
Acetaldehyde	&	G2	&	I, 0.01	&	I, 0.01	&	I, 0.01	&		&		&		&	1	&	1	&	1	& \\ \hline
Acethylene	&	G2	&	I, 0.01	&	I, 0.01	&	I, 0.01	&		&		&		&	1	&	1	&	1	& \\ \hline
Acetone	&	G2	&	I, 0.01	&	I, 0.01	&	I, 0.01	&		&		&		&	1	&	1	&	1	& \\ \hline
AlCl3	&	G2	&	I, 0.01	&	I, 0.01	&	W-AFQMC	&		&		&		&	1	&	1	&	10000	& \\ \hline
AlF3	&	G2	&	I, 0.01	&	I, 0.01	&	W-AFQMC	&		&		&		&	1	&	1	&	10000	& \\ \hline
Allene	&	G2	&	I, 0.01	&	I, 0.01	&	I, 0.01	&		&		&		&	1	&	1	&	1	& \\ \hline
Aziridine	&	G2	&	I, 0.01	&	I, 0.01	&	I, 0.01	&		&		&		&	1	&	1	&	1	& \\ \hline
BCl3	&	G2	&	I, 0.01	&	I, 0.01	&	I, 0.01	&		&		&		&	1	&	1	&	1	& \\ \hline
Benzene	&	G2	&	I, 0.01	&	I, 0.01	&	I, 0.01	&		&		&		&	5	&	5	&	5	& \\ \hline
BF3	&	G2	&	I, 0.01	&	I, 0.01	&	I, 0.01	&		&		&		&	1	&	1	&	1	& \\ \hline
Bicyclo-1-1-0-butane	&	G2	&	I, 0.01	&	III, 0.005	&	W-AFQMC	&	Yes	&		&		&	1	&	234	&	10000	& \\ \hline
CCl2CCl2	&	G2	&	I, 0.01	&	I, 0.01	&	I, 0.01	&		&		&		&	2	&	2	&	2	& \\ \hline
CCl4	&	G2	&	I, 0.01	&	I, 0.01	&	I, 0.01	&		&		&		&	1	&	1	&	1	& \\ \hline
CF2CF2	&	G2	&	I, 0.01	&	I, 0.01	&	W-AFQMC	&		&		&		&	1	&	1	&	10000	& \\ \hline
CF3-CN	&	G2	&	I, 0.01	&	I, 0.01	&	I, 0.01	&		&		&		&	1	&	1	&	1	& \\ \hline
CF4	&	G2	&	I, 0.01	&	I, 0.01	&	I, 0.01	&		&		&		&	1	&	1	&	1	& \\ \hline
CH2CH-CN	&	G2	&	I, 0.01	&	I, 0.01	&	W-AFQMC	&		&		&		&	2	&	2	&	10000	& \\ \hline
CH2Cl2	&	G2	&	I, 0.01	&	I, 0.01	&	I, 0.01	&		&		&		&	1	&	1	&	1	& \\ \hline
CH2F2	&	G2	&	I, 0.01	&	I, 0.01	&	I, 0.01	&		&		&		&	1	&	1	&	1	& \\ \hline
CH3-CH2-CH2-Cl	&	G2	&	I, 0.01	&	I, 0.01	&	I, 0.01	&		&		&		&	1	&	1	&	1	& \\ \hline
CH3-CH2-Cl	&	G2	&	I, 0.01	&	I, 0.01	&	I, 0.01	&		&		&		&	1	&	1	&	1	& \\ \hline
CH3-CH2-O-CH3	&	G2	&	I, 0.01	&	I, 0.01	&	I, 0.01	&		&		&		&	1	&	1	&	1	& \\ \hline
CH3-CH2-SH	&	G2	&	I, 0.01	&	I, 0.01	&	I, 0.01	&		&		&		&	1	&	1	&	1	& \\ \hline
CH3-CN	&	G2	&	I, 0.01	&	I, 0.01	&	I, 0.01	&		&		&		&	1	&	1	&	1	& \\ \hline
CH3-O-CH3	&	G2	&	I, 0.01	&	I, 0.01	&	I, 0.01	&		&		&		&	1	&	1	&	1	& \\ \hline
CH3-O-NO	&	G2	&	I, 0.01	&	I, 0.01	&	I, 0.01	&		&		&		&	2	&	2	&	2	& \\ \hline
CH3-S-CH3	&	G2	&	I, 0.01	&	I, 0.01	&	I, 0.01	&		&		&		&	1	&	1	&	1	& \\ \hline
CH3-SH	&	G2	&	I, 0.01	&	I, 0.01	&	I, 0.01	&		&		&		&	1	&	1	&	1	& \\ \hline
CH3-SiH3	&	G2	&	I, 0.01	&	I, 0.01	&	I, 0.01	&		&		&		&	1	&	1	&	1	& \\ \hline
CH3CFO	&	G2	&	I, 0.01	&	I, 0.01	&	I, 0.01	&		&		&		&	1	&	1	&	1	& \\ \hline
CH3Cl	&	G2	&	I, 0.01	&	I, 0.01	&	I, 0.01	&		&		&		&	1	&	1	&	1	& \\ \hline
CH3COCl	&	G2	&	I, 0.01	&	I, 0.01	&	I, 0.01	&		&		&		&	1	&	1	&	1	& \\ \hline
CH3CONH2	&	G2	&	I, 0.01	&	I, 0.01	&	I, 0.01	&		&		&		&	1	&	1	&	1	& \\ \hline
CH3COOH	&	G2	&	I, 0.01	&	I, 0.01	&	I, 0.01	&		&		&		&	1	&	1	&	1	& \\ \hline
CH3NO2	&	G2	&	I, 0.01	&	I, 0.01	&	I, 0.01	&		&		&		&	3	&	3	&	3	& \\ \hline
CH4	&	G2	&	I, 0.01	&	I, 0.01	&	I, 0.01	&		&		&		&	1	&	1	&	1	& \\ \hline
CHCl3	&	G2	&	I, 0.01	&	I, 0.01	&	I, 0.01	&		&		&		&	1	&	1	&	1	& \\ \hline
Cl2	&	G2	&	I, 0.01	&	I, 0.01	&	I, 0.01	&		&		&		&	2	&	2	&	2	& \\ \hline
ClF	&	G2	&	I, 0.01	&	I, 0.01	&	I, 0.01	&		&		&		&	2	&	2	&	2	& \\ \hline
CLF3	&	G2, MR	&	I, 0.01	&	II, 0.001	&	II, 0.001	&	Yes	&		&		&	2	&	1971	&	1971	& \\ \hline
ClNO	&	G2	&	I, 0.01	&	I, 0.01	&	I, 0.01	&		&		&		&	13	&	13	&	13	& \\ \hline
CO	&	G2	&	I, 0.01	&	I, 0.01	&	I, 0.01	&		&		&		&	1	&	1	&	1	& \\ \hline
CO2	&	G2	&	I, 0.01	&	I, 0.01	&	I, 0.01	&		&		&		&	1	&	1	&	1	& \\ \hline
CS	&	G2	&	I, 0.01	&	I, 0.01	&	I, 0.01	&		&		&		&	12	&	12	&	12	& \\ \hline
CS2	&	G2	&	I, 0.01	&	I, 0.01	&	I, 0.01	&		&		&		&	13	&	13	&	13	& \\ \hline
Cyanogen	&	G2	&	I, 0.01	&	I, 0.01	&	I, 0.01	&		&		&		&	12	&	12	&	12	& \\ \hline
Cyclobutane	&	G2	&	I, 0.01	&	I, 0.01	&	I, 0.01	&		&		&		&	1	&	1	&	1	& \\ \hline
Cyclobutene	&	G2	&	I, 0.01	&	I, 0.01	&	W-AFQMC	&		&		&		&	1	&	1	&	10000	& \\ \hline
Cyclopropane	&	G2	&	I, 0.01	&	I, 0.01	&	I, 0.01	&		&		&		&	1	&	1	&	1	& \\ \hline
Cyclopropene	&	G2	&	I, 0.01	&	I, 0.01	&	I, 0.01	&		&		&		&	1	&	1	&	1	& \\ \hline
Dimethylamine	&	G2	&	I, 0.01	&	I, 0.01	&	I, 0.01	&		&		&		&	1	&	1	&	1	& \\ \hline
Dimethylsulfoxide	&	G2	&	I, 0.01	&	I, 0.01	&	I, 0.01	&		&		&		&	1	&	1	&	1	& \\ \hline
Ethane	&	G2	&	I, 0.01	&	I, 0.01	&	I, 0.01	&		&		&		&	1	&	1	&	1	& \\ \hline
Ethanol	&	G2	&	I, 0.01	&	I, 0.01	&	I, 0.01	&		&		&		&	1	&	1	&	1	& \\ \hline
Ethenone	&	G2	&	I, 0.01	&	I, 0.01	&	I, 0.01	&		&		&		&	1	&	1	&	1	& \\ \hline
Ethylene	&	G2	&	I, 0.01	&	I, 0.01	&	I, 0.01	&		&		&		&	1	&	1	&	1	& \\ \hline
F2	&	G2	&	I, 0.01	&	II, 0.001	&	II, 0.001	&	Yes	&		&		&	2	&	13	&	13	& \\ \hline
F2O	&	G2	&	I, 0.01	&	II, 0.001	&	II, 0.001	&	Yes	&		&		&	1	&	257	&	257	& \\ \hline
Furan	&	G2	&	I, 0.01	&	I, 0.01	&	I, 0.01	&		&		&		&	1	&	1	&	1	& \\ \hline
Glyoxal	&	G2	&	I, 0.01	&	I, 0.01	&	I, 0.01	&		&		&		&	3	&	3	&	3	& \\ \hline
H2	&	G2	&	I, 0.01	&	I, 0.01	&	I, 0.01	&		&		&		&	1	&	1	&	1	& \\ \hline
H2CO	&	G2	&	I, 0.01	&	I, 0.01	&	I, 0.01	&		&		&		&	1	&	1	&	1	& \\ \hline
H2NNH2	&	G2	&	I, 0.01	&	I, 0.01	&	I, 0.01	&		&		&		&	1	&	1	&	1	& \\ \hline
H2O	&	G2	&	I, 0.01	&	I, 0.01	&	I, 0.01	&		&		&		&	1	&	1	&	1	& \\ \hline
HCF3	&	G2	&	I, 0.01	&	I, 0.01	&	I, 0.01	&		&		&		&	1	&	1	&	1	& \\ \hline
HCl	&	G2	&	I, 0.01	&	I, 0.01	&	I, 0.01	&		&		&		&	2	&	2	&	2	& \\ \hline
HCN	&	G2	&	I, 0.01	&	I, 0.01	&	I, 0.01	&		&		&		&	1	&	1	&	1	& \\ \hline
HCOOCH3	&	G2	&	I, 0.01	&	I, 0.01	&	I, 0.01	&		&		&		&	1	&	1	&	1	& \\ \hline
HCOOH	&	G2	&	I, 0.01	&	I, 0.01	&	I, 0.01	&		&		&		&	1	&	1	&	1	& \\ \hline
HF	&	G2	&	I, 0.01	&	I, 0.01	&	I, 0.01	&		&		&		&	1	&	1	&	1	& \\ \hline
HOCl	&	G2	&	I, 0.01	&	I, 0.01	&	I, 0.01	&		&		&		&	1	&	1	&	1	& \\ \hline
HOOH	&	G2	&	I, 0.01	&	I, 0.01	&	I, 0.01	&		&		&		&	1	&	1	&	1	& \\ \hline
Isobutane	&	G2	&	I, 0.01	&	I, 0.01	&	I, 0.01	&		&		&		&	1	&	1	&	1	& \\ \hline
Isobutene	&	G2	&	I, 0.01	&	I, 0.01	&	I, 0.01	&		&		&		&	1	&	1	&	1	& \\ \hline
Isopropyl-alcohol	&	G2	&	I, 0.01	&	I, 0.01	&	I, 0.01	&		&		&		&	1	&	1	&	1	& \\ \hline
Ketene	&	G2	&	I, 0.01	&	I, 0.01	&	I, 0.01	&		&		&		&	1	&	1	&	1	& \\ \hline
Li2	&	G2	&	I, 0.01	&	V, 1e-5	&	V, 1e-5	&	Yes	&		&		&	1	&	5	&	5	& \\ \hline
LiF	&	G2	&	I, 0.01	&	V, 1e-5	&	W-AFQMC	&	Yes	&		&		&	1	&	48	&	10000	& \\ \hline
LiH	&	G2	&	I, 0.01	&	I, 0.01	&	I, 0.01	&		&		&		&	1	&	1	&	1	& \\ \hline
Methanol	&	G2	&	I, 0.01	&	I, 0.01	&	I, 0.01	&		&		&		&	1	&	1	&	1	& \\ \hline
Methylamine	&	G2	&	I, 0.01	&	I, 0.01	&	I, 0.01	&		&		&		&	1	&	1	&	1	& \\ \hline
Methylene-cyclopropane	&	G2	&	I, 0.01	&	I, 0.01	&	I, 0.01	&		&		&		&	1	&	1	&	1	& \\ \hline
N2	&	G2	&	I, 0.01	&	I, 0.01	&	I, 0.01	&		&		&		&	1	&	1	&	1	& \\ \hline
Na2	&	G2	&	I, 0.01	&	I, 0.01	&	I, 0.01	&		&		&		&	2	&	2	&	2	& \\ \hline
NaCl	&	G2	&	I, 0.01	&	I, 0.01	&	I, 0.01	&		&		&		&	1	&	1	&	1	& \\ \hline
NF3	&	G2	&	I, 0.01	&	I, 0.01	&	I, 0.01	&		&		&		&	1	&	1	&	1	& \\ \hline
NH3	&	G2	&	I, 0.01	&	I, 0.01	&	I, 0.01	&		&		&		&	1	&	1	&	1	& \\ \hline
NNO	&	G2	&	I, 0.01	&	I, 0.01	&	I, 0.01	&		&		&		&	1	&	1	&	1	& \\ \hline
OCS-m1	&	G2	&	I, 0.01	&	II, 0.001	&	II, 0.001	&	Yes	&		&		&	1	&	439	&	439	& \\ \hline
Oxirane	&	G2	&	I, 0.01	&	I, 0.01	&	I, 0.01	&		&		&		&	1	&	1	&	1	& \\ \hline
Ozone	&	G2, MR	&	I, 0.01	&	III, 0.002, 98.5\%	&	W-AFQMC	&	Yes	&	Yes	&		&	3	&	2234	&	10000	& \\ \hline
P2	&	G2	&	I, 0.01	&	I, 0.01	&	I, 0.01	&		&		&		&	15	&	15	&	15	& \\ \hline
PF3	&	G2	&	I, 0.01	&	I, 0.01	&	I, 0.01	&		&		&		&	1	&	1	&	1	& \\ \hline
PH3	&	G2	&	I, 0.01	&	I, 0.01	&	I, 0.01	&		&		&		&	1	&	1	&	1	& \\ \hline
Propane	&	G2	&	I, 0.01	&	I, 0.01	&	I, 0.01	&		&		&		&	1	&	1	&	1	& \\ \hline
Propene-CS	&	G2	&	I, 0.01	&	I, 0.01	&	I, 0.01	&		&		&		&	1	&	1	&	1	& \\ \hline
Propyne	&	G2	&	I, 0.01	&	I, 0.01	&	I, 0.01	&		&		&		&	1	&	1	&	1	& \\ \hline
Pyridine	&	G2	&	I, 0.01	&	I, 0.01	&	I, 0.01	&		&		&		&	5	&	5	&	5	& \\ \hline
Pyrole	&	G2	&	I, 0.01	&	I, 0.01	&	I, 0.01	&		&		&		&	1	&	1	&	1	& \\ \hline
SH2	&	G2	&	I, 0.01	&	I, 0.01	&	I, 0.01	&		&		&		&	1	&	1	&	1	& \\ \hline
Si2H6	&	G2	&	I, 0.01	&	I, 0.01	&	I, 0.01	&		&		&		&	1	&	1	&	1	& \\ \hline
SiCl4	&	G2	&	I, 0.01	&	I, 0.01	&	I, 0.01	&		&		&		&	1	&	1	&	1	& \\ \hline
SiF4	&	G2	&	I, 0.01	&	I, 0.01	&	I, 0.01	&		&		&		&	1	&	1	&	1	& \\ \hline
SiH4	&	G2	&	I, 0.01	&	I, 0.01	&	I, 0.01	&		&		&		&	1	&	1	&	1	& \\ \hline
SiO	&	G2	&	I, 0.01	&	I, 0.01	&	I, 0.01	&		&		&		&	1	&	1	&	1	& \\ \hline
SO2	&	G2	&	I, 0.01	&	I, 0.01	&	W-AFQMC	&		&		&		&	3	&	3	&	10000	& \\ \hline
Spiropentane	&	G2	&	I, 0.01	&	I, 0.01	&	I, 0.01	&		&		&		&	1	&	1	&	1	& \\ \hline
Thiooxirane	&	G2	&	I, 0.01	&	II, 0.001	&	II, 0.001	&	Yes	&		&		&	1	&	24	&	24	& \\ \hline
Thiophene	&	G2	&	I, 0.01	&	I, 0.01	&	I, 0.01	&		&		&		&	1	&	1	&	1	& \\ \hline
Trans-1-3-butadiene	&	G2	&	I, 0.01	&	II, 0.001	&	II, 0.001	&	Yes	&		&		&	1	&	15	&	15	& \\ \hline
Trans-butane	&	G2	&	I, 0.01	&	I, 0.01	&	I, 0.01	&		&		&		&	1	&	1	&	1	& \\ \hline
Trans-ethylamine	&	G2	&	I, 0.01	&	I, 0.01	&	I, 0.01	&		&		&		&	1	&	1	&	1	& \\ \hline
Trimethyl-amine	&	G2	&	I, 0.01	&	I, 0.01	&	I, 0.01	&		&		&		&	1	&	1	&	1	& \\ \hline
Vinyl-chloride	&	G2	&	I, 0.01	&	I, 0.01	&	W-AFQMC	&		&		&		&	1	&	1	&	10000	& \\ \hline
Vynil-fluoride	&	G2	&	I, 0.01	&	I, 0.01	&	I, 0.01	&		&		&		&	1	&	1	&	1	& \\ \hline
1,3-cyclohexadiene	&	G3	&	I, 0.01	&	II, 0.001	&	II, 0.001	&	Yes	&		&		&	1	&	131	&	131	& \\ \hline
1,3-DiFluorobenzene	&	G3	&	I, 0.01	&	I, 0.01	&	I, 0.01	&		&		&		&	2	&	2	&	2	& \\ \hline
1,4-DiFluorobenzene	&	G3	&	I, 0.01	&	I, 0.01	&	I, 0.01	&		&		&		&	6	&	6	&	6	& \\ \hline
2-methyl	&	G3	&	I, 0.01	&	I, 0.01	&	I, 0.01	&		&		&		&	1	&	1	&	1	& \\ \hline
2,5-Dihydrothiophene	&	G3	&	I, 0.01	&	I, 0.01	&	I, 0.01	&		&		&		&	1	&	1	&	1	& \\ \hline
3-methyl	&	G3	&	I, 0.01	&	I, 0.01	&	I, 0.01	&		&		&		&	1	&	1	&	1	& \\ \hline
Acetic	&	G3	&	I, 0.01	&	II, 0.001	&	II, 0.001	&	Yes	&		&		&	1	&	1166	&	1166	& \\ \hline
azulene	&	G3	&	I, 0.01	&	II, 0.001, 98\%	&	W-AFQMC	&	Yes	&		&		&	12	&	2347	&	10000	& \\ \hline
benzoquinone	&	G3	&	I, 0.01	&	I, 0.01	&	W-AFQMC	&		&		&		&	6	&	6	&	10000	& \\ \hline
c2f6	&	G3	&	I, 0.01	&	I, 0.01	&	W-AFQMC	&		&		&		&	1	&	1	&	10000	& \\ \hline
C4H4N2	&	G3	&	I, 0.01	&	II, 0.001	&	W-AFQMC	&	Yes	&	Yes	&		&	7	&	736	&	10000	& \\ \hline
C4H6	&	G3	&	I, 0.01	&	I, 0.01	&	I, 0.01	&		&		&		&	1	&	1	&	1	& \\ \hline
C4H6O	&	G3	&	I, 0.01	&	I, 0.01	&	I, 0.01	&		&		&		&	1	&	1	&	1	& \\ \hline
C4H8O2	&	G3	&	I, 0.01	&	I, 0.01	&	I, 0.01	&		&		&		&	1	&	1	&	1	& \\ \hline
C5H8	&	G3	&	I, 0.01	&	I, 0.01	&	I, 0.01	&		&		&		&	1	&	1	&	1	& \\ \hline
C6H12	&	G3	&	I, 0.01	&	I, 0.01	&	I, 0.01	&		&		&		&	1	&	1	&	1	& \\ \hline
C6H5-CH3	&	G3	&	I, 0.01	&	I, 0.01	&	I, 0.01	&		&		&		&	1	&	1	&	1	& \\ \hline
C6H5-NH2	&	G3	&	I, 0.01	&	I, 0.01	&	I, 0.01	&		&		&		&	1	&	1	&	1	& \\ \hline
C6H5-OH	&	G3	&	I, 0.01	&	I, 0.01	&	I, 0.01	&		&		&		&	1	&	1	&	1	& \\ \hline
cf3cl	&	G3	&	I, 0.01	&	I, 0.01	&	I, 0.01	&		&		&		&	1	&	1	&	1	& \\ \hline
CH3\_2CH-CHO	&	G3	&	I, 0.01	&	I, 0.01	&	I, 0.01	&		&		&		&	1	&	1	&	1	& \\ \hline
CH3\_2CH-CN	&	G3	&	I, 0.01	&	I, 0.01	&	I, 0.01	&		&		&		&	1	&	1	&	1	& \\ \hline
CH3\_2CH-O-CH\_CH3\_2	&	G3	&	I, 0.01	&	I, 0.01	&	I, 0.01	&		&		&		&	1	&	1	&	1	& \\ \hline
CH3\_3C-NH2	&	G3	&	I, 0.01	&	I, 0.01	&	I, 0.01	&		&		&		&	1	&	1	&	1	& \\ \hline
CH3\_3C-O-CH3	&	G3	&	I, 0.01	&	I, 0.01	&	I, 0.01	&		&		&		&	1	&	1	&	1	& \\ \hline
CH3\_3C-SH	&	G3	&	I, 0.01	&	I, 0.01	&	I, 0.01	&		&		&		&	1	&	1	&	1	& \\ \hline
CH3-C\_O\_-CCH	&	G3	&	I, 0.01	&	III, 0.005,98.5\%	&	W-AFQMC	&	Yes	&	Yes	&	Yes	&	2	&	1701	&	10000	& \\ \hline
CH3-C\_O\_-O-CH\_CH3\_2	&	G3	&	I, 0.01	&	I, 0.01	&	I, 0.01	&		&		&		&	1	&	1	&	1	& \\ \hline
CH3-C\_O\_-OCH3	&	G3	&	I, 0.01	&	I, 0.01	&	I, 0.01	&		&		&		&	1	&	1	&	1	& \\ \hline
CH3-CH\_OCH3\_2	&	G3	&	I, 0.01	&	I, 0.01	&	I, 0.01	&		&		&		&	1	&	1	&	1	& \\ \hline
CH3-CH2-CH\_CH3\_-NO2	&	G3	&	I, 0.01	&	I, 0.01	&	I, 0.01	&		&		&		&	3	&	3	&	3	& \\ \hline
CH3-CH2-CO-CH2-CH3	&	G3	&	I, 0.01	&	II, 0.001	&	II, 0.001	&	Yes	&		&		&	1	&	2	&	2	& \\ \hline
CH3-CH2-O-CH2-CH3	&	G3	&	I, 0.01	&	I, 0.01	&	I, 0.01	&		&		&		&	1	&	1	&	1	& \\ \hline
CH3-CH2-S-S-CH2-CH3	&	G3	&	I, 0.01	&	I, 0.01	&	I, 0.01	&		&		&		&	1	&	1	&	1	& \\ \hline
CH3-CHCH-CHO	&	G3	&	I, 0.01	&	I, 0.01	&	I, 0.01	&		&		&		&	2	&	2	&	2	& \\ \hline
CH3-CO-CH2-CH3	&	G3	&	I, 0.01	&	I, 0.01	&	I, 0.01	&		&		&		&	1	&	1	&	1	& \\ \hline
Chlorobenzene	&	G3	&	I, 0.01	&	I, 0.01	&	I, 0.01	&		&		&		&	5	&	5	&	5	& \\ \hline
Cl2O2S	&	G3	&	I, 0.01	&	I, 0.01	&	W-AFQMC	&		&		&		&	2	&	2	&	10000	& \\ \hline
Cl2S2	&	G3	&	I, 0.01	&	I, 0.01	&	I, 0.01	&		&		&		&	6	&	6	&	6	& \\ \hline
cyclooctatetraene	&	G3	&	I, 0.01	&	I, 0.01	&	W-AFQMC	&		&		&		&	1	&	1	&	10000	& \\ \hline
cyclopentane	&	G3	&	I, 0.01	&	I, 0.01	&	I, 0.01	&		&		&		&	1	&	1	&	1	& \\ \hline
cyclopentanone	&	G3	&	I, 0.01	&	I, 0.01	&	I, 0.01	&		&		&		&	1	&	1	&	1	& \\ \hline
dimethyl	&	G3	&	I, 0.01	&	I, 0.01	&	I, 0.01	&		&		&		&	1	&	1	&	1	& \\ \hline
Fluorobenzene	&	G3	&	I, 0.01	&	I, 0.01	&	I, 0.01	&		&		&		&	1	&	1	&	1	& \\ \hline
n-Butyl	&	G3	&	I, 0.01	&	I, 0.01	&	I, 0.01	&		&		&		&	1	&	1	&	1	& \\ \hline
n-heptane	&	G3	&	I, 0.01	&	I, 0.01	&	I, 0.01	&		&		&		&	1	&	1	&	1	& \\ \hline
n-hexane	&	G3	&	I, 0.01	&	I, 0.01	&	I, 0.01	&		&		&		&	1	&	1	&	1	& \\ \hline
N-methyl	&	G3	&	I, 0.01	&	I, 0.01	&	I, 0.01	&		&		&		&	1	&	1	&	1	& \\ \hline
n-octane	&	G3	&	I, 0.01	&	I, 0.01	&	I, 0.01	&		&		&		&	1	&	1	&	1	& \\ \hline
n-pentane	&	G3	&	I, 0.01	&	I, 0.01	&	I, 0.01	&		&		&		&	1	&	1	&	1	& \\ \hline
Naphthalene	&	G3	&	I, 0.01	&	I, 0.01	&	W-AFQMC	&		&		&		&	7	&	7	&	10000	& \\ \hline
NC-CH2-CH2-CN	&	G3	&	I, 0.01	&	I, 0.01	&	I, 0.01	&		&		&		&	1	&	1	&	1	& \\ \hline
Neopentane	&	G3	&	I, 0.01	&	I, 0.01	&	I, 0.01	&		&		&		&	1	&	1	&	1	& \\ \hline
P4	&	G3	&	I, 0.01	&	I, 0.01	&	I, 0.01	&		&		&		&	1	&	1	&	1	& \\ \hline
para-cyclohexadiene	&	G3	&	I, 0.01	&	I, 0.01	&	I, 0.01	&		&		&		&	1	&	1	&	1	& \\ \hline
PCl3	&	G3	&	I, 0.01	&	I, 0.01	&	I, 0.01	&		&		&		&	1	&	1	&	1	& \\ \hline
PCl5	&	G3	&	I, 0.01	&	I, 0.01	&	I, 0.01	&		&		&		&	1	&	1	&	1	& \\ \hline
Perhydropyridine	&	G3	&	I, 0.01	&	I, 0.01	&	I, 0.01	&		&		&		&	1	&	1	&	1	& \\ \hline
pf5	&	G3	&	I, 0.01	&	I, 0.01	&	I, 0.01	&		&		&		&	1	&	1	&	1	& \\ \hline
POCl3	&	G3	&	I, 0.01	&	I, 0.01	&	I, 0.01	&		&		&		&	1	&	1	&	1	& \\ \hline
pyrimidine	&	G3	&	I, 0.01	&	II, 0.001	&	II, 0.001	&	Yes	&		&		&	6	&	1054	&	1054	& \\ \hline
SCl2	&	G3	&	I, 0.01	&	I, 0.01	&	I, 0.01	&		&		&		&	1	&	1	&	1	& \\ \hline
sf6	&	G3	&	I, 0.01	&	I, 0.01	&	W-AFQMC	&		&		&		&	1	&	1	&	10000	& \\ \hline
SiCl2	&	G3	&	I, 0.01	&	I, 0.01	&	I, 0.01	&		&		&		&	2	&	2	&	2	& \\ \hline
SO3	&	G3	&	I, 0.01	&	I, 0.01	&	I, 0.01	&		&		&		&	3	&	1	&	1	& \\ \hline
t-butanol	&	G3	&	I, 0.01	&	I, 0.01	&	I, 0.01	&		&		&		&	1	&	1	&	1	& \\ \hline
t-Butyl	&	G3	&	I, 0.01	&	I, 0.01	&	I, 0.01	&		&		&		&	1	&	1	&	1	& \\ \hline
tetrahydrofuran	&	G3	&	I, 0.01	&	I, 0.01	&	I, 0.01	&		&		&		&	1	&	1	&	1	& \\ \hline
Tetrahydropyran	&	G3	&	I, 0.01	&	I, 0.01	&	I, 0.01	&		&		&		&	1	&	1	&	1	& \\ \hline
Tetrahydropyrrole	&	G3	&	I, 0.01	&	I, 0.01	&	I, 0.01	&		&		&		&	1	&	1	&	1	& \\ \hline
Tetrahydrothiophene	&	G3	&	I, 0.01	&	I, 0.01	&	I, 0.01	&		&		&		&	1	&	1	&	1	& \\ \hline
Tetrahydrothiopyran	&	G3	&	I, 0.01	&	I, 0.01	&	I, 0.01	&		&		&		&	1	&	1	&	1	& \\ \hline
Tetramethylsilane	&	G3	&	I, 0.01	&	I, 0.01	&	I, 0.01	&		&		&		&	1	&	1	&	1	& \\ \hline
alcl	&	W4-11	&	I, 0.01	&	I, 0.01	&	I, 0.01	&		&		&		&	1	&	1	&	1	& \\ \hline
alf	&	W4-11	&	I, 0.01	&	I, 0.01	&	I, 0.01	&		&		&		&	1	&	1	&	1	& \\ \hline
alh	&	W4-11	&	I, 0.01	&	I, 0.01	&	I, 0.01	&		&		&		&	1	&	1	&	1	& \\ \hline
alh3	&	W4-11	&	I, 0.01	&	I, 0.01	&	I, 0.01	&		&		&		&	1	&	1	&	1	& \\ \hline
b2h6	&	W4-11	&	I, 0.01	&	I, 0.01	&	I, 0.01	&		&		&		&	1	&	1	&	1	& \\ \hline
bf	&	W4-11	&	I, 0.01	&	I, 0.01	&	I, 0.01	&		&		&		&	1	&	1	&	1	& \\ \hline
bh	&	W4-11	&	I, 0.01	&	I, 0.01	&	I, 0.01	&		&		&		&	1	&	1	&	1	& \\ \hline
bh3	&	W4-11	&	I, 0.01	&	I, 0.01	&	I, 0.01	&		&		&		&	1	&	1	&	1	& \\ \hline
bhf2	&	W4-11	&	I, 0.01	&	I, 0.01	&	I, 0.01	&		&		&		&	1	&	1	&	1	& \\ \hline
c-hono	&	W4-11	&	I, 0.01	&	I, 0.01	&	I, 0.01	&		&		&		&	2	&	2	&	2	& \\ \hline
c-n2h2	&	W4-11	&	I, 0.01	&	I, 0.01	&	I, 0.01	&		&		&		&	2	&	2	&	2	& \\ \hline
ch2nh	&	W4-11	&	I, 0.01	&	I, 0.01	&	I, 0.01	&		&		&		&	1	&	1	&	1	& \\ \hline
ch3f	&	W4-11	&	I, 0.01	&	I, 0.01	&	I, 0.01	&		&		&		&	1	&	1	&	1	& \\ \hline
clcn	&	W4-11	&	I, 0.01	&	I, 0.01	&	I, 0.01	&		&		&		&	1	&	1	&	1	& \\ \hline
dioxirane	&	W4-11	&	I, 0.01	&	I, 0.01	&	I, 0.01	&		&		&		&	1	&	1	&	1	& \\ \hline
f2co	&	W4-11	&	I, 0.01	&	II, 0.001	&	II, 0.001	&	Yes	&		&		&	1	&	25	&	25	& \\ \hline
fccf	&	W4-11	&	I, 0.01	&	I, 0.01	&	I, 0.01	&		&		&		&	1	&	1	&	1	& \\ \hline
hccf	&	W4-11	&	I, 0.01	&	I, 0.01	&	I, 0.01	&		&		&		&	1	&	1	&	1	& \\ \hline
hcno	&	W4-11	&	I, 0.01	&	I, 0.01	&	I, 0.01	&		&		&		&	1	&	1	&	1	& \\ \hline
hcof	&	W4-11	&	I, 0.01	&	I, 0.01	&	I, 0.01	&		&		&		&	1	&	1	&	1	& \\ \hline
hnco	&	W4-11	&	I, 0.01	&	II, 0.001	&	II, 0.001	&	Yes	&		&		&	1	&	50	&	50	& \\ \hline
hnnn	&	W4-11	&	I, 0.01	&	I, 0.01	&	I, 0.01	&		&		&		&	1	&	1	&	1	& \\ \hline
hno	&	W4-11	&	I, 0.01	&	I, 0.01	&	I, 0.01	&		&		&		&	2	&	2	&	2	& \\ \hline
hocn	&	W4-11	&	I, 0.01	&	I, 0.01	&	I, 0.01	&		&		&		&	1	&	1	&	1	& \\ \hline
hof	&	W4-11	&	I, 0.01	&	I, 0.01	&	I, 0.01	&		&		&		&	1	&	1	&	1	& \\ \hline
nh2cl	&	W4-11	&	I, 0.01	&	I, 0.01	&	I, 0.01	&		&		&		&	1	&	1	&	1	& \\ \hline
oxirene	&	W4-11	&	I, 0.01	&	I, 0.01	&	I, 0.01	&		&		&		&	1	&	1	&	1	& \\ \hline
s2o	&	W4-11	&	I, 0.01	&	I, 0.01	&	W-AFQMC	&		&		&		&	2	&	2	&	10000	& \\ \hline
sih3f&	W4-11	&	I, 0.01	&	I, 0.01	&	I, 0.01	&		&		&		&	1	&	1	&	1	& \\ \hline
t-hono	&	W4-11	&	I, 0.01	&	I, 0.01	&	I, 0.01	&		&		&		&	2	&	2	&	2	& \\ \hline
t-n2h2	&	W4-11	&	I, 0.01	&	I, 0.01	&	I, 0.01	&		&		&		&	1	&	1	&	1	& \\ \hline
c2h5f	&	W4-11	&	I, 0.01	&	I, 0.01	&	I, 0.01	&		&		&		&	1	&	1	&	1	& \\ \hline
bn	&	W4-11, MR	&	I, 0.01	&	IV, 0.001	&	IV, 0.001	&	Yes	&		&		&	2	&	1597	&	1597	& \\ \hline
c2	&	W4-11, MR	&	I, 0.01	&	III, 0.001	&	III, 0.001	&	Yes	&		&		&	2	&	771	&	771	& \\ \hline
cl2o	&	W4-11	&	I, 0.01	&	I, 0.01	&	I, 0.01	&		&		&		&	1	&	1	&	1	& \\ \hline
foof	&	W4-11, MR	&	I, 0.01	&	I, 0.01	&	W-AFQMC	&		&		&		&	13	&	13	&	10000	& \\ \hline
s3	&	W4-11, MR	&	I, 0.01	&	I, 0.01	&	W-AFQMC	&		&		&		&	3	&	3	&	10000	& \\ \hline
s4-c2v	&	W4-11, MR	&	I, 0.01	&	I, 0.01	&	W-AFQMC	&		&		&		&	10	&	10	&	10000	& \\ \hline
clf5	&	W4-17, MR	&	I, 0.01	&	II, 0.001, 98.5\%	&	II, 0.001, 98.5\%	&	Yes	&		&		&	2	&	2187	&	2187	& \\ \hline
cloocl	&	W4-17, MR	&	I, 0.01	&	I, 0.01	&	I, 0.01	&		&		&		&	5	&	5	&	5	& \\ \hline
beta-lactim	&	W4-17	&	I, 0.01	&	I, 0.01	&	I, 0.01	&		&		&		&	1	&	1	&	1	& \\ \hline
borole	&	W4-17	&	I, 0.01	&	I, 0.01	&	I, 0.01	&		&		&		&	1	&	1	&	1	& \\ \hline
c2cl2	&	W4-17	&	I, 0.01	&	I, 0.01	&	I, 0.01	&		&		&		&	1	&	1	&	1	& \\ \hline
c2cl6	&	W4-17	&	I, 0.01	&	I, 0.01	&	I, 0.01	&		&		&		&	1	&	1	&	1	& \\ \hline
c2clh	&	W4-17	&	I, 0.01	&	I, 0.01	&	I, 0.01	&		&		&		&	1	&	1	&	1	& \\ \hline
ccl2o	&	W4-17	&	I, 0.01	&	I, 0.01	&	I, 0.01	&		&		&		&	1	&	1	&	1	& \\ \hline
cf2cl2	&	W4-17	&	I, 0.01	&	I, 0.01	&	I, 0.01	&		&		&		&	1	&	1	&	1	& \\ \hline
ch2clf	&	W4-17	&	I, 0.01	&	I, 0.01	&	I, 0.01	&		&		&		&	1	&	1	&	1	& \\ \hline
ch3ph2	&	W4-17	&	I, 0.01	&	I, 0.01	&	I, 0.01	&		&		&		&	1	&	1	&	1	& \\ \hline
cis-c2f2cl2	&	W4-17	&	I, 0.01	&	I, 0.01	&	I, 0.01	&		&		&		&	1	&	1	&	1	& \\ \hline
clcof	&	W4-17	&	I, 0.01	&	II, 0.001	&	II, 0.001	&	Yes	&		&		&	1	&	481	&	481	& \\ \hline
cyclobutadiene	&	W4-17	&	I, 0.01	&	II, 0.001	&	II, 0.001	&	Yes	&		&		&	1	&	52	&	52	& \\ \hline
cyclopentadiene	&	W4-17	&	I, 0.01	&	I, 0.01	&	I, 0.01	&		&		&		&	1	&	1	&	1	& \\ \hline
dioxetan2one	&	W4-17	&	I, 0.01	&	II, 0.001	&	II, 0.001	&	Yes	&		&		&	1	&	191	&	191	& \\ \hline
dioxetane	&	W4-17	&	I, 0.01	&	I, 0.01	&	I, 0.01	&		&		&		&	1	&	1	&	1	& \\ \hline
dithiotane	&	W4-17	&	I, 0.01	&	I, 0.01	&	I, 0.01	&		&		&		&	1	&	1	&	1	& \\ \hline
fno	&	W4-17	&	I, 0.01	&	I, 0.01	&	I, 0.01	&		&		&		&	2	&	2	&	2	& \\ \hline
formamide	&	W4-17	&	I, 0.01	&	I, 0.01	&	I, 0.01	&		&		&		&	1	&	1	&	1	& \\ \hline
formic-anhydride	&	W4-17	&	I, 0.01	&	I, 0.01	&	I, 0.01	&		&		&		&	1	&	1	&	1	& \\ \hline
hclo4	&	W4-17	&	I, 0.01	&	II, 0.001, 99\%	&	II, 0.001, 99\%	&	Yes	&		&		&	1	&	2428	&	2428	& \\ \hline
hoclo2	&	W4-17	&	I, 0.01	&	II, 0.001	&	II, 0.001	&	Yes	&		&		&	1	&	3086	&	3086	& \\ \hline
hoclo	&	W4-17	&	I, 0.01	&	II, 0.001	&	II, 0.001	&	Yes	&		&		&	1	&	553	&	553	& \\ \hline
n2o4	&	W4-17, MR	&	I, 0.01	&	II, 0.002	&	W-AFQMC	&	Yes	&		&		&	70	&	3506	&	10000	& \\ \hline
nh2f	&	W4-17	&	I, 0.01	&	I, 0.01	&	I, 0.01	&		&		&		&	1	&	1	&	1	& \\ \hline
nh2oh	&	W4-17	&	I, 0.01	&	I, 0.01	&	I, 0.01	&		&		&		&	1	&	1	&	1	& \\ \hline
oxadiazole	&	W4-17	&	I, 0.01	&	I, 0.01	&	I, 0.01	&		&		&		&	3	&	3	&	3	& \\ \hline
oxetane	&	W4-17	&	I, 0.01	&	I, 0.01	&	I, 0.01	&		&		&		&	1	&	1	&	1	& \\ \hline
silole	&	W4-17	&	I, 0.01	&	III, 0.005, 99\%	&	III, 0.005, 99\%	&	Yes	&		&		&	2	&	1621	&	1621	& \\ \hline
tetrahedrane	&	W4-17	&	I, 0.01	&	I, 0.01	&	I, 0.01	&		&		&		&	1	&	1	&	1	& \\ \hline
trans-c2f2cl2	&	W4-17	&	I, 0.01	&	I, 0.01	&	I, 0.01	&		&		&		&	1	&	1	&	1	& \\ \hline\end{longtable}
}

\subsection{W-AFQMC}
\begin{table}
    \centering
    \begin{tabular}{|c|c|c|c|c|l|} \hline 
         Molecule&  Dataset&  AFQMC I&  AFQMC II&  DLPNO& CCSD(T)\\ \hline 
         AlCl3
&  G2&  1.10(48)&  -1.95(43)
&  0.77
& 0.73
\\ \hline 
         AlF3
&  G2&  1.78(58)&  -1.41(35)
&  0.11
& 0.14
\\ \hline 
         Bicyclo-1-1-0-butane&  G2&  -1.74(42)&  -1.50(43)&  -1.92
& -1.92
\\ \hline 
         CF2CF2
&  G2
&  0.33(58)&  -1.28(69)
&  0.32
& 0.40
\\ \hline 
         CH2CH-CN
&  G2
&  0.61(73)&  -0.83(35)
&  -0.79
& -0.74
\\ \hline 
         Cyclobutene
&  G2&  -0.52(46)&  -0.38(36)
&  -0.19
& -0.14
\\ \hline 
         LiF
&  G2
&  0.91(11)&  0.47(9)
&  1.17& 1.18
\\ \hline 
         Ozone
&  G2, MR&  -2.57(25)&  -0.96(35)
&  -3.24& -2.07
\\ \hline 
         SO2
&  G2
&  -1.88(74)&  -0.76(43)&  -1.62
& -1.24
\\ \hline 
 Vinyl-chloride
& G2& -1.17(47)& -1.26(27)
& -0.13
&-0.11\\ \hline 
 azulene
& G3& 1.73(99)& 0.99(105)
& 0.53
&N/A\\ \hline 
 benzoquinone
& G3
& -1.77(70)& -1.92(85)
& -1.40
&-0.69
\\ \hline 
 c2f6
& G3
& 1.72(85)& 0.55(91)
& 1.19&1.22
\\ \hline 
 3-butyn-2-one& G3
& -4.68(55)& -4.60(69)& -3.57&-3.35\\ \hline 
 C4H4N2
& G3
& -3.11(77)& -1.74(53)& -2.52
&-1.96
\\ \hline 
 Cl2O2S
& G3
& -1.98(69)& -1.99(78)& 
-3.43
&-3.53
\\ \hline 
 cyclooctatetraene
& G3
& -1.43(65)& -1.49(104)& -2.42
&-1.99\\ \hline 
 Naphthalene
& G3
& -0.37(89)& -1.15(113)& 
-1.02
&N/A\\ \hline 
 sf6& G3
& -1.46(89)& -0.60(88)& -1.61&-1.71\\ \hline 
 foof
& W4-11, MR& -0.69(78)& 0.60(52)& 
-1.12&-0.57\\ \hline 
 s2o
& W4-11& 0.17(52)& -1.48(35)& -0.21&0.52\\ \hline 
 S3& W4-11, MR& -0.71(47)& -0.34(86)& 
0.05&0.82\\ \hline 
 S4& W4-11, MR& 0.07(79)& -1.44(88)& -1.26&0.66\\ \hline
 n2o4& W4-17, MR& -1.47(79)& -1.09(79)& 
0.03&1.67\\ \hline \hline
 RMSD& & 1.74(14)& 1.54(15)& 1.67&1.55\\\hline
    \end{tabular}
    \caption{All of the molecules run using the W-AFQMC (these, combined with the rest of the molecules run using AFQMC I, form AFQMC II) and their deviations in kcal/mol, and statistical error in parentheses. The "AFQMC I" results are shown for comparison, a combination of AFQMC 0 and better trial wavefunctions. DLPNO-CCSD(T) and CCSD(T) deviations are also shown for comparison. Azulene and naphthalene CCSD(T) results are unavailable due to limitations in disk space.}
    \label{tab:W-AFQMC}
\end{table}

\begin{figure}
    \centering
    \includegraphics[width=0.5\textwidth]{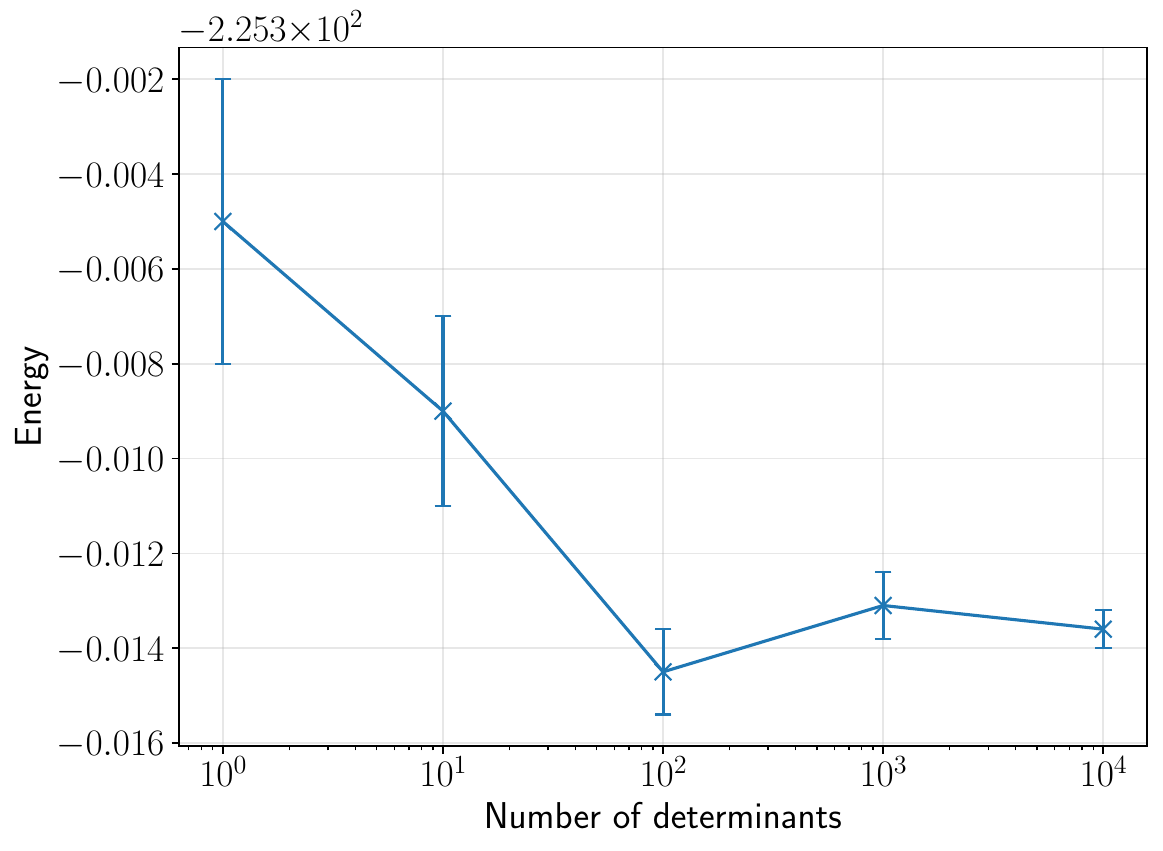}
    \caption{Convergence of the ground state energy of ozone with respect to the number of determinants in the trial calculated using W-AFQMC. We used the aug-cc-pVTZ-DK basis set and froze the core electrons. The trial active space obtained using the natural orbital procedure has the size (18e, 37o).}
    \label{fig:det_convergence}
\end{figure}

The procedure for the W-AFQMC is similar to that for the trials run using L-AFQMC discussed above, with differences accounting for the ability to run with more determinants. We also attempt to converge atom energies with this implementation rather than use the fitted atom energies, due to the ability to run upwards of 10 thousand determinants and in account of atomic energies being difficult to calculate with AFQMC~\cite{JoonhoTwenty}. Firstly, we use an active space with all valence electrons and the 100 lowest energy valence Hartree Fock orbitals for the first HCI calculation, rather than selecting based on a (generally smaller) orbital map as above. This space can be enlarged further for bigger or more correlated molecules, but we found our choice to be sufficient for the mostly single-reference molecules studied here. We used an \(\epsilon_1=10^{-4}\) threshold for the variational HCI calculations for molecules, while a tighter threshold of \(\epsilon_1=5\times10^{-5}\) for atoms. To choose active space for the second calculation, we use a NOON threshold of 0.001 for molecules and 0.0001 for atoms. For this study, this procedure produced active spaces of up to 60 orbitals. Finally, rather than HCI calculation as above, we perform an HCISCF calculation using this active space with the same \(\epsilon_1\) as in the first step. The goal here is not to converge the HCISCF calculation very well, but to generate a good set of orbitals to allow compact HCI expansions. We use up to \(10^4\) leading determinants from the HCI expansion in these optimized orbitals as trial states in AFQMC. This choice for the number of determinants was made based on tests of convergence for a set of molecules from the multireference subset, which requires more determinants for convergence. As an illustrative example, we show the convergence trace with respect to the number of determinants for the ozone molecule in Fig. \ref{fig:det_convergence}. The energy is converged within stochastic error bars at \(10^3\) determinants in this case. 

We used the following formula, due to Martin~\cite{MartinCBS}, for performing two-point CBS extrapolations:
\begin{equation}
   E_n = E_{\infty} + \frac{A}{(n+\frac{1}{2})^4},
\end{equation}
where \(n\) is the cardinality of the basis set. We denote the CBS limit obtained using basis sets of cardinality \(n_1\) and \(n_2\) using method \(M\) as \(E_{M}^{n_1n_2}\). aug-cc-PVnZ-DK basis sets were used for frozen core AFQMC and MP2 calculations. We add the MP2 basis set correction to AFQMC energies as
\begin{equation}
   E_{\text{AFQMC},\infty} \approx E_{\text{AFQMC}}^{DT} + E_{\text{MP2}}^{Q5} - E_{\text{MP2}}^{DT}.  
\end{equation}        
In a few cases where the 5Z basis set HF calculations could not be converged, we used the \(E_{\text{MP2}}^{QT}\) energy instead. 

The AFQMC II results involve taking select molecules (e.g. outliers, or molecules with experimental uncertainty) and running using the aforementioned protocol, and combining these results with the AFQMC I of molecules not run using this second implementation (see Table~\ref{tab:All_molecules_trials}). Again, AFQMC II progressively builds on AFQMC I. Note that these basis sets and CBS extrapolation schemes only apply to W-AFQMC, where we prioritize running with a large number of determinants. For AFQMC 0, AFQMC I, CCSD(T), and DLPNO-CCSD(T), see the main text for the basis sets and CBS. The full list of trials run using W-AFQMC is shown in Table~\ref{tab:W-AFQMC}. For this subspace of data, AFQMC, CCSD(T), and DLPNO-CCSD(T) perform very similarly,.

\newpage
\section{4. AFQMC Localization}
\addcontentsline{toc}{section}{4. AFQMC Localization}

For all molecules and trials for L-AFQMC, we use the compression threshold of $T_{\text{SVD}}$ 0.0001, which is shown to be close to convergence (of energy) to the non-compressed implementation~\cite{Weber_Vuong_Devlaminck_Shee_Lee_Reichman_Friesner_2022}. For a representative molecule, CCl4, as shown in Table~\ref{tab:ccl4_lo}, with AFQMC 0 (1 determinant in this case), the compression rate is 60\% for frozen core and 80\% for all-electron. The CBS energy difference between L-AFQMC with and without compression is -0.06 kcal/mol, for the frozen core calculation, which is one tenth of the statistical error. On the other hand, the all-electron compression error is on the order of -3 kcal/mol. For reference, the energy of the 200 determinant CCl4 trial also mentioned in the above section is -1884.5075(11), which is closer to the no LO value. As with the timestep error, we expect the LO error to decrease with increasing the active space in the trial. However, this is at  the expense of less compression.

\begin{table}
    \centering
    \begin{tabular}{|c|c|c|c|} \hline 
         &  LO CBS Energy (Ha)&  No LO CBS Energy (Ha)& Difference (kcal/mol)
\\ \hline 
         Frozen core&  -1882.7855(10)&  -1882.7856(10)& -0.06
\\ \hline 
 All electron& -1884.5015(12)& -1884.5062(12)&-2.98\\ \hline
    \end{tabular}
    \caption{CBS energy for frozen vs no frozen core for CCl4 single determinant trial.}
    \label{tab:ccl4_lo}
\end{table}

\newpage
\section{5. Frozen Core vs. All Electron Performance}
\addcontentsline{toc}{section}{5. Frozen Core vs. All Electron Performance}

As discussed in Section 2 above, the all electron calculations display a larger timestep error, which requires a larger number of determinants to converge and is not fully cancelled out by the atoms, even with atom fitting. Here, we look at the differences in overall performance between frozen core (with MP2 core-valence correction) and all-electron calculations. Table~\ref{tab:frozen_rmsd} lists the RMSD for the combined dataset for each method, with frozen and all-electron. Table~\ref{tab:frozen_outliers} lists the number of outliers for each method. While the RMSD using all-electron calculation is decent across the dataset, the number of outliers significantly increases compared to frozen core.

\begin{table}
    \centering
    \begin{tabular}{|c|c|c|c|l|} \hline 
         Methood&  AFQMC 0&  AFQMC I& CCSD(T) &DLPNO-CCSD(T)
\\ \hline 
         Frozen&  1.67&  0.98& 0.77 &0.87
\\ \hline 
         All electron&  2.16&  1.25& 0.84 &1.04\\ \hline
    \end{tabular}
    \caption{RMSD for the combined dataset of 259 molecules for frozen vs all electron calculations for each method, reported in kcal/mol.}
    \label{tab:frozen_rmsd}
\end{table}

\begin{table}
    \centering
    \begin{tabular}{|c|c|c|c|l|} \hline 
         Methood&  AFQMC 0&  AFQMC I&  CCSD(T)&DLPNO-CCSD(T)
\\ \hline 
         Frozen&  30&  3&  4&6
\\ \hline 
         All electron&  58&  15&  8&15\\ \hline
    \end{tabular}
    \caption{Total number of outliers for frozen vs all electron calculation for each method.}
    \label{tab:frozen_outliers}
\end{table}

For a detailed list of all deviations, frozen and all electron, as well as the statistical errors, the AFQMC 0 and AFQMC I details (trial, active space, determinants for both TZ and QZ basis sets), as well as lists of the atomic energies, refer to the .xslx spreadsheet also provided. Due to the difficulty of converging the all-electron trials, there are many cases where manual trials were selected in the way of chemical intuition and looking at orbital energies and symmetries to confirm the relatively larger deviation values are not due to the NOON selection procedure selecting qualitatively wrong CI expansions (AFQMC 0 remains as is without manual selection). For these manual selections, only one round of SHCI/SHCISCF is run with $\epsilon_1 =5\times10^{-5}$, and the orbitals are selected from the HOMO-LUMO gap in the canonical (RHF) basis.

\newpage
\section{6. Composite CCSD(T) CBS}
\addcontentsline{toc}{section}{6. Composite CCSD(T) CBS}

For the majority of G3 molecules and some G2 molecules (benzene, spiropentane, butane, trimethyl-amine), we are unable to run full CCSD(T) at the aug-cc-pVQZ-DK level due to limitations in disk space of the temporary directory of each of our computational nodes (2 TB). Therefore, we settle for a composite scheme where the CCSD(T) energy for aug-cc-p(C)VQZ-DK is extrapolated from DLPNO and cc-p(C)VQZ-DK CCSD(T).
\begin{equation}
   E_{\text{CCSD(T)}}^{\text{aug-cc-p(C)VXZ}} \approx E_{\text{CCSD(T)}}^{\text{cc-p(C)VXZ}} + E_{\text{DLPNO 6/7}}^{\text{aug-cc-p(C)VXZ}} - E_{\text{DLPNO 6/7}}^{\text{cc-p(C)VXZ}}  
\end{equation}   
where DLPNO 6/7 means DLPNO-CCSD(T) extrapolated with TCut6 and TCut7 PNO thresholds as discussed in main text Section 2.5 and 2.6. Overall, we do not get a significant difference using this composite CBS compared to full CCSD(T) aug-cc-p(C)VTZ/aug-cc-p(C)VQZ extrapolation. Table~\ref{tab:W417_composite} shows the difference in single point energy between regular T/Q and this composite scheme. The composite scheme is consistently higher in energy, which means for calculating atomization energy the atom fit will cancel out the difference further.
{
\footnotesize
\begin{longtable}{|p{4cm}p{4cm}p{0cm}|}
\caption{Difference in energy between the regular CCSD(T) T/Q CBS and CCSD(T) using the composite CBS scheme for the W4-17 set, reported in kcal/mol.} % needs to go inside longtable environment
\label{tab:W417_composite}
\endfirsthead
\endhead
\hline
Molecule	&	CBS energy diff (kcal/mol)	& \\ \hline
beta-lactim	&	-0.20	& \\ \hline
borole	&	-0.15	& \\ \hline
c2cl2	&	-0.12	& \\ \hline
c2cl6	&	N/A	& \\ \hline
c2clh	&	-0.09	& \\ \hline
ccl2o	&	-0.08	& \\ \hline
cf2cl2	&	-0.16	& \\ \hline
ch2clf	&	-0.15	& \\ \hline
ch3ph2	&	-0.10	& \\ \hline
cis-c2f2cl2	&	-0.24	& \\ \hline
clcof	&	-0.15	& \\ \hline
clf5	&	-0.82	& \\ \hline
cloocl	&	0.00	& \\ \hline
cyclobutadiene	&	-0.20	& \\ \hline
cyclopentadiene	&	-0.21	& \\ \hline
dioxetan2one	&	-0.27	& \\ \hline
dioxetane	&	-0.22	& \\ \hline
dithiotane	&	-0.22	& \\ \hline
fno	&	-0.24	& \\ \hline
formamide	&	-0.10	& \\ \hline
formic-anhydride	&	-0.15	& \\ \hline
hclo4	&	-0.09	& \\ \hline
hoclo2	&	-0.14	& \\ \hline
hoclo	&	-0.12	& \\ \hline
n2o4	&	-0.04	& \\ \hline
nh2f	&	-0.12	& \\ \hline
nh2oh	&	-0.13	& \\ \hline
oxadiazole	&	-0.41	& \\ \hline
oxetane	&	-0.12	& \\ \hline
silole	&	-0.18	& \\ \hline
tetrahedrane	&	-0.17	& \\ \hline
trans-c2f2cl2	&	-0.31	& \\ \hline
\end{longtable}
}

\newpage

\section{7. Effect of DKH2 or X2C}
\addcontentsline{toc}{section}{7. Effect of DKH2 or X2C}

\begin{table}
    \centering
    \small
    \begin{tabular}{|cccccc|} \hline
         Molecule&  HF no relham &  HF X2C &  HF DKH2 &  HF DKH2 - no relham &  HF DKH2 - X2C \\
 & (Ha)& (Ha)& (Ha)& (kcal/mol)&(kcal/mol)\\ \hline
         beta-lactim&  -245.86732&  -245.99934&  -245.99941&  -82.88765&  -0.04\\ \hline
         borole&  -179.06580&  -179.13416&  -179.13419&  -42.91142&  -0.01\\ \hline
         c2cl2&  -994.03679&  -997.54946&  -997.55081&  -2,205.04688&  -0.84\\ \hline
         c2cl6&  -2,830.82156&  -2,841.29977&  -2,841.30379&  -6,577.59784&  -2.52\\ \hline
         c2clh&  -535.44360&  -537.21538&  -537.21606&  -1,112.21485&  -0.43\\ \hline
         ccl2o&  -1,031.13482&  -1,034.68839&  -1,034.68977&  -2,230.73158&  -0.86\\ \hline
         cf2cl2&  -1,155.18237&  -1,158.86763&  -1,158.86911&  -2,313.43201&  -0.93\\ \hline
         ch2clf&  -597.69627&  -599.54651&  -599.54725&  -1,161.48998&  -0.47\\ \hline
         ch3ph2&  -381.36740&  -382.37077&  -382.37144&  -630.03995&  -0.42\\ \hline
         cis-c2f2cl2&  -1,193.01529&  -1,196.71563&  -1,196.71712&  -2,322.89740&  -0.93\\ \hline
         clcof&  -671.42697&  -673.33284&  -673.33362&  -1,196.42645&  -0.49\\ \hline
         clf5\_MR&  -956.06111&  -958.27092&  -958.27195&  -1,387.30234&  -0.64\\ \hline
         cloocl\_MR&  -1,067.90024&  -1,071.49489&  -1,071.49630&  -2,256.52842&  -0.88\\ \hline
         cyclobutadiene&  -153.69507&  -153.75658&  -153.75660&  -38.60874&  -0.01\\ \hline
         cyclopentadiene&  -192.85604&  -192.93287&  -192.93289&  -48.22382&  -0.02\\ \hline
         dioxetan2one&  -301.57987&  -301.77693&  -301.77705&  -123.72616&  -0.08\\ \hline
         dioxetane&  -227.81242&  -227.95389&  -227.95398&  -88.82412&  -0.05\\ \hline
         dithiotane&  -872.67266&  -875.34988&  -875.35130&  -1,680.84919&  -0.90\\ \hline
         fno&  -228.69739&  -228.87767&  -228.87779&  -113.20593&  -0.08\\ \hline
         formamide&  -168.99829&  -169.09958&  -169.09964&  -63.59908&  -0.04\\ \hline
         formic-anhydride&  -301.58866&  -301.78581&  -301.78593&  -123.78963&  -0.08\\ \hline
         hclo4&  -758.97780&  -760.93752&  -760.93834&  -1,230.23922&  -0.51\\ \hline
         hoclo2&  -684.15782&  -686.06447&  -686.06525&  -1,196.91129&  -0.49\\ \hline
         hoclo&  -609.34468&  -611.19691&  -611.19765&  -1,162.73901&  -0.47\\ \hline
         n2o4&  -408.18080&  -408.46418&  -408.46436&  -177.93450&  -0.11\\ \hline
         nh2f&  -155.01807&  -155.14244&  -155.14252&  -78.09545&  -0.05\\ \hline
         nh2oh&  -131.03832&  -131.12431&  -131.12436&  -53.98654&  -0.03\\ \hline
         oxadiazole&  -260.63036&  -260.77797&  -260.77805&  -92.67488&  -0.05\\ \hline
         oxetane&  -191.97620&  -192.07760&  -192.07765&  -63.65914&  -0.03\\ \hline
         silole&  -443.80478&  -444.58867&  -444.58927&  -492.26886&  -0.38\\ \hline
         tetrahedrane&  -153.65087&  -153.71213&  -153.71215&  -38.45463&  -0.01\\ \hline
         trans-c2f2cl2& -1,193.01679& -1,196.71705& -1,196.71853& -2,322.84354& -0.93\\ \hline
    \end{tabular}
    \caption{HF energies for aug-cc-pCVTZ-DK basis set with frozen core, without relativistic Hamiltonian, with X2C, and with DKH2..}
    \label{tab:Relham_hf}
\end{table}

\begin{table}
    \centering
    \small
    \begin{tabular}{|cccccc|} \hline
         Molecule&  MP2 no relham&  MP2 X2C &  MP2 DKH2 &  MP2 DKH2 - no relham&  MP2 DKH2 - X2C\\
 & (Ha)& (Ha)& (Ha)& (kcal/mol)&(kcal/mol)\\ \hline
         beta-lactim&  -246.81800&  -246.95052&  -246.95059&  -83.20&  -0.04\\ \hline
         borole&  -179.79723&  -179.86582&  -179.86585&  -43.06&  -0.01\\ \hline
         c2cl2&  -994.73057&  -998.24384&  -998.24519&  -2,205.42&  -0.85\\ \hline
         c2cl6&  -2,832.35764&  -2,842.83774&  -2,842.84175&  -6,578.78&  -2.52\\ \hline
         c2clh&  -535.94811&  -537.72024&  -537.72092&  -1,112.44&  -0.43\\ \hline
         ccl2o&  -1,031.92423&  -1,035.47865&  -1,035.48003&  -2,231.27&  -0.86\\ \hline
         cf2cl2&  -1,156.26245&  -1,159.94898&  -1,159.95047&  -2,314.23&  -0.93\\ \hline
         ch2clf&  -598.33102&  -600.18192&  -600.18267&  -1,161.91&  -0.47\\ \hline
         ch3ph2&  -381.71729&  -382.72089&  -382.72156&  -630.18&  -0.42\\ \hline
         cis-c2f2cl2&  -1,194.24248&  -1,197.94411&  -1,197.94560&  -2,323.71&  -0.93\\ \hline
         clcof&  -672.27791&  -674.18465&  -674.18543&  -1,196.97&  -0.49\\ \hline
         clf5\_MR&  -957.70307&  -959.91529&  -959.91632&  -1,388.81&  -0.64\\ \hline
         cloocl\_MR&  -1,068.79946&  -1,072.39521&  -1,072.39662&  -2,257.22&  -0.88\\ \hline
         cyclobutadiene&  -154.33868&  -154.40041&  -154.40043&  -38.75&  -0.01\\ \hline
         cyclopentadiene&  -193.67577&  -193.75286&  -193.75289&  -48.39&  -0.02\\ \hline
         dioxetan2one&  -302.61700&  -302.81485&  -302.81497&  -124.23&  -0.08\\ \hline
         dioxetane&  -228.63342&  -228.77545&  -228.77554&  -89.18&  -0.05\\ \hline
         dithiotane&  -873.37337&  -876.05125&  -876.05268&  -1,681.27&  -0.90\\ \hline
         fno&  -229.42315&  -229.60414&  -229.60426&  -113.65&  -0.08\\ \hline
         formamide&  -169.61996&  -169.72165&  -169.72171&  -63.85&  -0.04\\ \hline
         formic-anhydride&  -302.62283&  -302.82077&  -302.82089&  -124.28&  -0.08\\ \hline
         hclo4&  -760.72019&  -762.17228&  -762.17310&  -911.70&  -0.51\\ \hline
         hoclo2&  -685.60341&  -687.05390&  -687.05468&  -910.67&  -0.49\\ \hline
         hoclo&  -610.47327&  -611.92055&  -611.92129&  -908.63&  -0.47\\ \hline
         n2o4&  -409.94073&  -409.91656&  -409.91673&  15.06&  -0.11\\ \hline
         nh2f&  -155.51095&  -155.63576&  -155.63585&  -78.37&  -0.05\\ \hline
         nh2oh&  -131.52091&  -131.60722&  -131.60727&  -54.20&  -0.03\\ \hline
         oxadiazole&  -261.61319&  -261.76146&  -261.76153&  -93.09&  -0.05\\ \hline
         oxetane&  -192.73816&  -192.83994&  -192.84000&  -63.91&  -0.03\\ \hline
         silole&  -444.57097&  -445.35522&  -445.35583&  -492.49&  -0.38\\ \hline
         tetrahedrane&  -154.30672&  -154.36822&  -154.36824&  -38.60&  -0.01\\ \hline
         trans-c2f2cl2& -1,194.24329& -1,197.94484& -1,197.94633& -2,323.66& -0.93\\ \hline
    \end{tabular}
    \caption{MP2 energies for aug-cc-pCVTZ-DK basis set with frozen core without relativistic Hamiltonian, with X2C, and with DKH2.}
    \label{tab:Relham_mp2}
\end{table}

Displayed in Tables~\ref{tab:Relham_hf} and~\ref{tab:Relham_mp2} are the difference in total Hartree-Fock energies and MP2 energies for the basis set aug-cc-pCVTZ-DK i) without the use of a relativistic Hamiltonian, ii) with the use of the X2C Hamiltonian, and iii) with the use of the DKH2 Hamiltonian for the W4-17 dataset. The difference between the use of DKH2 and X2C in the total energy is small, especially for first-row atoms. For the second-row atoms, the difference between the use of DKH2 and X2C is approximately proportional to the number of each atom (for example, Cl atom contributes 0.4 to 0.5 kcal/mol to the difference) which will become cancelled out with the atom fitting procedure as described in the main text.

\newpage

\section{8. Alternative Geometries}
\addcontentsline{toc}{section}{8. Alternative Geometries}

DFT geometry optimization with a small basis set and single point energy calculation with a higher-level theory is a popular approach and is used successfully to derive accurate thermochemistry in  numerous methods such as G2/G3/G4 and ccCA, and is shown to provide accurate total energies~\cite{WangMolecular2023,BakowiesDensity2021} as well as vibrational properties~\cite{MartinW11999,BakowiesDensity2021}. While we do not use accurate wavefunction methods to optimize geometry for the entire dataset as this is generally limited to only very small systems~\cite{SemidalasCanG42024,ChanSimple2023,NelsonApproximating2023}, there are cases where we have found that an alternative geometry is useful which we list here.

To assist SCF convergence, for the W4 "multireference" (TAEMR, which is separate from our MR subset, refer Section 11) molecules (BN, C2, Cl2O, FOOF, S3, S4, ClF5, F2O and ClOOCl), as well as a few other exceptions that help converge either AFQMC, CCSD(T) or DLPNO-CCSD(T) (CCl2CCl2, SO3, HClO4, HOClO2, HOClO, S2O, and N2O4), we use the CCSD(T)/cc-pV(Q+d)Z geometries provided by the W4-11~\cite{Karton_W411} and W4-17~\cite{Karton_W417} sources for CCSD(T), DLPNO-CCSD(T) and AFQMC. In particular, we note, as shown in Table~\ref{tab:geometries}, cases where the CCSD(T) energy is noticeably improved (by > 2 kcal/mol) after using the CCSD(T)-optimized geometry. All geometries used in the correlated methods are provided in a .zip file.
% as well as a few other exceptions where the CCSD(T) values obtained with the B3LYP/6-31G* geometries does not agree with the values published by W4-17 prior to higher T correction, (CCl2CCl2, SO3, HClO4, HOClO2, HOClO, S2O, and N2O4), we use the CCSD(T)/cc-pV(Q+d)Z geometries provided by the W4-11~\cite{Karton_W411} and W4-17~\cite{Karton_W417} sources. Shown in Table \ref{tab:geometries} are such cases where there is a deviation of  > 1 kcal/mol from that optimized by Karton et al. 

%The W4 CCSD(T) deviation is from addition of higher order T corrections to CCSD(T).

\begin{table}
    \centering
    \begin{tabular}{|c|c|c|} \hline 
         Molecule& \thead{B3LYP/6-31G* \\ CCSD(T) deviation} & \thead{CCSD(T)/cc-pV(Q+d)Z \\ CCSD(T) deviation}  \\ \hline 
 SO3& -1.87&0.23\\\hline 
         HClO4& -6.75 & 1.11
 \\ \hline 
         HOClO2& -3.88  & 0.73
 \\ \hline 
 HOClO& -1.91  &0.28 \\ \hline
    \end{tabular}
    \caption{Comparison of CCSD(T) deviations from experiment for the molecules where the B3LYP/6-31G* optimized geometry shows a difference of > 2 kcal/mol in energy from that optimized by CCSD(T)/cc-pV(Q+d)Z~\cite{Karton_W417}.}
    \label{tab:geometries}
\end{table}

\newpage
\section{9. Atom Energies}
\addcontentsline{toc}{section}{9. Atom Energies}

\begin{table}
    \centering
    \begin{tabular}{|c|c|c|c|} \hline 
         CCSD(T) atom&  Fitted atom energies (Ha)&  Ab inito atom energies (Ha)& Diff (kcal/mol)\\ \hline 
         C&  -37.80495&  -37.80485& -0.059
\\ \hline 
         O&  -75.05708&  -75.05855& 0.919
\\ \hline 
         S&  -398.75915&  -398.75935& 0.124
\\ \hline 
         F&  -99.75609&  -99.75658& 0.304
\\ \hline 
         N&  -54.55929&  -54.55991& 0.387
\\ \hline 
         B&  -24.60865&  -24.60916& 0.321
\\ \hline 
         Na&  -162.39546&  -162.39489& -0.357
\\ \hline 
         Al&  -242.37215&  -242.37132& -0.517
\\ \hline 
         Si&  -289.54608&  -289.54398& -1.318
\\ \hline 
 P& -341.64853& -341.64883&0.188
\\ \hline 
 Cl& -461.12061& -461.11929&-0.827
\\ \hline 
 Li& -7.45643& -7.45327&-1.986\\ \hline
    \end{tabular}
    \caption{Frozen core atom energy without MP2 core correction  for CCSD(T). Ab initio energies are obtained the same way as for molecules in the main text (T/Q CBS extrapolation, see main text Section 2.6). Atomic energies fit to G2/G3 as discussed in Section 2.10. The difference between these energies is shown in kcal/mol.}
    \label{tab:ccsdt_atom_energies}
\end{table}
\begin{table}
    \centering
    \begin{tabular}{|c|c|c|c|} \hline 
         DLPNO atom&  Fitted atom energies (Ha)&  Ab inito atom energies (Ha)& Diff (kcal/mol)\\ \hline 
         C&  -37.80453&  -37.80477& 0.152
\\ \hline 
         O&  -75.05634&  -75.05763& 0.806
\\ \hline 
         S&  -398.75851&  -398.75867& 0.101
\\ \hline 
         F&  -99.75562&  -99.75537& -0.153
\\ \hline 
         N&  -54.55851&  -54.55947& 0.605
\\ \hline 
         B&  -24.60857&  -24.60927& 0.442
\\ \hline 
         Na&  -162.39518&  -162.39485& -0.208
\\ \hline 
         Al&  -242.37209&  -242.37148& -0.381
\\ \hline 
         Si&  -289.54591&  -289.54389& -1.272
\\ \hline 
 P& -341.64780& -341.64814&0.208
\\ \hline 
 Cl& -461.12023& -461.11867&-0.980
\\ \hline 
 Li& -7.45643& -7.45325&-1.996\\ \hline
    \end{tabular}
    \caption{Frozen core atom energy without MP2 core correction  for DLPNO-CCSD(T). Ab initio energies are obtained the same way as for molecules in the main text (T/Q CBS extrapolation with TCutPNO 6/7, see main text Section 2.5 and 2.6). Atomic energies fit to G2/G3 as discussed in Section 2.10. The difference between these energies is shown in kcal/mol.}
    \label{tab:dlpno_atom_energies}
\end{table}

\begin{table}
    \centering
    \begin{tabular}{|c|c|c|} \hline 
         CCSD(T) Atom&  W4-11& W4-17
\\ \hline 
         Fitted&  0.44& 0.79
\\ \hline 
         Ab initio&  0.94& 1.71\\ \hline
    \end{tabular}
    \caption{CCSD(T) RMSD of the W4 datasets in kcal/mol, using atom energies fitted from the G2/G3 experimental data compared to using \textit{ab initio} atom energies from CCSD(T) with spin-orbit coupling and CBS extrapolation (see main text Section 2).}
    \label{tab:CCSDT_fit}
\end{table}

Atom energies before and after fitting to G2/G3 for the CCSD(T) and DLPNO-CCSD(T) methods are shown in Tables~\ref{tab:ccsdt_atom_energies} and~\ref{tab:dlpno_atom_energies}. \textit{Ab initio} atom energies are obtained using the same method as for molecules (TZ/QZ CBS, see main text), and in addition with spin-orbit corrections~\cite{Curtiss_Raghavachari_Redfern_Pople_1997}. Atoms such as O and Si can be different on the order of 1 kcal/mol, while Li has a 2 kcal/mol in difference. Shown in Table~\ref{tab:CCSDT_fit} are the CCSD(T) RMSDs of the W4-11 and W4-17 datasets using the fitted atom energies (only using G2/G3 data) and using \textit{ab initio} aotm energies. There are no outliers (> 2kcal/mol) using the fitted atomic energies, while the outliers using \textit{ab initio} atom energies are N2O4 (2.78 kcal/mol), C2Cl6 (-6.95 kcal/mol) and FOOF (3.01 kcal/mol).

We also note that the cases where AFQMC I and AFQMC 0 have different fitted atom energies, as shown in Tables~\ref{tab:afqmc0afqmc1atom} and~\ref{tab:afqmc0afqmc1atom_nofrozen}. The all-electron (no frozen core) atom energies show more difference upon improving the results with AFQMC I, possibly due to the reduction of timestep error with using larger trials. The carbon atom energy being the most robust reflects the relative abundance of carbon in the G2/G3 datasets.

\begin{table}
    \centering
    \begin{tabular}{|c|c|c|c|} \hline 
         AFQMC atom,&  Fitted atom energies &  Fitted atom energies & Diff (kcal/mol)\\
 frozen& AFQMC 0 (Ha)& AFQMC I (Ha)&\\\hline  
         C&  -37.80593&  -37.80592
& -0.009
\\ \hline 
         O&  -75.05872&  -75.05884
& 0.079
\\ \hline 
         S&  -398.76035&  -398.76024
& -0.066
\\ \hline 
         F&  -99.75762&  -99.75756
& -0.039
\\ \hline 
         N&  -54.56073&  -54.56062
& -0.068
\\ \hline 
         B&  -24.60945&  -24.60962
& 0.104
\\ \hline 
         Na&  -162.39617&  -162.39618
& 0.006
\\ \hline 
         Al&  -242.37202&  -242.37209
& 0.046
\\ \hline 
         Si&  -289.54708&  -289.54712
& 0.028
\\ \hline 
 P& -341.64971& -341.64975
&0.022
\\ \hline 
 Cl& -461.12163& -461.12158
&-0.030
\\ \hline 
 Li& -7.45746& -7.45650&-0.605\\ \hline
    \end{tabular}
    \caption{Frozen core atom energy without MP2 core correction for AFQMC 0 and AFQMC I, fitted to G2/G3 heats of formation. The difference between these energies is shown in kcal/mol.}
    \label{tab:afqmc0afqmc1atom}
\end{table}

\begin{table}
    \centering
    \begin{tabular}{|c|c|c|c|} \hline 
         AFQMC atom,&  Fitted atom energies &  Fitted atom energies & Diff (kcal/mol)
\\
 no frozen& AFQMC 0 (Ha)& AFQMC I (Ha)&\\\hline  
         C&  -37.84282&  -37.84281
& -0.008
\\ \hline 
         O&  -75.10131&  -75.10148
& 0.107
\\ \hline 
         S&  -399.17722&  -399.17728
& 0.040
\\ \hline 
         F&  -99.81872&  -99.81892
& 0.128
\\ \hline 
         N&  -54.60029&  -54.60011
& -0.110
\\ \hline 
         B&  -24.64709&  -24.64622
& -0.547
\\ \hline 
         Na&  -162.45462&  -162.45432
& -0.189
\\ \hline 
         Al&  -242.77266&  -242.77236
& -0.191
\\ \hline 
         Si&  -289.95449&  -289.95407
& -0.260
\\ \hline 
 P& -342.06083& -342.06072
&-0.069
\\ \hline 
 Cl& -461.54156& -461.54194
&0.236
\\ \hline 
 Li& -7.45773& -7.45707&-0.412\\ \hline
    \end{tabular}
    \caption{All-electron atom energy without MP2 core correction for AFQMC 0 and AFQMC I, fitted to G2/G3 heats of formation. The difference between these energies is shown in kcal/mol.}
    \label{tab:afqmc0afqmc1atom_nofrozen}
\end{table}

\clearpage
\section{10. Computational Timings}
\addcontentsline{toc}{section}{10. Computational Timings}

\begin{table}[h!]
    \footnotesize
    \centering
    \begin{tabular}{|ccccccccc|} \hline 
         Molecule&  CCSD(T)&  DLPNO&  AFQMC 0&  AFQMC I&  W-AFQMC &  AFQMC 0&  AFQMC I& W-AFQMC \\ 
 & & & & & & dets& dets&dets\\ \hline 
 Unit& CPU& CPU& GPU& GPU& CPU& & &\\\hline 
         Bicyclobutane&  1.20&  1.44&  35.4&  347&  530&  1&  1098& 10000

\\ \hline 
         N2O4&  2.48&  1.68&  42.4&  904&  1175&  80&  3506& 10000

\\ \hline 
         Azulene&  >176&  30.9&  111&  3840&  5440&  12&  2347& 10000 \\ \hline
    \end{tabular}
    \caption{Timings of different benchmark methods for three molecules, for the basis aug-cc-pVTZ-DK.}
    \label{tab:timings}
\end{table}
Shown in Table~\ref{tab:timings} is the number of  computational hours required for each benchmark method for a select few molecules. DLPNO refers to DLPNO-CCSD(T). Units are reported in hours. The CPU models we use are AMD EPYC 7643 48-Core Processor with 2 threads per core (96 threads) and run with 8 threads per calculation. GPU calculations are run with AMD Instinct MI250X GPU which have 2 GCDs each and each calculation is run with 80 or 160 "GPUs" (GCDs). Therefore one CPU hour refers to one hour on one thread and one GPU hour refers to one hour on one GCD. The total walltime of the calculation is the CPU or GPU hour divided by the number of parallel processes which is 8 for CPU calculations and 80 or 160 for GPU calculations with AFQMC 0. For AFQMC I, we use up to 960 parallel processes resulting in a real-world wall-time of under 4 hours. In Table~\ref{tab:timings}, the azulene CCSD(T) time is the time elapsed before the calculation exhausts the temporary storage in the compute node and as a result we do not have a finished calculation for CCSD(T) energy. While we have not attempted to convert GPU hours and CPU hours to the same unit due to variations in hardware, GPU hours are usually much more expensive (on the order of 50$\times$~\cite{accessCI}). Therefore, for large determinant trials, W-AFQMC displays a cost advantage, although L-AFQMC runs much faster in real time due to GPU parallelization. However, we only use 250 walkers with W-AFQMC while we use 1920 walkers for L-AFQMC, and we reduce the amount of energy evaluation times by 2.5$\times$ for W-AFQMC compared to L-AFQMC. While the larger active space and determinants for the W-AFQMC trial compensates for the statistical error to an extent, to achieve the same statistical error as L-AFQMC (Table ~\ref{tab:sterr_comparison}) requires more sampling which will increase the computational time significantly. These factors make a quantitative comparison difficult. What is clear is the advantage of CCSD(T) and DLPNO-CCSD(T) for small systems, which is not surprising because of the prefactor in AFQMC as a statistical method that requires averaging over imaginary time. CCSD(T) however, scales prohibitively, as can be seen for azulene.

\begin{table}
    \centering
    \begin{tabular}{|c|c|c|l|} \hline 
          St. err (Ha)&  AFQMC 0&  AFQMC I&W-AFQMC
\\ \hline 
         Bicyclobutane&  0.0004&  0.0004&0.0005
\\ \hline 
         n2o4&  0.0007&  0.0005&0.0009
\\ \hline 
 azulene& 0.0007& 0.0007&0.0012\\ \hline
    \end{tabular}
    \caption{Statistical errors for the AFQMC protocols in only aug-cc-pVTZ-DK basis, reported in Hartrees. }
    \label{tab:sterr_comparison}
\end{table}

Although fluctuations in hardware (for example, filesystem load) make these timings approximate, the rough scaling of the DLPNO and L-AFQMC methods can be demonstrated. For example, we show the CPU and GPU hours for the entire W4-17 dataset in Tables~\ref{tab:W4-17_tz_timings} and~\ref{tab:W4-17_qz_timings}. While DLPNO scales as approximately $N^3$ or $N^2$ (if we compare the same system with the TZ and QZ basis sets), and L-AFQMC does not appear to scale worse than this with system size from TZ to QZ. However, the prefactor makes the difference in absolute computational cost very noticeable from DLPNO-CCSD(T) to AFQMC. On the other hand, CCSD(T) shows larger fluctuations in timings.

{
\small
\begin{longtable}{|cccccccc|}
    \caption{W4-17 dataset TZ (as used in the main text) timings (excluding Hartree-Fock times) in CPU hours for the coupled-cluster methods and GPU hours for AFQMC 0 and AFQMC I (AFQMC timings scaled to fixed 1 mHa statistical error). DLPNO-CCSD(T) was parallelized with 8 proceses and AFQMC 0 with 80 GPUs. AFQMC I is only shown if the trial was different from AFQMC 0.}
    \label{tab:W4-17_tz_timings}
    \endfirsthead
    \endhead
    \hline
         W4-17 TZ&  Electrons&  Basis functions&  DLPNO 7&  DLPNO 6&  CCSD(T)&  AFQMC 0& AFQMC I\\ \hline
beta-lactim	&	38	&	345	&	1.52	&	0.56	&	2.89	&	5.16	&	\\ \hline
borole	&	34	&	345	&	1.42	&	0.55	&	2.38	&	5.71	&	\\ \hline
c2cl2	&	46	&	242	&	0.38	&	0.17	&	0.5	&	6.02	&	\\ \hline
c2cl6	&	114	&	542	&	7.45	&	4.21	&	44.35	&	13.00	&	\\ \hline
c2clh	&	30	&	190	&	0.2	&	0.07	&	0.16	&	2.90	&	\\ \hline
ccl2o	&	48	&	242	&	0.48	&	0.2	&	0.66	&	3.19	&	\\ \hline
cf2cl2	&	58	&	314	&	0.89	&	0.59	&	2.34	&	4.07	&	\\ \hline
ch2clf	&	34	&	226	&	0.24	&	0.1	&	0.38	&	1.35	&	\\ \hline
ch3ph2	&	26	&	236	&	0.2	&	0.07	&	0.29	&	1.47	&	\\ \hline
cis-c2f2cl2	&	64	&	360	&	1.84	&	0.87	&	4.98	&	6.99	&	\\ \hline
clcof	&	40	&	226	&	0.33	&	0.17	&	0.42	&	4.76	&	8.93\\ \hline
clf5\_MR	&	62	&	370	&	2.55	&	1.27	&	8.32	&	11.87	&	315.75\\ \hline
cloocl\_MR	&	50	&	242	&	0.53	&	0.22	&	0.7	&	5.05	&	\\ \hline
cyclobutadiene	&	28	&	276	&	0.73	&	0.24	&	0.75	&	3.84	&	4.51\\ \hline
cyclopentadiene	&	36	&	368	&	1.72	&	0.67	&	3.84	&	3.88	&	\\ \hline
dioxetan2one	&	38	&	276	&	0.88	&	0.38	&	1.22	&	5.87	&	13.95\\ \hline
dioxetane	&	32	&	276	&	0.61	&	0.25	&	1	&	3.79	&	\\ \hline
dithiotane	&	48	&	334	&	0.98	&	0.46	&	2.18	&	2.87	&	\\ \hline
fno	&	24	&	151	&	0.18	&	0.05	&	0.09	&	2.93	&	\\ \hline
formamide	&	24	&	207	&	0.29	&	0.09	&	0.26	&	1.54	&	\\ \hline
formic-anhydride	&	38	&	276	&	0.79	&	0.33	&	1.14	&	4.96	&	\\ \hline
hclo4	&	50	&	282	&	1.14	&	0.48	&	1.62	&	6.17	&	123.91\\ \hline
hoclo2	&	42	&	236	&	0.67	&	0.23	&	0.7	&	5.42	&	85.46\\ \hline
hoclo	&	34	&	190	&	0.3	&	0.14	&	0.25	&	4.95	&	15.64\\ \hline
n2o4	&	46	&	276	&	1.12	&	0.57	&	2.48	&	8.91	&	109.30\\ \hline
nh2f	&	18	&	151	&	0.11	&	0.05	&	0.07	&	1.36	&	\\ \hline
nh2oh	&	18	&	161	&	0.1	&	0.04	&	0.08	&	1.50	&	\\ \hline
oxadiazole	&	36	&	276	&	1.18	&	0.4	&	1.09	&	5.07	&	\\ \hline
oxetane	&	32	&	322	&	0.77	&	0.43	&	1.8	&	4.38	&	\\ \hline
silole	&	44	&	397	&	1.82	&	0.89	&	5.14	&	3.07	&	48.05\\ \hline
tetrahedrane	&	28	&	276	&	0.63	&	0.23	&	0.76	&	2.83	&	\\ \hline
trans-c2f2cl2	&	64	&	360	&	1.9	&	0.95	&	5.61	&	7.42	&	\\ \hline
\end{longtable}
}

{
\small
\begin{longtable}{|cccccccc|}
        \caption{W4-17 dataset QZ (as used in the main text) timings (excluding Hartree-Fock times) in CPU hours for the coupled-cluster methods and GPU hours for AFQMC 0 and AFQMC I (AFQMC timings scaled to fixed 1 mHa statistical error). DLPNO-CCSD(T) was parallelized with 8 proceses and AFQMC 0 with 160 GPUs. AFQMC I is only shown if the trial was different from AFQMC 0.}
        \label{tab:W4-17_qz_timings}
        \endfirsthead
        \endhead
        \hline    
         W4-17 QZ&  Electrons&  Basis functions&  DLPNO 7&  DLPNO 6&  CCSD(T)&  AFQMC 0& AFQMC I\\ \hline
beta-lactim	&	38	&	630	&	6.04	&	2.61	&	36.18	&	14.31	&	\\ \hline
borole	&	34	&	630	&	5.37	&	2.24	&	33.65	&	22.84	&	\\ \hline
c2cl2	&	46	&	428	&	1.68	&	0.65	&	7.99	&	10.03	&	\\ \hline
c2cl6	&	114	&	964	&	35.74	&	23.25	&	N/A	&	66.02	&	\\ \hline
c2clh	&	30	&	340	&	0.7	&	0.25	&	10.02	&	4.97	&	\\ \hline
ccl2o	&	48	&	428	&	1.78	&	0.76	&	106.64	&	4.63	&	\\ \hline
cf2cl2	&	58	&	566	&	4.03	&	2.18	&	5.65	&	9.96	&	\\ \hline
ch2clf	&	34	&	415	&	1.01	&	0.42	&	11.92	&	3.78	&	\\ \hline
ch3ph2	&	26	&	444	&	0.9	&	0.32	&	49.67	&	5.67	&	\\ \hline
cis-c2f2cl2	&	64	&	646	&	7.82	&	4.3	&	5.52	&	24.46	&	\\ \hline
clcof	&	40	&	403	&	1.34	&	0.6	&	86.72	&	12.63	&	4.40\\ \hline
clf5\_MR	&	62	&	679	&	9.99	&	4.95	&	9.13	&	73.68	&	266.89\\ \hline
cloocl\_MR	&	50	&	428	&	1.97	&	0.78	&	34.75	&	12.44	&	\\ \hline
cyclobutadiene	&	28	&	504	&	2.51	&	0.93	&	72.91	&	15.75	&	7.74\\ \hline
cyclopentadiene	&	36	&	676	&	6.86	&	3.6	&	6.91	&	12.86	&	\\ \hline
dioxetan2one	&	38	&	492	&	3.29	&	1.48	&	39.06	&	22.19	&	13.92\\ \hline
dioxetane	&	32	&	504	&	2.3	&	1.02	&	237.1	&	10.53	&	\\ \hline
dithiotane	&	48	&	612	&	4.04	&	2.22	&	0.79	&	12.08	&	\\ \hline
fno	&	24	&	269	&	0.59	&	0.18	&	3.11	&	5.88	&	\\ \hline
formamide	&	24	&	378	&	1	&	0.34	&	28.26	&	6.15	&	\\ \hline
formic-anhydride	&	38	&	492	&	2.88	&	1.21	&	41.62	&	12.54	&	\\ \hline
hclo4	&	50	&	500	&	3.9	&	1.75	&	8.95	&	23.66	&	78.80\\ \hline
hoclo2	&	42	&	420	&	2.12	&	0.82	&	2.44	&	12.84	&	102.42\\ \hline
hoclo	&	34	&	340	&	0.98	&	0.3	&	34.54	&	7.45	&	34.87\\ \hline
n2o4	&	46	&	480	&	5.22	&	2.12	&	28.52	&	22.23	&	45.96\\ \hline
nh2f	&	18	&	281	&	0.39	&	0.11	&	0.83	&	3.04	&	\\ \hline
nh2oh	&	18	&	298	&	0.47	&	0.12	&	33.38	&	2.71	&	\\ \hline
oxadiazole	&	36	&	492	&	4.35	&	1.48	&	22.44	&	13.48	&	\\ \hline
oxetane	&	32	&	596	&	3.57	&	1.68	&	155.49	&	18.10	&	\\ \hline
silole	&	44	&	730	&	7.41	&	3.67	&	47.21	&	10.49	&	22.62\\ \hline
tetrahedrane	&	28	&	504	&	2.66	&	1.03	&	49.44	&	9.48	&	\\ \hline
trans-c2f2cl2	&	64	&	646	&	8.03	&	4.34	&	51.49	&	13.35	&	\\ \hline
\end{longtable}
}

\begin{table}
    \centering
    \small
    \begin{tabular}{|ccccccc|} \hline
         TZ&  Correlated electrons&  Basis functions&  DLPNO 7&  DLPNO 6&  CCSD(T)&  AFQMC 0\\  \hline
         methane&  8&  138&  0.04&  0.01&  0.01&  0.35\\ \hline
         ethane&  14&  230&  0.12&  0.04&  0.12&  1.20\\ \hline
         propane&  20&  322&  0.31&  0.13&  7.74&  1.62\\ \hline
         butane&  26&  414&  0.77&  0.43&  11.86&  3.13\\ \hline
         pentane&  32&  506&  1.51&  0.81&  19.97&  3.74\\ \hline
         hexane&  38&  598&  2.45&  1.63&  34.90&  10.44\\ \hline
         heptane&  44&  690&  4.14&  2.65&  N/A&  14.01\\ \hline
         octane&  50&  782&  5.27&  3.31&  N/A&  9.12\\ \hline
    \end{tabular}
    \caption{Linear hydrocarbons TZ (as used in the main text) timings (excluding Hartree-Fock times) in CPU hours for the coupled-cluster methods and GPU hours for AFQMC 0 (scaled to fixed 1 mHa statistical error). DLPNO-CCSD(T) was parallelized with 8 proceses and AFQMC 0 with 80 GPUs.}
    \label{tab:linear_hydrocarbon_timings_tz}
\end{table}

\begin{table}
    \centering
    \small
    \begin{tabular}{|ccccccc|} \hline
         QZ&  Correlated electrons&  Basis functions&  DLPNO 7&  DLPNO 6&  CCSD(T)&  AFQMC 0\\  \hline
         methane&  8&  264
&  0.15
&  0.03
&  8.90
&  0.45
\\ \hline
         ethane&  14&  436
&  0.42
&  0.16
&  17.19
&  1.97
\\ \hline
         propane&  20&  608
&  1.39
&  0.73
&  279.44&  2.42
\\ \hline
         butane&  26&  780
&  3.48
&  2.26
&  &  5.57
\\ \hline
         pentane&  32&  952
&  7.97
&  4.42
&  &  14.23
\\ \hline
         hexane&  38&  1124
&  14.11
&  10.15
&  &  38.66
\\ \hline
         heptane&  44&  1296
&  21.96
&  13.57
&  &  75.81
\\ \hline
         octane&  50&  1468&  16.46&  9.90&  &  191.78\\ \hline
    \end{tabular}
    \caption{Linear hydrocarbons QZ (as used in the main text) timings (excluding Hartree-Fock times) in CPU hours for the coupled-cluster methods and GPU hours for AFQMC 0 (scaled to fixed 1 mHa statistical error). DLPNO-CCSD(T) was parallelized with 8 processes (except octane which uses 4) and AFQMC 0 with 160 GPUs.}
    \label{tab:linear_hydrocarbon_timings_qz}
\end{table}

\begin{figure}
    \centering
    \includegraphics[width=0.8\textwidth]{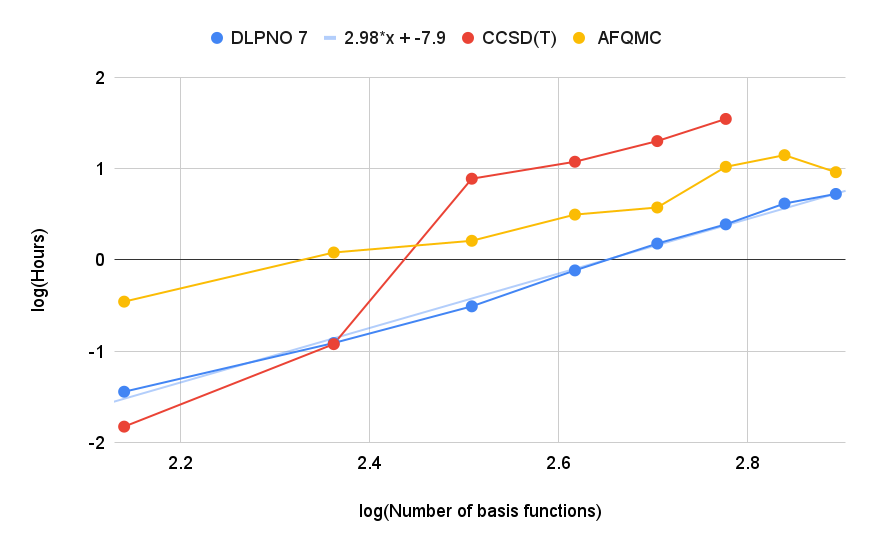}
    \caption{Timings (on a $\log_{10}$ scale) of DLPNO-CCSD(T) with TCutPNO = 7, CCSD(T), and AFQMC 0 in hours (CPU or GPU hours), excluding HF reference generation, of a the series of linear hydrocarbons as written in Table~\ref{tab:linear_hydrocarbon_timings_tz}. DLPNO-CCSD(T) shows approximately cubic scaling.}
    \label{fig:linear_hydrocarbon_timings_tz}
\end{figure}

\begin{figure}
    \centering
    \includegraphics[width=0.8\textwidth]{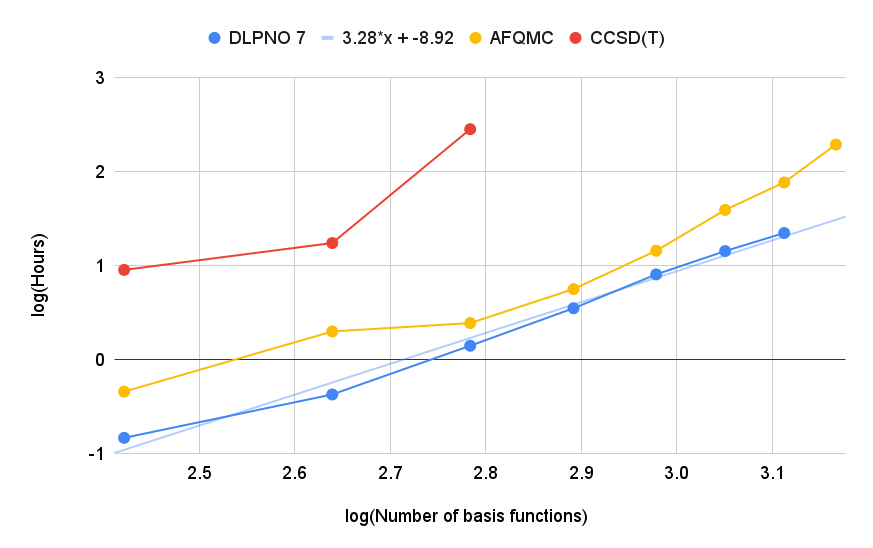}
    \caption{Timings (on a $\log_{10}$ scale) of DLPNO-CCSD(T) with TCutPNO = 7, CCSD(T), and AFQMC 0 in hours (CPU or GPU hours), excluding HF reference generation, of a the series of linear hydrocarbons as written in Table~\ref{tab:linear_hydrocarbon_timings_qz}. DLPNO-CCSD(T) shows approximately cubic scaling. Octane is not included in the DLPNO-CCSD(T) fit because it required more memory per process so we used 4 parallel threads instead of 8 (as for the rest) as well as allocated more memory.}
    \label{fig:linear_hydrocarbon_timings_qz}
\end{figure}

Tables~\ref{tab:linear_hydrocarbon_timings_tz} and \ref{tab:linear_hydrocarbon_timings_qz} show CPU and GPU hours for a series of linear hydrocarbons of increasing size using the TZ and QZ basis set as described in the main text. DLPNO-CCSD(T) scales approximately cubicly with respect to the number of basis functions (as shown in Figures~\ref{fig:linear_hydrocarbon_timings_tz} and~\ref{fig:linear_hydrocarbon_timings_qz}). CCSD(T) shows a less smooth scaling trajectory possibly due to disk or memory requirement differences. In particular, for hexane we had to reduce the number of parallel processes from 8 to 2 to allow for more memory per process which is expected to decrease the total CPU hours taken. The AFQMC scaling is also more noisy as compression and batching of walkers can cause non-linear changes in calculation speed.

\clearpage

\section{11. Multireference Diagnostics}
\addcontentsline{toc}{section}{11. Multireference Diagnostics}

In categorizing our datasets, as described in main text Section 2.1, we further identify a multireference (MR) subset from the collection of 259 molecules. The criteria for classifying molecules into the MR subset incorporate a suite of diagnostics: the TAE(T) diagnostic as cited by Karton et al.~\cite{Karton_W417}, the $T_1$ ~\cite{T1_diagnostic}and $D_1$~\cite{D1_diagnostic} diagnostics, the measure of spin contamination (expressed as $\Delta \langle S^2\rangle = \langle S^2\rangle_{\text{unrestricted calculation}} - \langle S^2\rangle_{\text{exact}}$)~\cite{Pulay1988,Lee2019,neugebauer2023toward}, and the $1-c_0^2$~\cite{Langhoff1974} metric from CASCI calculations. A molecule is typically classified into the MR category if it demonstrates a TAE(T) > 10, $T_1$ > 0.02, $D_1$ > 0.05, regularized $\Delta \expectation{S^2}{} > 0.05$~\cite{neugebauer2023toward} and $1-c_0^2$ > 0.1. Although no single diagnostic can be solely relied upon to identify the multireference character of a molecule, together, these diagnostics offer detailed insight, with each one emphasizing unique aspects. %\textbf{There is no one definitive diagnostic and the above multireference diagnostics describe physically different properties are are often seem as complementary}. 
For example, for the coupled cluster based diagnostics, while $T_1$ and $D_1$ are both based on single excitations and are well correlated~\cite{Wang2015}, $D_1$ emphasizes correlation in more local regions of the molecule. Meanwhile, TAE(T) correlates better with CCSDTQ5 energy from connected quadruple and quintuple excitations~\cite{Karton_W411}. On the other hand, spin contamination refers to the mixing of low-lying excited states with higher multiplicity into the ground state, and $1-c_0^2$ is an indication of the insufficiency of the reference wavefunction for capturing static correlation in CAS methods but limited by the active space.
$T_1$ and $D_1$ multireference diagnostics are calculated with CCSD and the aug-cc-pVTZ-DK/aug-cc-pCVTZ-DK basis set with frozen core (the same as mentioned in main text Section 2.6 for "TZ"). Spin contamination, $\Delta \langle S^2\rangle = \langle S^2\rangle_\text{PBE0} - \langle S^2\rangle_{\text{exact}}$, is carried out using unrestricted Kohn-Sham DFT using the PBE0 functional\cite{weigend2005a} and def2-SVP\cite{Adamo1999} basis set, as shown by Neugebauer et. al~\cite{neugebauer2023toward} to be negligibly different from that obtained by a larger basis set. We use the same regularization as Neugebauer, which for closed shell is a factor of $\frac{1}{0.75}$. The $1-c_0^2$ diagnostic is obtained from the AFQMC 0 leading CI coefficient contribution ($c_0$) calculated using aug-cc-pVTZ-DK/aug-cc-pCVTZ-DK basis sets (same basis set combination as mentioned in the main text Section 2.6).

\begin{table}
    \footnotesize 
    \centering
    \begin{tabular}{|c|l|l|c|c|c|c|c|c|c|c|} \hline 
         Molecule&   Dataset& AFQMC I&AFQMC  II&  DLPNO&  CCSD(T)&  TAE(T)&  $T_1$&  $D_1$&   $\Delta \expectation{S^2}{}$& $1-c_0^2$\\ \hline 
         BN&   W4-11& -1.13&-1.13&  -0.53&  0.85&  18.8&  0.072&  0.197&  1.36& 0.09
\\ \hline 
         O3&   G2& -2.57&-0.96&  -3.24&  -2.07&  17.4&  0.027&  0.075&  0.59& 0.10
\\ \hline 
         FOOF&   W4-11& -0.69&0.60&  -1.12&  -0.57&  16.9&  0.027&  0.090&  0.00& 0.08
\\ \hline 
         ClF5&   W4-17& -1.30&-1.30&  -0.93&  0.12&  14.8&  0.017&  0.054&  0& 0.01
\\ \hline 
         C2&   W4-11& -1.07&-1.07&  -1.40&  -0.54&  13.3&  0.038&  0.085&  1.27& 0.18
\\ \hline 
         ClF3&   G2& -1.80&-1.80&  -1.25&  -0.85&  12.8&  0.018&  0.056&  0& 0.01
\\ \hline 
         ClOOCl&   W4-17& -1.38&-1.38&  0.47& 0.66&  12.2&  0.019&  0.058&  0& 0.01
\\ \hline 
         S4&   W4-11& 0.07&-1.44&  -1.26&  0.66&  12.2&  0.023&  0.089&  0.79& 0.15
\\ \hline 
         S3&   W4-11& -0.71&-0.34&  0.05&  0.82&  10.2&  0.022&  0.053&  0.09& 0.07
\\ \hline 
 N2O4&  W4-17& -1.47&-1.09& 0.03& 1.67& 9.1& 0.021& 0.069& 0&0.12\\ \hline
    \end{tabular}
    \caption{List of the molecules we have determined to be MR, their deviations in each respective method AFQMC I, AFQMC II, DLPNO-CCSD(T), and CCSD(T) against reference heat of formation, in kcal/mol. In comparison, the Multireference diagnostics TAE(T), $T_1$, $D_1$, $\Delta S^2$ (regularized)\cite{neugebauer2023toward}, and $1-c_0^2$ are also shown for each molecule. Multireference molecules are considered as TAE(T) > 10~\cite{Karton_W417}, $T_1$ > 0.02, $D_1$ > 0.05, regularized $\Delta \expectation{S^2}{} > 0.05$~ and $1-c_0^2$ > 0.1. The list is ordered according to TAE(T) as in Karton et al.~\cite{Karton_W417}.}
    \label{tab:MR_molecules}
\end{table}

To identify the multireference (MR) subset, we initially utilized the TAEMR set as defined by Karton et al.~\cite{Karton_W411,Karton_W417} , having a threshold of TAE(T) > 10. From this set, we excluded F2O and Cl2O due to their failure to meet the specified thresholds for across diagnostics, exhibiting relatively low values. Conversely, we included N2O4 in the set because it demonstrated high values for the other criteria. Table~\ref{tab:MR_molecules} presents the finalized MR subset comprising 10 molecules. It details the original dataset to which each molecule belongs (categorized as explained in main text Section 2.1 and listed in the .xlsx document provided), the deviations observed using wavefunction methods, and the values for each MR diagnostic. Although TAE(T) was our primary diagnostic in alignment with the approach of Karton et al., for three molecules in our MR subset (ClF5, ClF3, and ClOOCl), only one additional diagnostic ($D_1$) surpassed the threshold. For the remainder, at least two diagnostics exceeded the established thresholds. Table~\ref{tab:MR_all} lists the MR diagnostics for the entire molecule list.

{
\footnotesize
\begin{longtable}{|p{4cm}p{2.5cm}p{2.5cm}p{2.5cm}p{2.5cm}p{0cm}|}
\caption{MR diagnostics for the entire combined dataset.} % needs to go inside longtable environment
\label{tab:MR_all}
\endfirsthead
\endhead
\hline
Molecule	&	$T_1$&  $D_1$&   $\Delta \expectation{S^2}{}$& $1-c_0^2$	& \\ \hline
2-butyne	&	0.0106	&	0.026	&	0	&	0	& \\ \hline
Acetaldehyde	&	0.0146	&	0.047	&	0	&	0	& \\ \hline
Acethylene	&	0.0130	&	0.028	&	0	&	0	& \\ \hline
Acetone	&	0.0136	&	0.049	&	0	&	0	& \\ \hline
AlCl3	&	0.0081	&	0.023	&	0	&	0	& \\ \hline
AlF3	&	0.0130	&	0.034	&	0	&	0	& \\ \hline
Allene	&	0.0121	&	0.028	&	0	&	0	& \\ \hline
Aziridine	&	0.0092	&	0.022	&	0	&	0	& \\ \hline
BCl3	&	0.0112	&	0.045	&	0	&	0	& \\ \hline
Benzene	&	0.0101	&	0.027	&	0	&	0.0140	& \\ \hline
BF3	&	0.0123	&	0.043	&	0	&	0	& \\ \hline
Bicyclo-1-1-0-butane	&	0.0084	&	0.016	&	0	&	0	& \\ \hline
CCl2CCl2	&	0.0108	&	0.034	&	0	&	0.0156	& \\ \hline
CCl4	&	0.0106	&	0.026	&	0	&	0	& \\ \hline
CF2CF2	&	0.0132	&	0.041	&	0	&	0	& \\ \hline
CF3-CN	&	0.0136	&	0.034	&	0	&	0	& \\ \hline
CF4	&	0.0120	&	0.032	&	0	&	0	& \\ \hline
CH2CH-CN	&	0.0137	&	0.038	&	0	&	0.0087	& \\ \hline
CH2Cl2	&	0.0091	&	0.023	&	0	&	0	& \\ \hline
CH2F2	&	0.0118	&	0.029	&	0	&	0	& \\ \hline
CH3-CH2-CH2-Cl	&	0.0087	&	0.024	&	0	&	0	& \\ \hline
CH3-CH2-Cl	&	0.0086	&	0.023	&	0	&	0	& \\ \hline
CH3-CH2-O-CH3	&	0.0102	&	0.027	&	0	&	0	& \\ \hline
CH3-CH2-SH	&	0.0095	&	0.022	&	0	&	0	& \\ \hline
CH3-CN	&	0.0126	&	0.029	&	0	&	0	& \\ \hline
CH3-O-CH3	&	0.0102	&	0.026	&	0	&	0	& \\ \hline
CH3-O-NO	&	0.0213	&	0.073	&	0	&	0.0188	& \\ \hline
CH3-S-CH3	&	0.0095	&	0.022	&	0	&	0	& \\ \hline
CH3-SH	&	0.0096	&	0.021	&	0	&	0	& \\ \hline
CH3-SiH3	&	0.0100	&	0.019	&	0	&	0	& \\ \hline
CH3CFO	&	0.0147	&	0.049	&	0	&	0	& \\ \hline
CH3Cl	&	0.0081	&	0.021	&	0	&	0	& \\ \hline
CH3COCl	&	0.0147	&	0.046	&	0	&	0	& \\ \hline
CH3CONH2	&	0.0147	&	0.054	&	0	&	0	& \\ \hline
CH3COOH	&	0.0151	&	0.054	&	0	&	0	& \\ \hline
CH3NO2	&	0.0186	&	0.067	&	0	&	0.0266	& \\ \hline
CH4	&	0.0076	&	0.012	&	0	&	0	& \\ \hline
CHCl3	&	0.0100	&	0.024	&	0	&	0	& \\ \hline
Cl2	&	0.0090	&	0.022	&	0	&	0.0098	& \\ \hline
ClF	&	0.0129	&	0.032	&	0	&	0.0068	& \\ \hline
CLF3	&	0.0182	&	0.056	&	0	&	0.0125	& \\ \hline
ClNO	&	0.0217	&	0.062	&	0	&	0.0606	& \\ \hline
CO	&	0.0188	&	0.039	&	0	&	0	& \\ \hline
CO2	&	0.0183	&	0.047	&	0	&	0	& \\ \hline
CS	&	0.0255	&	0.050	&	0	&	0.0303	& \\ \hline
CS2	&	0.0200	&	0.050	&	0	&	0.0395	& \\ \hline
Cyanogen	&	0.0151	&	0.029	&	0	&	0.0468	& \\ \hline
Cyclobutane	&	0.0084	&	0.018	&	0	&	0	& \\ \hline
Cyclobutene	&	0.0102	&	0.029	&	0	&	0	& \\ \hline
Cyclopropane	&	0.0078	&	0.017	&	0	&	0	& \\ \hline
Cyclopropene	&	0.0100	&	0.029	&	0	&	0	& \\ \hline
Dimethylamine	&	0.0090	&	0.023	&	0	&	0	& \\ \hline
Dimethylsulfoxide	&	0.0158	&	0.047	&	0	&	0	& \\ \hline
Ethane	&	0.0079	&	0.014	&	0	&	0	& \\ \hline
Ethanol	&	0.0099	&	0.025	&	0	&	0	& \\ \hline
Ethenone	&	0.0168	&	0.046	&	0	&	0	& \\ \hline
Ethylene	&	0.0108	&	0.031	&	0	&	0	& \\ \hline
F2	&	0.0125	&	0.028	&	0	&	0.0153	& \\ \hline
F2O	&	0.0165	&	0.042	&	0	&	0	& \\ \hline
Furan	&	0.0135	&	0.043	&	0	&	0	& \\ \hline
Glyoxal	&	0.0164	&	0.049	&	0	&	0.0150	& \\ \hline
H2	&	0.0053	&	0.008	&	0	&	0	& \\ \hline
H2CO	&	0.0157	&	0.045	&	0	&	0	& \\ \hline
H2NNH2	&	0.0094	&	0.021	&	0	&	0	& \\ \hline
H2O	&	0.0102	&	0.022	&	0	&	0	& \\ \hline
HCF3	&	0.0123	&	0.031	&	0	&	0	& \\ \hline
HCl	&	0.0064	&	0.012	&	0	&	0	& \\ \hline
HCN	&	0.0144	&	0.028	&	0	&	0	& \\ \hline
HCOOCH3	&	0.0158	&	0.057	&	0	&	0	& \\ \hline
HCOOH	&	0.0168	&	0.054	&	0	&	0	& \\ \hline
HF	&	0.0102	&	0.018	&	0	&	0	& \\ \hline
HOCl	&	0.0120	&	0.026	&	0	&	0	& \\ \hline
HOOH	&	0.0127	&	0.025	&	0	&	0	& \\ \hline
Isobutane	&	0.0084	&	0.016	&	0	&	0	& \\ \hline
Isobutene	&	0.0100	&	0.031	&	0	&	0	& \\ \hline
Isopropyl-alcohol	&	0.0100	&	0.027	&	0	&	0	& \\ \hline
Ketene	&	0.0168	&	0.046	&	0	&	0	& \\ \hline
Li2	&	0.0112	&	0.027	&	0	&	0	& \\ \hline
LiF	&	0.0139	&	0.028	&	0	&	0	& \\ \hline
LiH	&	0.0070	&	0.014	&	0	&	0	& \\ \hline
Methanol	&	0.0100	&	0.023	&	0	&	0	& \\ \hline
Methylamine	&	0.0087	&	0.020	&	0	&	0	& \\ \hline
Methylene-cyclopropane	&	0.0101	&	0.031	&	0	&	0	& \\ \hline
N2	&	0.0134	&	0.026	&	0	&	0	& \\ \hline
Na2	&	0.0104	&	0.041	&	0	&	0.0157	& \\ \hline
NaCl	&	0.0060	&	0.017	&	0	&	0	& \\ \hline
NF3	&	0.0167	&	0.044	&	0	&	0	& \\ \hline
NH3	&	0.0084	&	0.019	&	0	&	0	& \\ \hline
NNO	&	0.0202	&	0.048	&	0	&	0	& \\ \hline
OCS-m1	&	0.0193	&	0.049	&	0	&	0	& \\ \hline
Oxirane	&	0.0113	&	0.028	&	0	&	0	& \\ \hline
Ozone	&	0.0274	&	0.075	&	0.59	&	0.1024	& \\ \hline
P2	&	0.0176	&	0.033	&	0	&	0.0521	& \\ \hline
PF3	&	0.0146	&	0.038	&	0	&	0	& \\ \hline
PH3	&	0.0137	&	0.022	&	0	&	0	& \\ \hline
Propane	&	0.0082	&	0.015	&	0	&	0	& \\ \hline
Propene-CS	&	0.0102	&	0.031	&	0	&	0	& \\ \hline
Propyne	&	0.0114	&	0.027	&	0	&	0	& \\ \hline
Pyridine	&	0.0119	&	0.034	&	0	&	0.0160	& \\ \hline
Pyrole	&	0.0112	&	0.031	&	0	&	0	& \\ \hline
SH2	&	0.0102	&	0.018	&	0	&	0	& \\ \hline
Si2H6	&	0.0129	&	0.024	&	0	&	0	& \\ \hline
SiCl4	&	0.0091	&	0.023	&	0	&	0	& \\ \hline
SiF4	&	0.0119	&	0.030	&	0	&	0	& \\ \hline
SiH4	&	0.0111	&	0.017	&	0	&	0	& \\ \hline
SiO	&	0.0264	&	0.055	&	0	&	0	& \\ \hline
SO2	&	0.0226	&	0.059	&	0	&	0.0238	& \\ \hline
Spiropentane	&	0.0084	&	0.019	&	0	&	0	& \\ \hline
Thiooxirane	&	0.0100	&	0.022	&	0	&	0	& \\ \hline
Thiophene	&	0.0131	&	0.037	&	0	&	0	& \\ \hline
Trans-1-3-butadiene	&	0.0113	&	0.035	&	0	&	0	& \\ \hline
Trans-butane	&	0.0084	&	0.016	&	0	&	0	& \\ \hline
Trans-ethylamine	&	0.0090	&	0.023	&	0	&	0	& \\ \hline
Trimethyl-amine	&	0.0096	&	0.027	&	0	&	0	& \\ \hline
Vinyl-chloride	&	0.0106	&	0.029	&	0	&	0	& \\ \hline
Vynil-fluoride	&	0.0124	&	0.029	&	0	&	0	& \\ \hline
1,3-cyclohexadiene	&	0.0108	&	0.034	&	0	&	0	& \\ \hline
1,3-DiFluorobenzene	&	0.0120	&	0.030	&	0	&	0.0080	& \\ \hline
1,4-DiFluorobenzene	&	0.0118	&	0.033	&	0	&	0.0188	& \\ \hline
2-methyl	&	0.0125	&	0.038	&	0	&	0	& \\ \hline
2,5-Dihydrothiophene	&	0.0112	&	0.029	&	0	&	0	& \\ \hline
3-methyl	&	0.0087	&	0.017	&	0	&	0	& \\ \hline
Acetic	&	0.0157	&	0.052	&	0	&	0	& \\ \hline
azulene	&	0.0116	&	0.042	&	0	&	0.0606	& \\ \hline
benzoquinone	&	0.0158	&	0.061	&	0	&	0.0300	& \\ \hline
c2f6	&	0.0127	&	0.038	&	0	&	0	& \\ \hline
C4H4N2	&	0.0115	&	0.028	&	0	&	0.0358	& \\ \hline
C4H6	&	0.0113	&	0.029	&	0	&	0	& \\ \hline
C4H6O	&	0.0133	&	0.040	&	0	&	0	& \\ \hline
C4H8O2	&	0.0117	&	0.033	&	0	&	0	& \\ \hline
C5H8	&	0.0109	&	0.033	&	0	&	0	& \\ \hline
C6H12	&	0.0088	&	0.017	&	0	&	0	& \\ \hline
C6H5-CH3	&	0.0100	&	0.027	&	0	&	0.0035	& \\ \hline
C6H5-NH2	&	0.0113	&	0.034	&	0	&	0	& \\ \hline
C6H5-OH	&	0.0116	&	0.034	&	0	&	0	& \\ \hline
cf3cl	&	0.0120	&	0.033	&	0	&	0	& \\ \hline
CH3\_2CH-CHO	&	0.0128	&	0.048	&	0	&	0	& \\ \hline
CH3\_2CH-CN	&	0.0112	&	0.029	&	0	&	0	& \\ \hline
CH3\_2CH-O-CH\_CH3\_2	&	0.0102	&	0.031	&	0	&	0	& \\ \hline
CH3\_3C-NH2	&	0.0092	&	0.025	&	0	&	0	& \\ \hline
CH3\_3C-O-CH3	&	0.0102	&	0.029	&	0	&	0	& \\ \hline
CH3\_3C-SH	&	0.0096	&	0.024	&	0	&	0	& \\ \hline
CH3-C\_O\_-CCH	&	0.0150	&	0.048	&	0	&	0.0090	& \\ \hline
CH3-C\_O\_-O-CH\_CH3\_2	&	0.0135	&	0.058	&	0	&	0	& \\ \hline
CH3-C\_O\_-OCH3	&	0.0146	&	0.057	&	0	&	0	& \\ \hline
CH3-CH\_OCH3\_2	&	0.0113	&	0.034	&	0	&	0	& \\ \hline
CH3-CH2-CH\_CH3\_-NO2	&	0.0155	&	0.068	&	0	&	0.0248	& \\ \hline
CH3-CH2-CO-CH2-CH3	&	0.0124	&	0.049	&	0	&	0	& \\ \hline
CH3-CH2-O-CH2-CH3	&	0.0101	&	0.028	&	0	&	0	& \\ \hline
CH3-CH2-S-S-CH2-CH3	&	0.0112	&	0.029	&	0	&	0	& \\ \hline
CH3-CHCH-CHO	&	0.0145	&	0.052	&	0	&	0.0096	& \\ \hline
CH3-CO-CH2-CH3	&	0.0129	&	0.049	&	0	&	0	& \\ \hline
Chlorobenzene	&	0.0102	&	0.026	&	0	&	0.0144	& \\ \hline
Cl2O2S	&	0.0182	&	0.061	&	0	&	0.0069	& \\ \hline
Cl2S2	&	0.0185	&	0.056	&	0	&	0.0131	& \\ \hline
cyclooctatetraene	&	0.0116	&	0.033	&	0	&	0	& \\ \hline
cyclopentane	&	0.0088	&	0.018	&	0	&	0	& \\ \hline
cyclopentanone	&	0.0129	&	0.051	&	0	&	0	& \\ \hline
dimethyl	&	0.0164	&	0.050	&	0	&	0	& \\ \hline
Fluorobenzene	&	0.0112	&	0.028	&	0	&	0.0049	& \\ \hline
n-Butyl	&	0.0088	&	0.024	&	0	&	0	& \\ \hline
n-heptane	&	0.0083	&	0.017	&	0	&	0	& \\ \hline
n-hexane	&	0.0086	&	0.016	&	0	&	0	& \\ \hline
N-methyl	&	0.0110	&	0.033	&	0	&	0	& \\ \hline
n-octane	&	0.0083	&	0.017	&	0	&	0	& \\ \hline
n-pentane	&	0.0085	&	0.016	&	0	&	0	& \\ \hline
Naphthalene	&	0.0100	&	0.032	&	0	&	0.0293	& \\ \hline
NC-CH2-CH2-CN	&	0.0130	&	0.029	&	0	&	0	& \\ \hline
Neopentane	&	0.0086	&	0.017	&	0	&	0	& \\ \hline
P4	&	0.0184	&	0.038	&	0	&	0	& \\ \hline
para-cyclohexadiene	&	0.0107	&	0.028	&	0	&	0	& \\ \hline
PCl3	&	0.0131	&	0.035	&	0	&	0	& \\ \hline
PCl5	&	0.0116	&	0.033	&	0	&	0	& \\ \hline
Perhydropyridine	&	0.0096	&	0.027	&	0	&	0	& \\ \hline
pf5	&	0.0125	&	0.037	&	0	&	0	& \\ \hline
POCl3	&	0.0143	&	0.044	&	0	&	0	& \\ \hline
pyrimidine	&	0.0146	&	0.045	&	0	&	0.0183	& \\ \hline
SCl2	&	0.0127	&	0.033	&	0	&	0	& \\ \hline
sf6	&	0.0128	&	0.032	&	0	&	0	& \\ \hline
SiCl2	&	0.0133	&	0.037	&	0	&	0.0106	& \\ \hline
SO3	&	0.0183	&	0.056	&	0	&	0.0138	& \\ \hline
t-butanol	&	0.0099	&	0.028	&	0	&	0	& \\ \hline
t-Butyl	&	0.0092	&	0.027	&	0	&	0	& \\ \hline
tetrahydrofuran	&	0.0104	&	0.028	&	0	&	0	& \\ \hline
Tetrahydropyran	&	0.0104	&	0.031	&	0	&	0	& \\ \hline
Tetrahydropyrrole	&	0.0096	&	0.026	&	0	&	0	& \\ \hline
Tetrahydrothiophene	&	0.0101	&	0.025	&	0	&	0	& \\ \hline
Tetrahydrothiopyran	&	0.0100	&	0.024	&	0	&	0	& \\ \hline
Tetramethylsilane	&	0.0093	&	0.020	&	0	&	0	& \\ \hline
alcl	&	0.012	&	0.026	&	0	&	0	& \\ \hline
alf	&	0.016	&	0.029	&	0	&	0	& \\ \hline
alh	&	0.014	&	0.026	&	0	&	0	& \\ \hline
alh3	&	0.008	&	0.015	&	0	&	0	& \\ \hline
b2h6	&	0.010	&	0.018	&	0	&	0	& \\ \hline
bf	&	0.016	&	0.030	&	0	&	0	& \\ \hline
bh	&	0.014	&	0.026	&	0.33	&	0	& \\ \hline
bh3	&	0.006	&	0.010	&	0	&	0	& \\ \hline
bhf2	&	0.013	&	0.043	&	0	&	0	& \\ \hline
c-hono	&	0.022	&	0.065	&	0	&	0.0214	& \\ \hline
c-n2h2	&	0.013	&	0.033	&	0	&	0.0118	& \\ \hline
ch2nh	&	0.012	&	0.034	&	0	&	0	& \\ \hline
ch3f	&	0.010	&	0.021	&	0	&	0	& \\ \hline
clcn	&	0.014	&	0.028	&	0	&	0	& \\ \hline
dioxirane	&	0.015	&	0.035	&	0	&	0	& \\ \hline
f2co	&	0.015	&	0.050	&	0	&	0	& \\ \hline
fccf	&	0.013	&	0.030	&	0	&	0	& \\ \hline
hccf	&	0.013	&	0.026	&	0	&	0	& \\ \hline
hcno	&	0.021	&	0.052	&	0	&	0	& \\ \hline
hcof	&	0.016	&	0.048	&	0	&	0	& \\ \hline
hnco	&	0.018	&	0.052	&	0	&	0	& \\ \hline
hnnn	&	0.020	&	0.054	&	0	&	0	& \\ \hline
hno	&	0.016	&	0.042	&	0.37	&	0.0200	& \\ \hline
hocn	&	0.015	&	0.033	&	0	&	0	& \\ \hline
hof	&	0.014	&	0.035	&	0	&	0	& \\ \hline
nh2cl	&	0.010	&	0.024	&	0	&	0	& \\ \hline
oxirene	&	0.014	&	0.032	&	0	&	0	& \\ \hline
s2o	&	0.023	&	0.061	&	0	&	0.0450	& \\ \hline
sih3f	&	0.011	&	0.023	&	0	&	0	& \\ \hline
t-hono	&	0.022	&	0.062	&	0	&	0.0203	& \\ \hline
t-n2h2	&	0.013	&	0.033	&	0	&	0	& \\ \hline
c2h5f	&	0.010	&	0.023	&	0	&	0	& \\ \hline
bn	&	0.072	&	0.197	&	1.36	&	0.0923	& \\ \hline
c2	&	0.038	&	0.085	&	1.27	&	0.1779	& \\ \hline
cl2o	&	0.015	&	0.044	&	0	&	0	& \\ \hline
foof	&	0.027	&	0.090	&	0	&	0.0812	& \\ \hline
s3	&	0.022	&	0.053	&	0.09	&	0.0685	& \\ \hline
s4-c2v	&	0.023	&	0.089	&	0.79	&	0.1518	& \\ \hline
clf5	&	0.017	&	0.054	&	0	&	0.0055	& \\ \hline
cloocl	&	0.019	&	0.058	&	0	&	0.0132	& \\ \hline
beta-lactim	&	0.013	&	0.045	&	0	&	0	& \\ \hline
borole	&	0.012	&	0.034	&	0	&	0	& \\ \hline
c2cl2	&	0.011	&	0.025	&	0	&	0	& \\ \hline
c2cl6	&	0.011	&	0.027	&	0	&	0	& \\ \hline
c2clh	&	0.012	&	0.026	&	0	&	0	& \\ \hline
ccl2o	&	0.015	&	0.047	&	0	&	0	& \\ \hline
cf2cl2	&	0.012	&	0.033	&	0	&	0	& \\ \hline
ch2clf	&	0.011	&	0.028	&	0	&	0	& \\ \hline
ch3ph2	&	0.012	&	0.024	&	0	&	0	& \\ \hline
cis-c2f2cl2	&	0.012	&	0.038	&	0	&	0	& \\ \hline
clcof	&	0.015	&	0.049	&	0	&	0	& \\ \hline
cyclobutadiene	&	0.012	&	0.040	&	0.21	&	0	& \\ \hline
cyclopentadiene	&	0.011	&	0.033	&	0	&	0	& \\ \hline
dioxetan2one	&	0.016	&	0.058	&	0	&	0	& \\ \hline
dioxetane	&	0.011	&	0.029	&	0	&	0	& \\ \hline
dithiotane	&	0.011	&	0.025	&	0	&	0	& \\ \hline
fno	&	0.022	&	0.062	&	0	&	0.0256	& \\ \hline
formamide	&	0.016	&	0.053	&	0	&	0	& \\ \hline
formic-anhydride	&	0.017	&	0.049	&	0	&	0	& \\ \hline
hclo4	&	0.019	&	0.062	&	0	&	0.0086	& \\ \hline
hoclo2	&	0.024	&	0.077	&	0	&	0.0094	& \\ \hline
hoclo	&	0.024	&	0.085	&	0	&	0.0083	& \\ \hline
n2o4	&	0.021	&	0.069	&	0	&	0.1193	& \\ \hline
nh2f	&	0.012	&	0.031	&	0	&	0	& \\ \hline
nh2oh	&	0.011	&	0.024	&	0	&	0	& \\ \hline
oxadiazole	&	0.018	&	0.063	&	0	&	0.0219	& \\ \hline
oxetane	&	0.010	&	0.029	&	0	&	0	& \\ \hline
silole	&	0.013	&	0.032	&	0	&	0.0075	& \\ \hline
tetrahedrane	&	0.009	&	0.017	&	0	&	0	& \\ \hline
trans-c2f2cl2	&	0.012	&	0.039	&	0	&	0	& \\ \hline
\end{longtable}
}

\newpage
\section{12. CCSD(T) DLPNO-CCSD(T) Correlation}
\addcontentsline{toc}{section}{12. CCSD(T) DLPNO-CCSD(T) Correlation}

\begin{figure}
    \centering
    \includegraphics[width=0.8\textwidth]{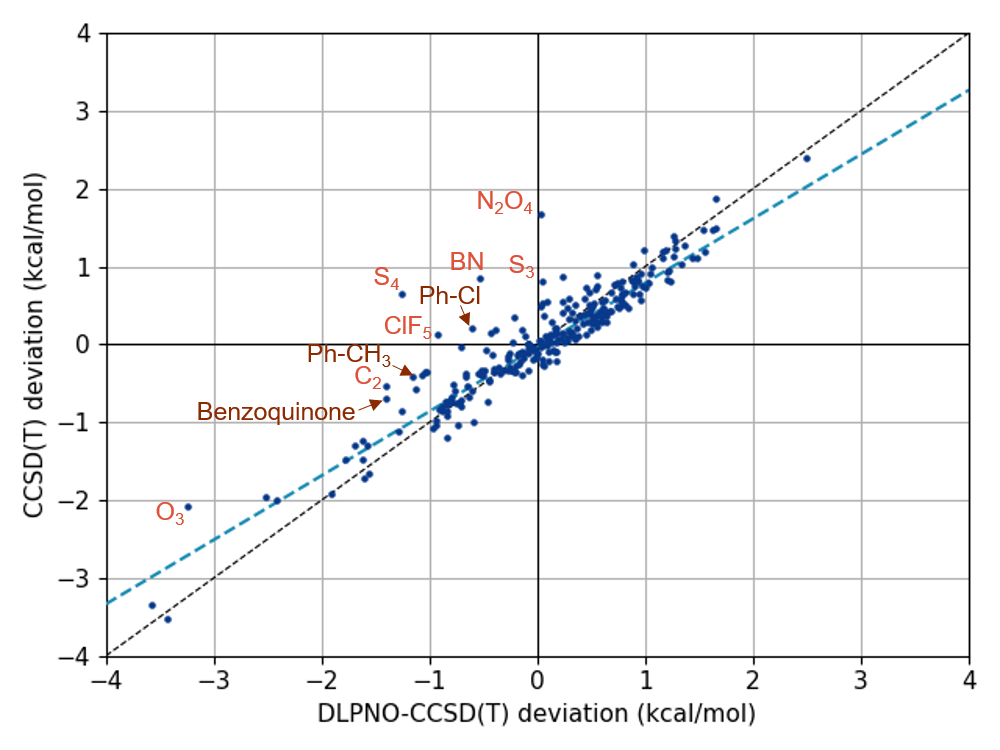}
    \caption{Correlation between DLPNO-CCSD(T) and CCSD(T) deviation of heat of formation from the reference value. DLPNO-CCSD(T) results show overall higher unsigned deviation than CCSD(T), as the line of best fit (blue dashed line) shows a smaller slope (0.83) than the 1:1 line (black dashed line). The coefficient of determination for the DLPNO-CCSD(T) correlation is $R^2=0.86$. The molecules with the top 10 largest differences between the CCSD(T) and DLPNO deviations for heat of formation are marked on the graph, where light red indicates multireference molecules (see Table~\ref{tab:MR_molecules}) and dark red indicates single-reference molecules. These 10 molecules all above below the line $y=x$, which means that the DLPNO-CCSD(T) heat of formation is higher than the CCSD(T) heat of formation.}
    \label{dlpno-ccsdt-correlation}
\end{figure}

Despite the overall accuracy of DLPNO-CCSD(T), there are cases where it does not capture all of the correlation compared to CCSD(T) such as multireference molecules and molecules where electrons are strongly delocalized. The DLPNO-CCSD(T) and CCSD(T) results show a strong positive correlation with a coefficient of determination of the line of best fit $R^2=0.86$, as depicted in Figure~\ref{dlpno-ccsdt-correlation} which plots CCSD(T) deviations against DLPNO-CCSD(T) deviations for the entire dataset.  In the top 10 cases where DLPNO-CCSD(T) deviates from CCSD(T), 7 of them are from our MR subset of 10 molecules (refer to Table~\ref{tab:MR_molecules}), and the remaining 3 are molecules with considerable electron delocalization. Table~3 in the main text lists these molecules in order of absolute atomization energy difference (or difference in deviation) of DLPNO-CCSD(T) and CCSD(T) methods. These cases possibly suggest components of correlation that DLPNO, as an approximation to CCSD(T), is unable to capture. Additionally, the largest deviations of DLPNO-CCSD(T) from CCSD(T) lie above the equivalence line (black dashed line), meaning that the atomization energy of DLPNO-CCSD(T) (opposite in sign to heat of formation) is lower than that of CCSD(T), i.e. the molecule single point energy (effectively, the correlation energy) is underestimated, even after accounting for differences in atomic energies.

\newpage
\section{13. AFQMC 0 Outliers}
\addcontentsline{toc}{section}{13. AFQMC 0 Outliers}

\begin{table}
    \scriptsize
    \centering
    \begin{tabular}{|c|l|c|c|c|c|c|c|} \hline 
         Molecule &Dataset&  Deviation&  First CI AS&  TZ final AS&  QZ final AS& \thead{TZ final \\ \#dets} & \thead{QZ final \\ \#dets}
\\ \hline 
         Bicyclo-1-1-0-butane &G2&  -2.49(48)
& 8e+8e,16o& 1e+1e,1o& 1e+1e,1o&  1& 1
\\ \hline 
         CLF3&G2, MR&  2.42(65)
& 14e+14e,16o& 1e+1e,2o& 1e+1e,2o&  2& 2
\\ \hline 
         F2 &G2&  -2.80(43)
& 7e+7e,8o& 1e+1e,2o& 1e+1e,2o&  2& 2
\\ \hline 
         F2O &G2&  -2.56(70)
& 10e+10e,12o& 1e+1e,1o& 1e+1e,1o&  1& 1
\\ \hline 
         Li2 &G2&  -2.28(19)
& 1e+1e,2o& 1e+1e,1o& 1e+1e,1o&  1& 1
\\ \hline 
         LiF &G2&  3.38(39)
& 4e+4e,5o& 1e+1e,1o& 1e+1e,1o&  1& 1
\\ \hline 
         OCS-m1 &G2&  3.25(67)
& 8e+8e,12o& 1e+1e,1o& 1e+1e,1o&  1& 1
\\ \hline 
         Ozone&G2, MR&  -5.05(51)
& 9e+9e,12o& 2e+2e,3o& 2e+2e,3o&  3& 3
\\ \hline 
         Thiooxirane &G2&  2.04(46)
& 7e+7e,12o& 1e+1e,1o& 1e+1e,1o&  1& 1
\\ \hline
 Trans-1-3-butadiene &G2& -2.03(55)
& 8e+8e,16o& 1e+1e,1o& 1e+1e,1o& 1&1
\\\hline
 1,3-cyclohexadiene &G3& -2.11(61)
& 12e+12e,24o& 1e+1e,1o& 1e+1e,1o& 1&1
\\\hline
 Acetic &G3& 2.10(87)
& 17e+17e,28o& 1e+1e,1o& 1e+1e,1o& 1&1
\\\hline
 azulene &G3& 2.65(74)
& 20e+20e,40o& 3e+3e,5o& 3e+3e,5o& 12&11
\\\hline
 C4H4N2 &G3& -2.39(65)
& 13e+13e,24o& 2e+2e,4o& 2e+2e,4o& 7&6
\\\hline
 3-butyn-2-one  &G3& -4.27(57)
& 11e+11e,20o& 1e+1e,2o& 1e+1e,2o& 2&2
\\\hline
 CH3-CH2-CO-CH2-CH3 &G3& 2.22(65)
& 13e+13e,24o& 1e+1e,1o& 1e+1e,1o& 1&1
\\\hline
 pyrimidine &G3& 3.51(53)
& 13e+13e,24o& 2e+2e,4o& 1e+1e,3o& 6&1
\\\hline
    \end{tabular}
    \caption{Details (G2 and G3 set) for the outliers of AFQMC 0 protocol. Deviation from experimental heat of formation is listed in kcal/mol with statistical error in parentheses. After the first CI is run with an active space based on orbital maps to the atoms of the molecules (refer Table~\ref{tab:orbital maps}) that returns the `First CI AS' listed, the second AS (shown here as `TZ final AS' and `QZ final AS', as the NOONs have a slight basis set dependency due to approximations such as the SHCI solver) is chosen from those orbitals from the first AS that have NOONs of between 0.01 and 1.99.  The final number of determinants is determined by the number of determinants to get to 99.5\% CI coefficient. Statistical errors are shown using parentheses.}
    \label{tab:init_afqmc_outliers_G2G3}
\end{table}

\begin{table}
    \scriptsize
    \centering
    \begin{tabular}{|c|l|c|c|c|c|c|c|} \hline 
         Molecule &Dataset&  Deviation&  First CI AS&  TZ final AS&  QZ final AS&  \thead{TZ final \\  \#dets}& \thead{QZ final \\ \#dets}\\ \hline 
         f2co &W4-11&  2.32(62)& 12e+12e,16o& 1e+1e,1o& 1e+1e,1o&  1& 1
\\ \hline 
         hnco &W4-11&  2.63(70)& 7e+7e,12o& 1e+1e,1o& 1e+1e,1o&  1& 1
\\ \hline 
         bn&W4-11, MR&  -10.51(61)& 4e+4e,8o& 1e+1e,2o& 1e+1e,2o&  2& 2
\\ \hline 
         c2&W4-11, MR&  -14.62(43)& 4e+4e,8o& 1e+1e,2o& 1e+1e,2o&  2& 2
\\ \hline 
         clf5&W4-17, MR&  3.13(104)& 21e+21e,24o& 1e+1e,2o& 1e+1e,1o&  2& 1
\\ \hline 
 clcof &W4-17& 3.00(72)& 12e+12e,16o& 1e+1e,1o& 1e+1e,1o& 1&1
\\ \hline 
 cyclobutadiene &W4-17& -2.04(68)& 8e+8e,16o& 1e+1e,1o& 1e+1e,1o& 1&1
\\ \hline 
 dioxetan2one &W4-17& 3.04(80)& 13e+13e,20o& 1e+1e,1o& 1e+1e,1o& 1&1
\\ \hline 
 hclo4& W4-17& 4.16(82)&  15e+15e,20o
&  1e+1e,2o&  1e+1e,1o& 1&1\\ \hline 
 hoclo2& W4-17& 2.65(71)&  12e+12e,16o
&  1e+1e,2o&  1e+1e,1o& 1&1\\ \hline 
 hoclo& W4-17& 2.15(64)& 9e+9e,12o& 1e+1e,2o& 1e+1e,1o& 1&1\\\hline
 n2o4&W4-17, MR& -2.48(76)& 17e+17e,24o& 7e+7e,10o& 7e+7e,10o& 70&60
\\ \hline 
 silole &W4-17& -2.19(52)& 10e+10e,20o& 1e+1e,2o& 1e+1e,1o& 2&1\\ \hline
    \end{tabular}
    \caption{Details (W4 set) for AFQMC 0 outliers where after the first CI is run, the second AS is chosen from those orbitals from the first AS that have NOONs of between 0.01 and 1.99. The final number determinants is determined by the number of determinants to get to 99.5\% CI coefficient. Refer to Table~\ref{tab:init_afqmc_outliers_G2G3} for molecules in the G2 and G3 datasets.}
    \label{tab:init_afqmc_outliers_W4}
\end{table}

In Table~\ref{tab:init_afqmc_outliers_G2G3}, we list all the outliers from the G2 and G3 datasets for heat of formation, using the AFQMC 0 protocol, and their deviations from the experimental references in kcal/mol, alongside the active spaces of the first CI, and the second (and final) CI after applying the NOON cutoff. The final active spaces determine the number of determinants, which we pick as the number of determinants required to retain 99.5\% of the CI weight. The final active spaces, and determinants chosen, for the TZ and QZ basis sets (see Section 2.6 for the exact "TZ" and "QZ" basis sets used) can be slightly different due to approximations such as the semistochastic heat bath, as well as basis set artefacts such as linear dependencies. Similarly, Table~\ref{tab:init_afqmc_outliers_W4} shows the list of W4 molecules and their deviations. 

\newpage
\bibliography{bibli}